\documentclass[
aps,
pra,
reprint,
superscriptaddress,
longbibliography
]{revtex4-2}

\usepackage{graphicx}
\usepackage{tikz}
\usepackage{xcolor}
\usepackage{dcolumn}
\usepackage{times}
\usepackage{xspace}
\usepackage{amsmath}
\usepackage{amssymb}
\usepackage{bm}
\usepackage{mathrsfs}
\usepackage{upgreek}
\usepackage{physics}
\usepackage[section]{placeins}
\usepackage[normalem]{ulem}
\usepackage{mathtools} %

\usepackage[caption=false]{subfig}
\usepackage[
colorlinks=true,
citecolor=blue,
linkcolor=blue,
urlcolor=blue
]{hyperref}
\usepackage{orcidlink}
\newcommand{\ASTRA}{\textsc{astra}}
\newcommand{\UKRmolplus}{UKRmol+\xspace}
\newcommand{\panelgraphic}[4][0.26,0.94]{%
  \begin{tikzpicture}
    \node[anchor=south west,inner sep=0] (panelimage) at (0,0)
      {\includegraphics[width=#2]{#4}};
    \begin{scope}[x={(panelimage.south east)},y={(panelimage.north west)}]
      \node[anchor=north west,inner sep=1pt,fill=white,
            font=\bfseries\fontsize{8}{9}\selectfont] at (#1) {#3};
    \end{scope}
  \end{tikzpicture}%
}

\begin{document}

\title{Electron Scattering with \ASTRA{}: Helium and Molecular Hydrogen}

\newcommand{\nist}{Theory, Modelling, and Simulation Group, Material Data Division, \\
National Institute of Standards and Technology, 100 Bureau Drive, Gaithersburg, Maryland 20899, USA}
\newcommand{\ucf}{Department of Physics, Chemistry \textit{\&}  CREOL, University of Central Florida, Orlando, Florida 32816, USA}
\newcommand{\ucfphys}{Department of Physics, University of Central Florida, Orlando, Florida 32816, USA}

\newcommand{\georgetown}{Institute for Soft Matter Synthesis and Metrology,
Georgetown University, Washington, DC 20057, USA}
\author{Siddhartha Chattopadhyay\,\orcidlink{0000-0002-2436-3477}}
\affiliation{\nist}
\affiliation{\georgetown}

\author{Jan Dvo\v{r}\'ak\,\orcidlink{0000-0002-3013-8831}}
\affiliation{\ucfphys}

\author{Carlos Marante\,\orcidlink{0000-0002-8254-7805}}
\affiliation{\ucfphys}

\author{Thomas C. Allison\,\orcidlink{0000-0003-4848-4043}}
\affiliation{\nist}

\author{Luca Argenti\,\orcidlink{0000-0003-1093-4379}}
\affiliation{\ucf}
\email{luca.argenti@ucf.edu}

\date{\today}

\begin{abstract}
We present an implementation of electron-impact scattering in \ASTRA{}~(AttoSecond TRAnsitions)~[J. M. Randazzo {\it et al.}, Phys. Rev. Research {\bf 5}, 043115 (2023)], a general close-coupling method for the molecular electronic continuum based on transition density matrices and already employed for molecular photoionization. The scattering states are obtained with a pole-free Galerkin method with variational surface matching, which yields the scattering matrix and, from it, the elastic and electronically inelastic integral cross sections and the molecular-frame differential cross sections. For helium, \ASTRA{} reproduces elastic and inelastic cross sections over a broad energy range, resolves the $\mathrm{He}^{-}(1s2s^2\,{}^2S^e)$ Feshbach resonance in both energy and angle, and reproduces the interference features in the angular distributions for $1s^2\to1s 2\ell$ excitations observed experimentally. For molecular hydrogen, the elastic and inelastic cross sections closely reproduce the benchmark results obtained with \UKRmolplus{} for the same close-coupling expansion. We also report orientation-resolved, three-dimensional molecular-frame angular distributions for state-resolved electron-impact excitation, which are shown to depend strongly on the molecular orientation. These results validate the method in view of its application to larger molecular targets.
\end{abstract}

\maketitle

\section{Introduction\label{sec:introduction}}

Accurate electron--molecule collision data are essential for modeling low-temperature plasmas~\cite{Christophorou2012,Bartschat-PNAS-16}, radiation-driven chemistry~\cite{Alizadeh-ARPC-15,Gaches-AA-24}, planetary and astrophysical environments~\cite{McConkey-PhysRep-08, Campbell2016, Gaches-AA-24}, and electron transport in gaseous, liquid, and condensed media~\cite{Vialetto-Plasma-24,Boyle-FDST-25,Solovyov-ChemRev-24,Taioli-SSR-25}. They are particularly important in semiconductor manufacturing, where electron-impact excitation, ionization, attachment, and dissociation generate the ions and reactive species that control plasma chemistry, energy deposition, and surface modification~\cite{Christophorou2012,Bartschat-PNAS-16}. Predictive plasma simulations therefore require internally consistent sets of elastic, momentum-transfer, excitation, dissociation, attachment, and ionization cross sections over broad ranges of incident energy~\cite{Tennyson-PSST-22,Alves-PSST-23}. For many technologically important species, however, available datasets remain incomplete or contain estimated and swarm-derived effective cross sections introduced to account for missing processes~\cite{Pancheshnyi-ChemPhys-12,Pitchford-PPP-17,Carbone-Atoms-21,Alves-PSST-23}. The problem is severe for complex molecules, for which state-resolved \emph{ab initio} calculations have long been prohibitive~\cite{Bartschat-PPP-17,Song-EPJD-20}.

Electron scattering is also central to strong-field and attosecond science. In the recollision picture of strong-field ionization, an electron released by tunnel ionization can be driven back to its parent system and scatter elastically or inelastically~\cite{Corkum-PRL-93,Schafer-PRL-93}. The resulting high-energy photoelectron distributions encode field-free electron-scattering amplitudes, on which quantitative rescattering theory and laser-induced electron diffraction are based~\cite{Chen-PRA-09,Lin-JPB-10,Meckel-Science-08,Blaga-Nature-12,Wolter-Science-16}.
Electron--neutral scattering is also relevant to photodetachment from negative ions~\cite{Lindroth-PRA-17,Rescigno-PRA-24}, which is sensitive to short-range correlation, molecular anisotropy, and shape and dipole resonances.

The theoretical description of electron scattering from neutral targets is challenging at low to moderate collision energies~\cite{Tennyson-JPB-24,Tennyson-PhysRep-10,Bartschat-PPP-17}. A quantitatively reliable treatment must account for electron exchange, target polarization, short- and long-range correlation, interchannel coupling, the correct asymptotic behavior of the scattered electron~\cite{Tennyson-PhysRep-10,daCosta-EPJD-15,Masin-CPC-20}, the formation of transient negative-ion states and, for molecular targets, nuclear motion~\cite{Domcke-PhysRep-91,Tennyson-JPB-24}. The computational cost increases rapidly with the size of the molecule, because both the number of correlated target states accessible in a given energy range and the number of partial waves required to represent the scattered electron grow~\cite{Bartschat-PPP-17,Masin-CPC-20,Zammit-JPB-17}. 

Several first-principles and model approaches address these challenges. Close-coupling methods based on the $R$-matrix formalism are routinely employed for electron scattering from atoms and polyatomic molecules~\cite{Burke2011,Tennyson-PhysRep-10,Zatsarinny-JPB-13,Masin-CPC-20,Masin-JCTC-26,Gorfinkiel-JPB-05}. The convergent close-coupling (CCC) method~\cite{Bray-PRA-94,Fursa-JPB-97,Bray-JPB-02} is an alternative realization of the close-coupling approach, which has been extended to electron-impact ionization and to fully differential $(e,2e)$ observables~\cite{Bray-PRA-96,Bray-PRL-97,Bray-PRA-01}. Its molecular extension, the molecular convergent close-coupling (MCCC) method~\cite{Zammit-PRA-13,Zammit-PRA-14,Zammit-PRL-16, Zammit-PRA-17,Zammit-JPB-17}, treats elastic scattering, electronic excitation, and ionization of molecular targets on the same footing. Other nonperturbative variational formulations are the Schwinger multichannel method~\cite{Takatsuka-PRA-81,Takatsuka-PRA-84,Lucchese-PhysRep-86,daCosta-EPJD-15} and the complex Kohn method~\cite{Schneider-PRA-88, Rescigno-JPB-88, McCurdy-PRA-89, Rescigno-PRL-89,Schneider-PRL-91,Rescigno-PRA-92, Greenman-PRA-17}, both of which have been applied extensively to elastic and electronically inelastic scattering from small molecules. At higher collision energies, less expensive optical-potential and distorted-wave approaches provide useful cross sections for larger molecules~\cite{Gianturco-PhysRep-86,Staszewska-PRA-83,Fliflet-PRA-80}.

In the present work, we extend \ASTRA{}~\footnote{Certain software is identified in this paper to foster understanding. Such identification does not imply recommendation or endorsement by the National Institute of Standards and Technology, nor does it imply that the materials or equipment identified are necessarily the best available for the purpose.}, a general molecular close-coupling code originally developed for molecular ionization~\cite{Randazzo-PRR-23}, to electron-target scattering. \ASTRA{} differs from \UKRmolplus{}~\cite{Masin-CPC-20} and CCC/MCCC~\cite{Fursa-JPB-97,Zammit-JPB-17} in that the close-coupling matrix elements are evaluated through transition density matrices (TDMs) between correlated target states and, unlike \UKRmolplus{}, it does not employ the $R$-matrix formalism to compute the scattering states. These are constructed in the finite close-coupling space and matched to their asymptotic form with the pole-free Galerkin method with variational surface matching (PFG-VSM)~\cite{Marante2026}, in which the surface derivative used for the matching is stationary with respect to errors in the interior wave function. From the on-shell scattering matrix, we calculate elastic and electronically inelastic integral cross sections and molecular-frame differential cross sections. The method is validated against benchmark calculations and measurements for helium and molecular hydrogen.

Electron--helium scattering is an ideal benchmark for the atomic case, because extensive measurements~\cite{Buckman-JPB-86, Shigemura-PRA-14} and highly converged calculations~\cite{Bartschat-JPB-98,McEachran-Atoms-21,Bartschat-Atoms-25} are available for elastic scattering and for the excitation of the low-lying states, including energy regions in which resonances and thresholds strongly modify the cross sections~\cite{Lange-JPB-06,Hoshino-JPB-09,Zeng-PRA-15}. Molecular hydrogen is the corresponding molecular benchmark. Its two-electron target structure can be represented with high accuracy, and benchmark calculations are available for both elastic and inelastic scattering, together with measurements for the elastic channel and for the lowest excitation~\cite{Branchett-JPB-90,Wrkich-JPB-02,Yoon-JPCRD-08,Zammit-PRL-16,Zammit-PRA-17,Zawadzki-PRA-18,Meltzer-JPB-20,Scarlett-ADNDT-21,Scarlett-PRA-23}. Our electron--helium calculations reproduce the principal features of the CCC and atomic $R$-matrix benchmarks and the measured angular distributions. For electron--$\mathrm{H}_2$ scattering, the close-coupling calculations, which use the target expansion of the \UKRmolplus{} benchmark~\cite{Meltzer-JPB-20}, are in excellent agreement with it, both for elastic scattering and for the excitation cross sections at higher energies. For H$_2$ we also present molecular-frame scattered-electron angular distributions (MFSAD), whose dependence on the orientation of the molecule is hidden by rotational averaging~\cite{Colgan-PRL-08,Bellm-PRL-10}. The laboratory-frame differential cross sections and the scattering from polyatomic targets are the subject of a separate work.

This paper is organized as follows. Section~\ref{sec:astra-overview} summarizes the close-coupling formulation of \ASTRA{}, the computation of the scattering states, and the expressions for the integral and molecular-frame differential cross sections. Section~\ref{sec:computational-details} lists the parameters of the calculations. Section~\ref{sec:results_he} compares the results for helium with benchmarks and analyzes the $\mathrm{He}^{-}(1s2s^2\,{}^2S^e)$ resonance. Section~\ref{sec:results_h2} presents the elastic and inelastic integral cross sections and the MFSAD for H$_2$. Section~\ref{sec:conclusions} offers our conclusions and outlines future extensions of the method. Appendix~\ref{app:mf-derivation} provides a self-contained derivation of the formulas for the electron-scattering observables implemented in ASTRA and explored in this work.

\section{Overview of \ASTRA{}}
\label{sec:astra-overview}

\ASTRA{} is a close-coupling code for the multichannel electronic continuum of polyatomic molecules. It combines correlated descriptions of bound target states, currently at the complete-active-space (CAS) level, with flexible numerical representations of one electron in the continuum~\cite{Randazzo-PRR-23}. An $(N+1)$-electron close-coupling state is expressed as a linear combination of spin-coupled antisymmetrized products of $N$-electron target states and single-electron orbitals,
\begin{equation}
\begin{split}
|\Psi^{(N+1)}\rangle &=
\sum_{a,P} c_{a P}\,\left[\hat{a}_P^\dagger\otimes
|a\rangle\right]_{S,\Sigma}
\end{split}
\label{eq:close-coupling-ansatz}
\end{equation}
Here, $|a\rangle$ is a correlated state of the $N$-electron target, $\hat{a}^\dagger_P$ is the creation operator of an electron in a bound or continuum spin orbital $P=(p,\sigma)$, where $p$ and $\sigma$ identify the spatial part of the orbital and the spin projection, respectively, $c_{a P}\in\mathbb{C}$ are expansion coefficients, and $[M\otimes N]_{S\Sigma} \equiv \sum_{\sigma,\sigma'} C_{S_M \sigma, S_N \sigma'}^{S\Sigma} M_{S_M\sigma} N_{S_N\sigma'}$ is the spin coupling between two spin tensors. The target states are expressed entirely in terms of Gaussian molecular orbitals $\varphi_i$, whereas the electron created by $\hat{a}^\dagger_P$ can reside either in an active molecular orbital or in any element $\psi_P$ of a hybrid orthonormal Gaussian--B-spline set of the form
\begin{equation}
\begin{split}
\psi_P(\mathbf{x}) = {}^2\chi_{\sigma}(\zeta)\Bigg[&\sum_{i} \varphi_i A_{ip} +\\
&+ \sum_{j\ell m} \frac{B_j(r)}{r}X_{\ell m}(\hat{r})\,C_{j\ell m,p}\Bigg],
\end{split}
\end{equation}
where $\mathbf{x}=(\vec{r},\zeta)$ collects the position and spin coordinates of an electron, $^2\chi_\sigma(\zeta)$ is a spin-$1/2$ function, and $X_{\ell m}(\hat{r})$ are real spherical harmonics. In a molecular context, real spherical harmonics are convenient to define angular channels because each of them belongs to an irreducible representation of $D_{2h}$ and of its
subgroups, the only point groups used in \ASTRA{} and in most quantum-chemistry programs, including \textsc{lucia}~\cite{Olsen2000a}, the general configuration-interaction code used here to generate the correlated target states and the TDMs between them. The set of $X_{\ell m}$ is truncated at a maximum value of the angular momentum, $\ell \leq \ell_{\max}$. 

Equation~\eqref{eq:close-coupling-ansatz} reproduces the correct asymptotic form of the single-electron continuum of the $(N+1)$-electron system. With sufficiently complete target-state and active-electron orbital spaces, it describes bound states, autoionizing Rydberg series (for positively charged targets), and interchannel coupling.
In \ASTRA{}, the matrix elements of the relevant operators in the close-coupling space are computed by contracting TDMs between correlated target states, which may have different multiplicity and spatial symmetry~\cite{Randazzo-PRR-23}, with one- and two-electron hybrid Gaussian--B-spline integrals evaluated with the \texttt{PRISM} library~\cite{PRISM-repository}. This approach avoids the explicit construction of the full many-electron configuration-interaction representation in the hybrid orbital basis. The TDMs are computed efficiently by \textsc{lucia}. 

\subsection{Scattering states and integral cross sections}
\label{sec:scattering-states}
The scattering states are computed using the PFG-VSM method of Ref.~\cite{Marante2026}. Box eigenstates, which vanish at the outer boundary, describe the interior solution, while surface functions supply its boundary values. Their coefficients are determined by projecting the Schr\"odinger equation onto the box states. Near a box eigenenergy, a direct spectral expansion contains denominators that approach zero. The pole-free auxiliary representation treats the corresponding states jointly, including clustered or degenerate levels. A residual correction makes the surface derivative stationary with respect to errors in the interior wave function. For the neutral targets considered here, matching uses zero-charge reference functions at $R_{\max}$, beyond which residual channel interactions are neglected.

Within a given total spin $S$ and spatial symmetry $\Gamma$, an asymptotic channel is denoted by $\alpha=(a,\ell,m)$, where $a$ identifies the target state and $(\ell,m)$ are the angular quantum numbers of the electron in the continuum. For total energy $E$, the corresponding projectile energy and momentum are
\begin{equation}
\varepsilon_\alpha=E-E_a,
\qquad
k_\alpha=\sqrt{2\varepsilon_\alpha}.
\end{equation}
Channels with $\varepsilon_\alpha>0$ are open.
The stationary scattering states
$\ket{\Psi_{\alpha E}^{S\Gamma(\pm)}}$,
with outgoing $(+)$ or incoming $(-)$ boundary
conditions, are normalized in energy,
\begin{equation}
\braket{\Psi_{\alpha E}^{S\Gamma(\pm)}}
        {\Psi_{\beta E'}^{S'\Gamma'(\pm)}}
=
\delta_{SS'}\delta_{\Gamma\Gamma'}
\delta_{\alpha\beta}\delta(E-E'),
\label{eq:scattering-normalization}
\end{equation}
and the \emph{on-shell} scattering matrix $S_{\beta\alpha}^{S\Gamma}(E)$ is defined by
\begin{equation}
\braket{\Psi_{\beta E}^{S\Gamma(-)}}
        {\Psi_{\alpha E'}^{S\Gamma(+)}}
=
S_{\beta\alpha}^{S\Gamma}(E)\,
\delta(E-E').
\label{eq:s-matrix-definition}
\end{equation}
All the scattering observables considered here follow from the \emph{on-shell} scattering matrix. Appendix~\ref{app:mf-derivation} derives the corresponding formulas, which we summarize here.
Within the fixed-nuclei approximation, the state-to-state integral scattering cross section for the process 
\begin{equation}
M(a)+e^-_{\varepsilon_i}
\longrightarrow
M(b)+e^-_{\varepsilon_f},
\label{eq:scattering-process}
\end{equation}
is given by
\begin{equation}
\sigma_{b\gets a}(\varepsilon_i)
=
\sum_{S\Gamma}
\sigma_{b\gets a}^{S\Gamma}(\varepsilon_i),
\label{eq:ics-total}
\end{equation}
where $E=E_a+\varepsilon_i=E_b+\varepsilon_f$ by energy conservation, and the symmetry-specific cross sections are
\begin{equation}
\sigma_{b\gets a}^{S\Gamma}(\varepsilon_i)
=
\frac{\pi(2S+1)}
     {4\varepsilon_i(2S_a+1)}
\sum_{\substack{\alpha\in\mathcal O_a^{S\Gamma}\\
                \beta\in\mathcal O_b^{S\Gamma}}}
\left|
S_{\beta\alpha}^{S\Gamma}
(E_a+\varepsilon_i)
-\delta_{\beta\alpha}
\right|^2,
\label{eq:ics-symmetry-resolved}
\end{equation}
where $\mathcal O_a^{S\Gamma}$ is the set of open channels of symmetry $S\Gamma$ associated with the target state $a$.

Consider an electron incident with momentum $\mathbf{k}_i$ on a target initially in state $a$ and emerging in state $b$ together with a scattered electron with momentum $\mathbf{k}_f$, where $|\mathbf{k}_f|^2/2=E_a+|\mathbf{k}_i|^2/2-E_b$, with the molecular orientation held fixed. For spin-unpolarized particles, the molecular-frame differential cross section is
\begin{equation}
\begin{split}
\frac{d\sigma_{b\gets a}^{\mathrm{MF}}}{d\Omega_f}
&=
\sum_S
\frac{\pi^2}{\varepsilon_i}
\frac{2S+1}{2S_a+1}\,\Bigg|
\sum_\Gamma
\sum_{\ell m\ell'm'}
i^{\ell-\ell'}\,\times\\
&\times\,X_{\ell m}(\hat{\mathbf{k}}_i)
X_{\ell'm'}(\hat{\mathbf{k}}_f)\,
\left[
S_{\beta\alpha}^{S\Gamma}(E)
-\delta_{\beta\alpha}
\right]
\Bigg|^2,
\end{split}
\label{eq:mf-dcs-main}
\end{equation}
where $\alpha=(a,\ell,m)$ and $\beta=(b,\ell',m')$. For spherically symmetric atomic targets, of course, the molecular-frame and the laboratory-frame differential cross sections coincide.
For details on the derivation of the continuum normalization, spin averaging, and angular transformations, see Appendix~\ref{app:mf-derivation}.

\section{Computational details}
\label{sec:computational-details}

Here we briefly summarize the numerical details. The calculations are performed with partial-wave cutoffs $\ell_{\max}=4$ and $6$. The Gaussian bases are t-aug-cc-pVQZ for He and t-aug-cc-pVTZ for H$_2$~\cite{Dunning-JCP-89,Kendall-JCP-92,Woon-JCP-94}, and the B-splines have order $8$~\cite{Bachau-RPP-01}. In both cases, the inner region, in which the B-splines are orthonormalized to the molecular orbitals, has radius $R_{\mathrm{in}}=60\,a_0$ and contains $95$ B-splines (approximately 1.5 per Bohr radius)~\footnote{1 Bohr = 1 $a_0$ = $5.291\,772\,105\,44\times10^{-11}$ m}. For He, the outer radius is $R_{\max}=200\,a_0$. For H$_2$, the $\ell_{\max}=4$ calculation uses $R_{\max}=200\,a_0$, whereas the $\ell_{\max}=6$ calculation uses $150\,a_0$. The H$_2$ calculations use the fixed-nuclei approximation at $R_0=1.448\,a_0$, the mean bond length in the vibrational ground state~\cite{Meltzer-JPB-20}. Atomic units are used throughout the rest of the paper unless stated otherwise.

\subsection{Helium target and scattering model}
\label{sec:he-computational-details}
Although helium is spherically symmetric, the calculations are performed using the $D_{2h}$ point group employed by the molecular implementation of \ASTRA{}. 
The largest helium calculations include 59 target states resolved in $D_{2h}$ symmetry, which correspond to 19 atomic $LS$ terms, coupled to partial waves up to $\ell_{\max}=6$.
The helium target states are constructed in two stages. First, Hartree-Fock orbitals are generated with \textsc{dalton}~\cite{DALTON2014} in the t-aug-cc-pVQZ Gaussian basis. Starting from these orbitals, a state-averaged MCSCF calculation with equal weights is performed with \textsc{lucia}~\cite{Olsen2000a,Olsen1988} over the low-lying singlet and triplet states through the $n=3$ manifold. The two electrons are distributed in a complete active space (CAS) of $(14,8,8,5,8,5,5,2)$ orbitals in the irreducible representations $(a_g,b_{3u},b_{2u},b_{1g},b_{1u},b_{2g},b_{3g},a_u)$ of $D_{2h}$, in the ordering of \textsc{dalton}. The state-averaged orbitals are then used in a multi-root CAS-CI calculation in the same orbital space, which generates the larger set of singlet and triplet target states required for the scattering calculation, including states of the $n=4$ manifold. Atomic terms are assigned based on root degeneracy.

The physical targets comprise all the singlet and triplet singly-excited states of the form $1sn\ell$, with $n\leq 3$. The $1s4\ell$ states are coupled to the internal orbitals only~\cite{Randazzo-PRR-23} to improve the short-range description of correlation without introducing additional asymptotic channels. For helium, polarization and correlation are represented through these CAS-CI roots and their interchannel couplings. For the elastic cross section below approximately 4~eV, the close-coupling expansion is restricted to first five target states, as discussed in Sec.~\ref{sec:he-ics}. 

\section{Helium benchmarks}
\label{sec:results_he}

Table~\ref{tab:he-target-states} compares the present target-state energies with those obtained with BSR-69 by Stepanovi\'c \textit{et al.}~\cite{Stepanovic-JPB-06}, the $R$-matrix Set~5 results of Zeng \textit{et al.}~\cite{Zeng-PRA-15}, and the recommended NIST excitation energies~\cite{NIST_ASD}. 
\begin{table}[!hbtp]
\caption{\label{tab:he-target-states}
Excitation energies, in eV~\footnote{1 eV = 1.602 176 634 $\times$ 10$^{-19}$ J}, of the lowest eleven helium target states in the present close-coupling calculation, relative to the $1s^2\,{}^1S^e$ ground state. Additional closed states belonging to the $n=4$ manifold are included in the internal close-coupling space but are not listed. The BSR-69 values were obtained from the absolute target-state energies reported by Stepanovi\'c \textit{et al.}~\cite{Stepanovic-JPB-06} by subtracting their calculated ground-state energy. The $R$-matrix Set~5 values of Zeng \textit{et al.}~\cite{Zeng-PRA-15} were converted from Rydbergs to eV.}
\begin{ruledtabular}
\begin{tabular}{lcccc}
State
& \ASTRA{}
& BSR-69~\cite{Stepanovic-JPB-06}
& $R$-matrix~\cite{Zeng-PRA-15}
& NIST~\cite{NIST_ASD} \\
\colrule
\rule{0pt}{2.8ex}
$1s^2\,{}^1S^e$   & 0.0000  & 0.0000  & 0.0000  & 0.0000  \\
$1s2s\,{}^3S^e$   & 19.7900 & 19.8168 & 19.7487 & 19.8196 \\
$1s2s\,{}^1S^e$   & 20.5973 & 20.6148 & 20.5650 & 20.6157 \\
$1s2p\,{}^3P^o$   & 20.9341 & 20.9616 & 20.8915 & 20.9640 \\
$1s2p\,{}^1P^o$   & 21.1903 & 21.2157 & 21.1473 & 21.2180 \\
$1s3s\,{}^3S^e$   & 22.7151 & 22.7158 & 22.6453 & 22.7184 \\
$1s3s\,{}^1S^e$   & 22.9463 & 22.9183 & 22.8616 & 22.9203 \\
$1s3p\,{}^3P^o$   & 23.0928 & 23.0045 & 22.9324 & 23.0071 \\
$1s3d\,{}^3D^e$   & 23.3471 & 23.0713 & 22.9977 & 23.0736 \\
$1s3d\,{}^1D^e$   & 23.3481 & 23.0717 & 22.9977 & 23.0740 \\
$1s3p\,{}^1P^o$   & 23.2364 & 23.0846 & 23.0154 & 23.0870 \\
\end{tabular}
\end{ruledtabular}
\end{table}

The excitation observables reported below resolve $n=2$ final states, whose \ASTRA{} excitation energies lie within 18 to 30~meV from the NIST values. The $1s3s\,{}^3S^e$ and $1s3s\,{}^1S^e$ thresholds are similarly accurate, differing from the NIST values by $3.3$~meV and $26.0$~meV, respectively. The deviations are larger for the other $n=3$ states. The $1s3p\,{}^3P^o$ and $1s3p\,{}^1P^o$ thresholds are too high by approximately $86$ and $149$~meV, respectively, and the $1s3d\,{}^3D^e$ and $1s3d\,{}^1D^e$ thresholds by approximately $274$~meV, resulting in the deviation from the CCC excitation cross sections at higher energies, as discussed below.

\begin{figure}[!t]
\centering
\includegraphics[width=\columnwidth]{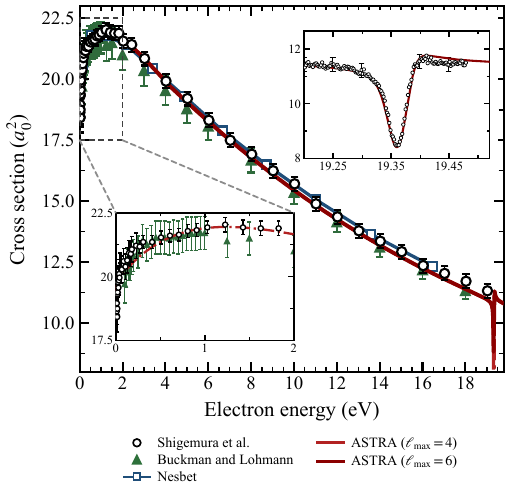}
\caption{\label{fig:he-elastic-ics}
Elastic integral cross section for electron scattering from $\mathrm{He}(1s^2\,{}^1S^e)$. The \ASTRA{} results obtained with partial-wave cutoffs $\ell_{\max}=4$ and $6$ are compared with the experimental total cross sections of Shigemura \textit{et al.}~\cite{Shigemura-PRA-14} and Buckman and Lohmann~\cite{Buckman-JPB-86}, and with the convergent-variational calculation of Nesbet~\cite{Nesbet-PRA-79}. The narrow structure immediately below the first excitation threshold is associated with the $\mathrm{He}^{-}(1s2s^2\,{}^2S^e)$ Feshbach resonance. The inset shows the region near threshold, between 0 and 2~eV, on an expanded scale. Below 4~eV, the results are computed with $\ell_{\max}=4$ and using the reduced five-target model discussed in the text. The upper-right inset compares Fig.~4(a) of Ref.~\cite{Shigemura-PRA-14} (circles) with the $\ell_{\max}=4$ \ASTRA{} resonance profile (solid line). See text for details. The experimental error bars are not accompanied by a specification of the associated confidence level.}
\end{figure}

\subsection{Integral cross sections\label{sec:he-ics}}

\begin{figure*}[hbtp!]
\centering
\captionsetup[subfloat]{farskip=0pt,nearskip=0pt,captionskip=0pt}
\subfloat{\panelgraphic[0.88,0.25]{0.48\textwidth}{(a)}
{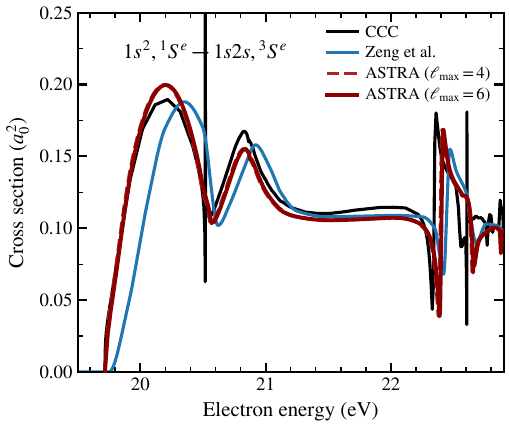}%
\label{fig:he-triplet-s-ics}}
\hfill
\subfloat{\panelgraphic[0.88,0.25]{0.48\textwidth}{(b)}
{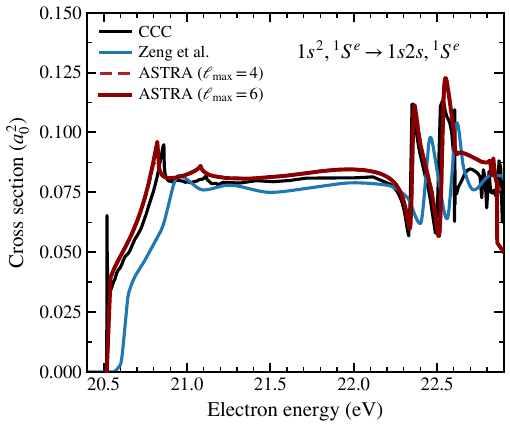}%
\label{fig:he-singlet-s-ics}}

\vspace{0.5em}

\subfloat{\panelgraphic[0.88,0.25]{0.48\textwidth}{(c)}
{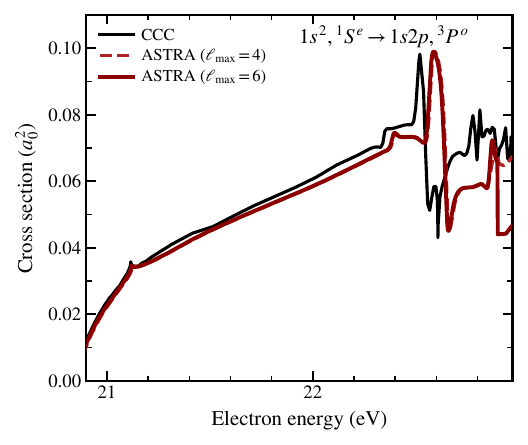}%
\label{fig:he-triplet-p-ics}}
\hfill
\subfloat{\panelgraphic[0.88,0.25]{0.48\textwidth}{(d)}
{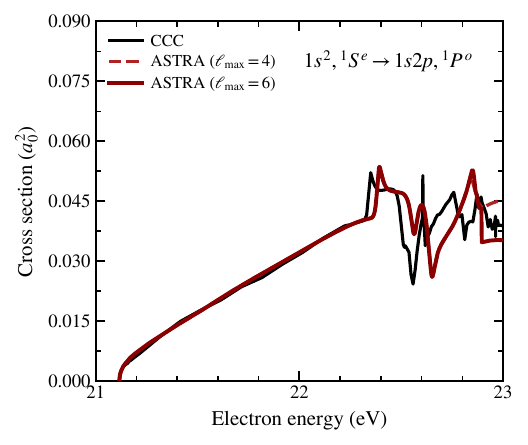}%
\label{fig:he-singlet-p-ics}}

\caption{\label{fig:he-inelastic-ics}
Integral cross sections for electron-impact excitation of helium from the $1s^2\,{}^1S^e$ ground state to (a) $1s2s\,{}^3S^e$, (b) $1s2s\,{}^1S^e$, (c) $1s2p\,{}^3P^o$, and (d) $1s2p\,{}^1P^o$. The \ASTRA{} results obtained with partial-wave cutoffs $\ell_{\max}=4$ and $6$ are compared with the CCC calculations of Fursa and Bray~\cite{Fursa-PRA-95}. For the $1s2s\,{}^3S^e$ and $1s2s\,{}^1S^e$ channels, the Set~5 $R$-matrix results of Zeng \textit{et al.}~\cite{Zeng-PRA-15} are also shown. \ASTRA{} results have been shifted so that their $n=2$ excitation thresholds coincide with the corresponding CCC thresholds.}
\end{figure*}

Figure~\ref{fig:he-elastic-ics} compares the elastic integral cross section obtained with \ASTRA{} with the measured total cross sections of Shigemura \textit{et al.}~\cite{Shigemura-PRA-14} and of Buckman and Lohmann~\cite{Buckman-JPB-86}, and with the convergent-variational results of Nesbet~\cite{Nesbet-PRA-79}. Note that below the first excitation threshold of helium, near $19.82$~eV, only the elastic channel is open.
Over the energy interval where both the $\ell_{\max}=4$ and $6$ calculations are shown, their cross sections are nearly indistinguishable, indicating that convergence is achieved already for $\ell_{\max}=4$.

The calculated cross section reaches a broad maximum of approximately $22\,a_0^2$ near $1$~eV and decreases smoothly at higher energies, following the measurements and Nesbet's calculation. Below approximately $4$~eV, the $\ell_{\max}=4$ curve is obtained with a reduced model comprising only the first five target-states listed in Table~\ref{tab:he-target-states}. The full 59-state model, which is appropriate for the inelastic cross sections and for higher energies, gives rise to numerical instabilities in the elastic channel near threshold. Without highly-excited target states, which play minimal role at low energies, the instability disappears and the agreement with the experiment is excellent, as shown in the inset of Fig.~\ref{fig:he-elastic-ics}.

Immediately below the first excitation threshold, the $\mathrm{He}^{-}(1s2s^2\,{}^2S^e)$ Feshbach resonance produces a narrow structure in the elastic cross section. Shigemura~\textit{et al.} observed it as an asymmetric dip near $19.37$~eV. The \ASTRA{} calculation has a sharp minimum at approximately $19.3$~eV, which is narrower than the measured one, since the latter is broadened by the finite energy resolution of the experiment. To compare the two profiles, we convolve the calculated cross section with a Gaussian with a full width at half maximum of $\sim33$~meV, chosen to reproduce the width of the measured dip. We also shift it by $37$~meV toward higher energy and add to it $0.56\,a_0^2$, about 5\% of the cross section, an offset comparable with the experimental error bars. The resulting profile, shown in the upper-right inset of Fig.~\ref{fig:he-elastic-ics}, is in excellent agreement with the measured one over the whole resonance region. The resonance is examined further in Sec.~\ref{subsec:he-feshbach}.

Figure~\ref{fig:he-inelastic-ics} shows the integral cross sections for excitation of the four $n=2$ states, $1s2s\,{}^3S^e$, $1s2s\,{}^1S^e$, $1s2p\,{}^3P^o$, and $1s2p\,{}^1P^o$. The \ASTRA{} results for $\ell_{\max}=4$ and $6$ are compared with the CCC calculations of Fursa and Bray~\cite{Fursa-PRA-95} and, for the two $1s2s$ channels, with the Set~5 $R$-matrix calculation of Zeng \textit{et al.}~\cite{Zeng-PRA-15}. To separate differences in the scattering dynamics from offsets in the target excitation energies, the \ASTRA{} results are shifted, independently for each channel, so that the corresponding $n=2$ threshold coincides with the CCC one. The cross sections are thus compared at the same energy above threshold. While this procedure aligns the cross sections at low energies, it amplifies the differences in the region of the resonances in proximity of the higher thresholds.

For the $1s2s$ states, Figs.~\ref{fig:he-triplet-s-ics} and \ref{fig:he-singlet-s-ics}, the threshold-aligned \ASTRA{} results reproduce the broad structures and the magnitude of the benchmark cross sections. The $1s2s\,{}^3S^e$ cross section rises rapidly above threshold to a maximum of approximately $0.20~a_0^2$ near $20.2$~eV, decreases to a minimum near $20.5$~eV, and has a second maximum near $20.8$~eV before settling to a slowly varying background of approximately $0.11 a_0^2$. The two maxima are the $^2P^o$ and $^2D^e$ resonances of $\mathrm{He}^-$~\cite{Buckman-RMP-94}. The first maximum is slightly higher in \ASTRA{} than in CCC, and the second is higher in CCC. The narrow spike in the CCC cross section near $20.5$~eV is absent from both the \ASTRA{} and the $R$-matrix results. The $1s2s\,{}^1S^e$ cross section, Fig.~\ref{fig:he-singlet-s-ics}, varies more smoothly. After the threshold rise and a first maximum near $20.8$~eV, the \ASTRA{} cross section remains close to $(0.08-0.085)~a_0^2$ from approximately $21.0$~eV to $22.2$~eV, in close agreement with CCC and slightly above the calculation of Zeng \textit{et al.} over part of the interval. 

The $1s2p$ excitation cross sections are shown in Figs.~\ref{fig:he-triplet-p-ics} and \ref{fig:he-singlet-p-ics}. For the $1s2p\,{}^3P^o$ state, the \ASTRA{} cross section follows the CCC result throughout the smooth region above threshold, increasing from approximately $0.01\,a_0^2$ to about $(0.07-0.075)a_0^2$ at the onset of the resonance structures and it is slightly smaller than the CCC one over part of this interval. For the dipole-allowed $1s^2\,{}^1S^e\rightarrow1s2p\,{}^1P^o$ transition, the threshold-aligned \ASTRA{} and CCC cross sections are nearly indistinguishable from threshold to approximately $22.3$~eV, where both reach about $0.04~a_0^2$. In the resonance region the two calculations have peaks and minima of comparable amplitude, slightly displaced from one another, and different backgrounds above the resonances. Since the $n=2$ thresholds are aligned, these displacements reflect differences in the energies of the higher target states, in the close-coupling expansions, and in the resulting resonance positions.

For all four excitation channels, the $\ell_{\max}=4$ and $6$ results are nearly superimposed in the smooth portions of the cross sections, and differ only slightly in the narrow resonance region associated with the higher thresholds. The excitation cross sections are therefore close to convergence with respect to the partial-wave expansion over this energy range. The remaining differences between \ASTRA{} and CCC are concentrated in the resonance profiles, which are affected by the target-state representation and by the composition of the close-coupling expansion.

For the integral cross sections, therefore, \ASTRA{} is on par with the benchmark calculations. The angular distributions, to which we now turn, are a more demanding and stringent test, and there \ASTRA{} is, in most channels, the closest to the measurements.

\begin{figure}[!t]
\centering
\includegraphics[width=\columnwidth]
{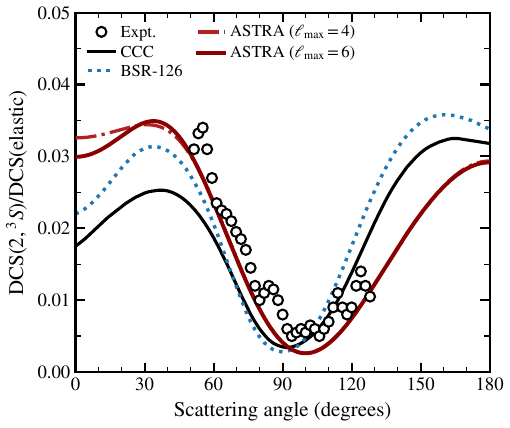}
\caption{\label{fig:he-dcs-ratio-20p3}
Ratio of the differential cross section for excitation of the $1s2s\,{}^3S^e$ state to the elastic differential cross section at an incident energy of $20.3~\mathrm{eV}$. The \ASTRA{} results obtained with projectile partial-wave cutoffs $\ell_{\max}=4$ and $6$ are compared with the experimental data and with the CCC and BSR-126 calculations reported by Lange \textit{et al.}~\cite{Lange-JPB-06}.}
\end{figure}

\subsection{Differential cross sections\label{sec:he-dcs}}

The angular dependence of electron-impact excitation is examined through the ratio of the inelastic to the elastic differential cross section,
\begin{equation}\label{eq:PDCS_ratio}
R_f(\theta;E)
=
\frac{d\sigma_f(E)/d\Omega}
     {d\sigma_{\mathrm{el}}(E)/d\Omega},
\end{equation}
where $f$ denotes the final excited state. This is the quantity measured by Lange~\textit{et al.}~\cite{Lange-JPB-06}, who reported ratios of inelastic to elastic scattering intensities at incident energies of $20.30$, $22.00$, and $23.48~\mathrm{eV}$, together with CCC and BSR-126 calculations. All helium differential-cross-section comparisons presented here, including both excitation-to-elastic ratios and absolute differential cross sections, use the unshifted \ASTRA{} target energies and excitation thresholds. The calculations are evaluated at the stated incident electron energies, without applying threshold alignment or empirical energy shifts.

Differential cross sections are more sensitive to the partial-wave expansion than integral cross sections, particularly at small angles. In Fig.~\ref{fig:he-dcs-ratio-20p3}, we illustrate this dependence at an incident energy of $20.30~\mathrm{eV}$ by comparing $\ell_{\max}=4$ and $6$. At this energy, the \(1s2s\,{}^3S^e\) state is the only energetically accessible excited target state. Its excitation angular distribution is strongly influenced by the broad \(\mathrm{He}^{-}(1s2s2p\,{}^2P^o)\) resonance in the region between the \(1s2s\,{}^3S^e\) and \(1s2s\,{}^1S^e\) thresholds. %
Both \ASTRA{} calculations, which differ only at small angles and coincide over the whole measured angular range, are in markedly better agreement with the experiment than CCC and BSR-126. They place the minimum of the ratio at $100^\circ$, at the center of the broad measured minimum, whereas CCC and BSR-126 place it near $90^\circ$. On the descending side of the distribution, between $50^\circ$ and $90^\circ$, \ASTRA{} is the closest to the data at every measured angle, with a deviation half that of the other two calculations.

\begin{figure*}[!t]
\centering
\begin{minipage}[t]{0.48\textwidth}
\centering
\panelgraphic[0.26,0.94]{\linewidth}{(a)}
{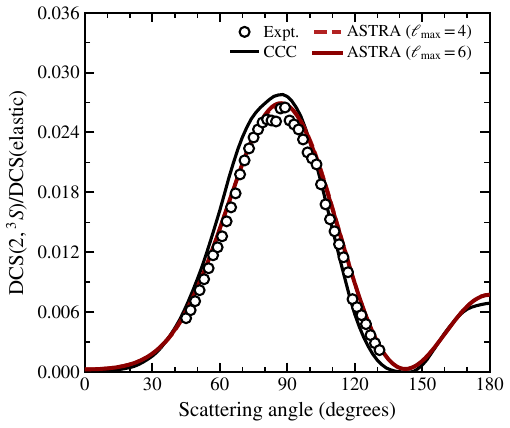}
\end{minipage}\hfill
\begin{minipage}[t]{0.48\textwidth}
\centering
\panelgraphic[0.26,0.94]{\linewidth}{(b)}
{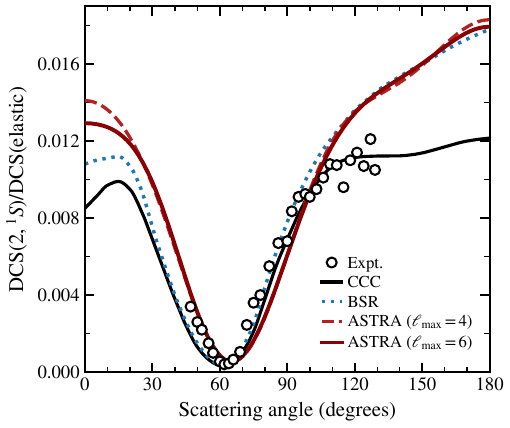}
\end{minipage}

\vspace{1.0ex}

\begin{minipage}[t]{0.48\textwidth}
\centering
\panelgraphic[0.68,0.94]{\linewidth}{(c)}
{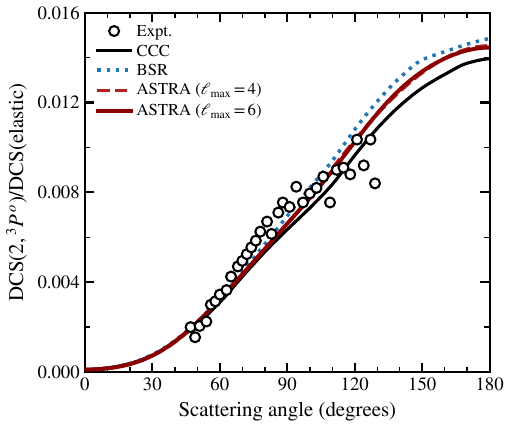}
\end{minipage}\hfill
\begin{minipage}[t]{0.48\textwidth}
\centering
\panelgraphic[0.26,0.94]{\linewidth}{(d)}
{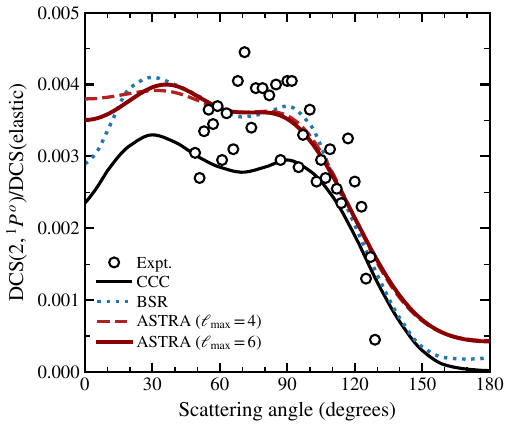}
\end{minipage}

\caption{\label{fig:he-dcs-ratios-22}
Ratio of differential cross sections for electron-impact excitation
of the helium $n=2$ states to the elastic differential cross section
at an incident energy of $22~\mathrm{eV}$:
(a) $1s^2\,{}^1S^e\rightarrow1s2s\,{}^3S^e$,
(b) $1s^2\,{}^1S^e\rightarrow1s2s\,{}^1S^e$,
(c) $1s^2\,{}^1S^e\rightarrow1s2p\,{}^3P^o$, and
(d) $1s^2\,{}^1S^e\rightarrow1s2p\,{}^1P^o$.
The \ASTRA{} results for both partial-wave cutoffs, $\ell_{\max}=4$ and $6$, are compared with the experimental data and CCC calculations reported by Lange \textit{et al.}~\cite{Lange-JPB-06}. The BSR-126 results~\cite{Lange-JPB-06} are additionally shown in panels (b)-(d).}
\end{figure*}

At $22.00~\mathrm{eV}$ all four $n=2$ excitation channels are open, and the energy lies below the resonances associated with the $n=3$ manifold. Figure~\ref{fig:he-dcs-ratios-22} compares the ratio~\eqref{eq:PDCS_ratio} computed with \ASTRA{}, CCC, and BSR-126 (when available) with the experiment.
For the $1s2s\,{}^3S^e$ state, Fig.~\ref{fig:he-dcs-ratios-22}(a), \ASTRA{}, CCC, and the experiment agree over the measured angular range. A broad maximum near $85^\circ$-$90^\circ$ is followed by a deep minimum at approximately $145^\circ$ and by a rise toward backward scattering. The \ASTRA{} maximum is slightly lower than the CCC one and closer to the experimental peak. For the $1s2s\,{}^1S^e$ state, Fig.~\ref{fig:he-dcs-ratios-22}(b), all calculations reproduce the minimum near $60^\circ$-$65^\circ$ and the subsequent increase, and \ASTRA{} follows the experiment from the minimum to approximately $110^\circ$. At small angles, where there are no data, the calculations differ, predicting ratios of about $0.013$ (\ASTRA{}, $\ell_{\max}= 6$), $0.011$ (BSR-126), and $0.008$ (CCC). Beyond $100^\circ$, \ASTRA{} and BSR-126 track each other and keep increasing, whereas CCC levels off beyond approximately $120^\circ$. %

For the $1s2p\,{}^3P^o$ state, Fig.~\ref{fig:he-dcs-ratios-22}(c), the ratio increases smoothly from small values at forward angles to approximately $0.01$ near $120^\circ$, and \ASTRA{} reproduces both the slope and the magnitude of the measured ratio throughout the experimental range. The three calculations are close to one another, with \ASTRA{} between CCC and BSR-126 at most angles. They separate slightly only toward backward scattering, where BSR-126 gives the largest ratio and CCC the smallest. 

For the dipole-allowed $1s2p\,{}^1P^o$ transition, Fig.~\ref{fig:he-dcs-ratios-22}(d), \ASTRA{} and BSR-126 predict similar ratios from forward to intermediate angles, with a plateau between approximately $50^\circ$ and $90^\circ$. Both lie above CCC over much of this interval and closer to the experimental points. Beyond approximately $100^\circ$, all three calculations decrease rapidly, as the measurements do. The \ASTRA{} curve lies above BSR-126 and CCC at both the smallest and largest angles. At forward angles the dipole-allowed singlet excitation prevails, by a factor of about five at $30^\circ$, whereas beyond approximately $60^\circ$ the exchange-driven triplet excitation takes over, and above $140^\circ$ it exceeds the singlet by more than an order of magnitude. %

\begin{figure*}[!t]
\centering
\begin{minipage}[t]{0.48\textwidth}
\centering
\panelgraphic[0.26,0.94]{\linewidth}{(a)}
{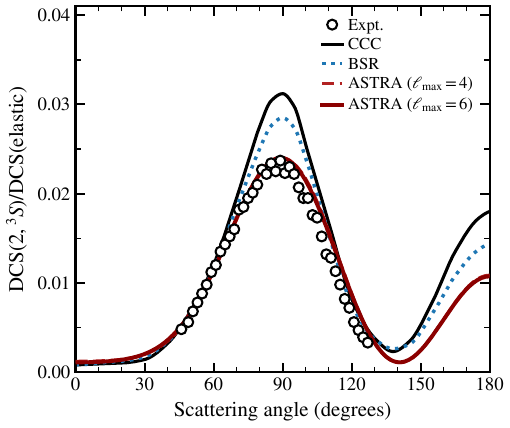}
\end{minipage}\hfill
\begin{minipage}[t]{0.48\textwidth}
\centering
\panelgraphic[0.6,0.94]{\linewidth}{(b)}
{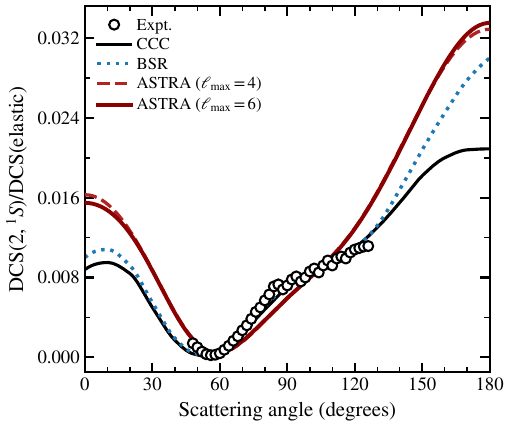}
\end{minipage}

\vspace{1.0ex}

\begin{minipage}[t]{0.48\textwidth}
\centering
\panelgraphic[0.26,0.94]{\linewidth}{(c)}
{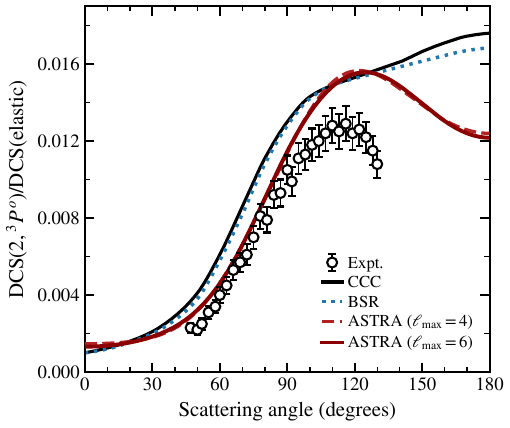}
\end{minipage}\hfill
\begin{minipage}[t]{0.48\textwidth}
\centering
\panelgraphic[0.42,0.94]{\linewidth}{(d)}
{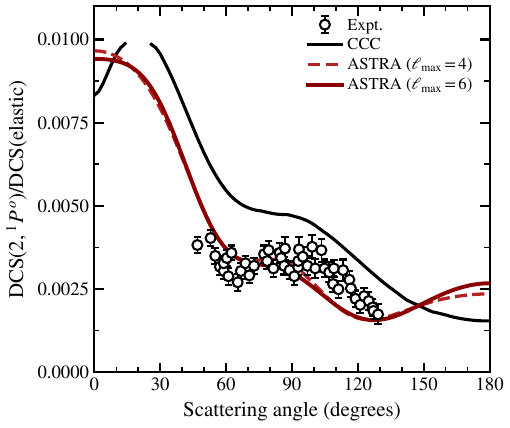}
\end{minipage}

\caption{\label{fig:he-dcs-ratios-23p48}
Ratio of differential cross sections for electron-impact excitation of the helium $n=2$ states to the elastic differential cross section at an incident energy of $23.48~\mathrm{eV}$:
(a) $1s^2\,{}^1S^e\rightarrow1s2s\,{}^3S^e$,
(b) $1s^2\,{}^1S^e\rightarrow1s2s\,{}^1S^e$,
(c) $1s^2\,{}^1S^e\rightarrow1s2p\,{}^3P^o$, and
(d) $1s^2\,{}^1S^e\rightarrow1s2p\,{}^1P^o$.
The \ASTRA{} results for both partial-wave cutoffs, $\ell_{\max}=4$ and $6$, are compared with the experimental data and CCC calculations reported by Lange \textit{et al.}~\cite{Lange-JPB-06}. The BSR-126 results are additionally shown in panels (a)-(c). Where shown, error bars represent the experimental uncertainties reported by Lange \textit{et al.}~\cite{Lange-JPB-06}.}
\end{figure*}

The comparison at $23.48~\mathrm{eV}$, Fig.~\ref{fig:he-dcs-ratios-23p48}, is a more stringent test than that at $22~\mathrm{eV}$, because several narrow negative-ion resonances occur in this region. Lange \textit{et al.}~\cite{Lange-JPB-06} showed that shifts of a few tens of meV can change some of the ratios substantially, particularly for the $P^o$ states. 
For the $1s2s\,{}^3S^e$ state, Fig.~\ref{fig:he-dcs-ratios-23p48}(a), all calculations reproduce the broad maximum centered near $90^\circ$. CCC and BSR-126 overestimate its magnitude by 32\% and 20\%, respectively, whereas \ASTRA{} coincides with the experimental peak. The decrease toward the minimum near $140^\circ$ is reproduced by all three calculations. The \ASTRA{} minimum is deeper, and the subsequent rise at backward angles correspondingly smaller. For the $1s2s\,{}^1S^e$ state, Fig.~\ref{fig:he-dcs-ratios-23p48}(b), all calculations reproduce the minimum near $55^\circ$ and the subsequent increase through the measured range. From approximately $55^\circ$ to $125^\circ$ the \ASTRA{} result follows the experimental points, with CCC and BSR-126 in slightly better agreement. At forward angles \ASTRA{} predicts a larger ratio than the other two calculations, and beyond approximately $130^\circ$ the three calculations separate progressively.

\begin{figure*}[hbtp!]
\centering

\begin{minipage}[t]{0.48\textwidth}
    \centering
    \panelgraphic[0.26,0.94]{\linewidth}{(a)}
    {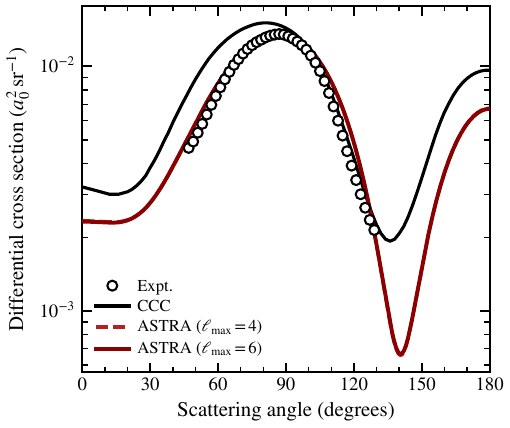}
\end{minipage}
\hfill
\begin{minipage}[t]{0.48\textwidth}
    \centering
    \panelgraphic[0.42,0.94]{\linewidth}{(b)}
    {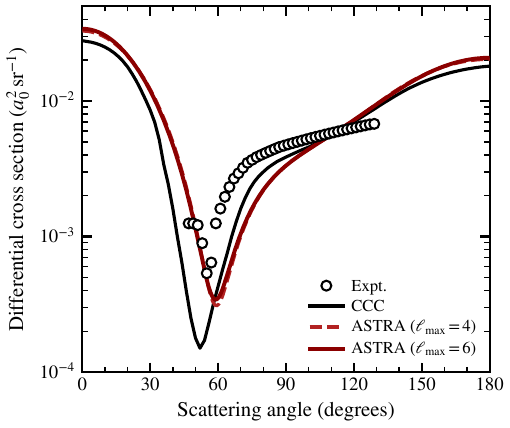}
\end{minipage}

\vspace{0.5em}

\begin{minipage}[t]{0.48\textwidth}
    \centering
    \panelgraphic[0.7,0.94]{\linewidth}{(c)}
    {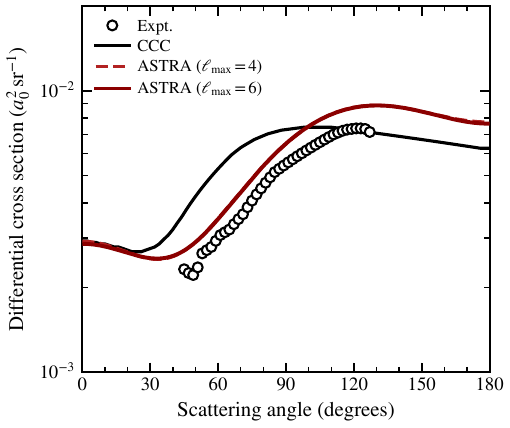}
\end{minipage}
\hfill
\begin{minipage}[t]{0.48\textwidth}
    \centering
    \panelgraphic[0.42,0.94]{\linewidth}{(d)}
    {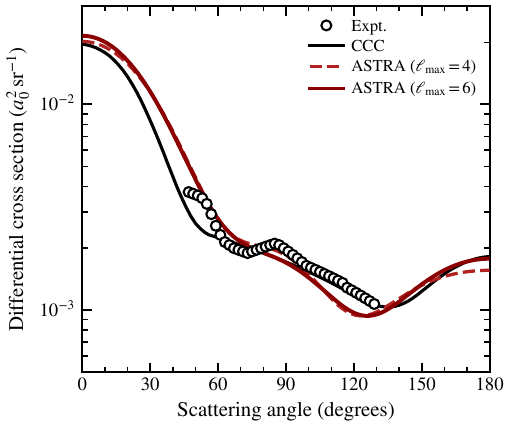}
\end{minipage}

\caption{\label{fig:he-hoshino-dcs-23p5}
Differential cross sections for electron-impact excitation of helium
near an incident energy of $23.5$~eV:
(a) $1s^2\,{}^1S^e\rightarrow1s2s\,{}^3S^e$,
(b) $1s^2\,{}^1S^e\rightarrow1s2s\,{}^1S^e$,
(c) $1s^2\,{}^1S^e\rightarrow1s2p\,{}^3P^o$, and
(d) $1s^2\,{}^1S^e\rightarrow1s2p\,{}^1P^o$.
The \ASTRA{} results obtained with partial-wave cutoffs
$\ell_{\max}=4$ and $6$ correspond to an incident energy of
$23.48$~eV. They are compared with the $23.5$~eV CCC calculations
and the $23.48$~eV measurements of Lange \textit{et al.}%
~\cite{Lange-JPB-06}, as reported by Hoshino \textit{et al.}%
~\cite{Hoshino-JPB-09}. The differential cross sections are shown
on a logarithmic scale.}
\end{figure*}

The measured ratio for the $1s2p\,{}^3P^o$ state, Fig.~\ref{fig:he-dcs-ratios-23p48}(c), increases smoothly from forward angles to a broad maximum near $110^\circ$-$120^\circ$. Here \ASTRA{} agrees with the experiment between 50$^\circ$ and 100$^\circ$ better than the other two calculations, and it is the only one that reproduces the measured maximum, whereas both BSR-126 and CCC increase monotonically across the whole angular range. 
For the dipole-allowed $1s2p\,{}^1P^o$ transition, Fig.~\ref{fig:he-dcs-ratios-23p48}(d), for which only CCC is available for comparison, the two calculations differ even more. CCC lies above the experiment over much of the measured angular range, whereas \ASTRA{} reproduces it up to $90^\circ$. Beyond this angle \ASTRA{} predicts a shallow minimum followed by an increase toward backward scattering, whereas CCC continues to decrease.

At $23.48$~eV, therefore, \ASTRA{} reproduces the measured angular structures in all four $n=2$ channels, and it is the closest to the experiment in three of them. For the $1s2s\,{}^3S^e$ state it reproduces the measured peak more accurately than CCC and BSR-126, for the $1s2p\,{}^3P^o$ state it reproduces the turnover at large angles, which the other two calculations do not have, and for the $1s2p\,{}^1P^o$ state it follows the measurements up to $90^\circ$, where the CCC calculation of Ref.~\cite{Lange-JPB-06} exceeds them by about 50\%. For the $1s2s\,{}^1S^e$ state all three calculations follow the measurements, with CCC and BSR-126 slightly closer.

\subsubsection{Absolute differential cross sections}

\begin{figure*}[hbtp!]
\centering

\begin{minipage}[t]{0.48\textwidth}
    \centering
    \panelgraphic[0.26,0.94]{\linewidth}{(a)}
    {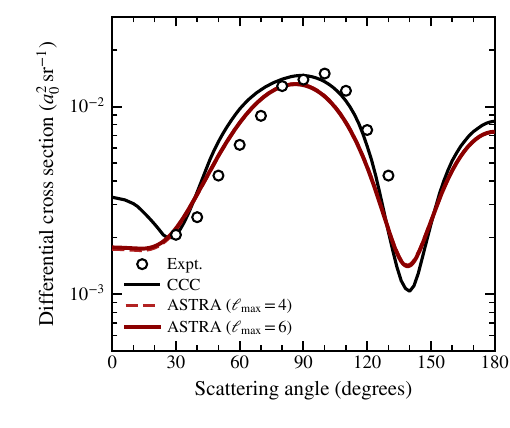}
\end{minipage}
\hfill
\begin{minipage}[t]{0.48\textwidth}
    \centering
    \panelgraphic[0.42,0.94]{\linewidth}{(b)}
    {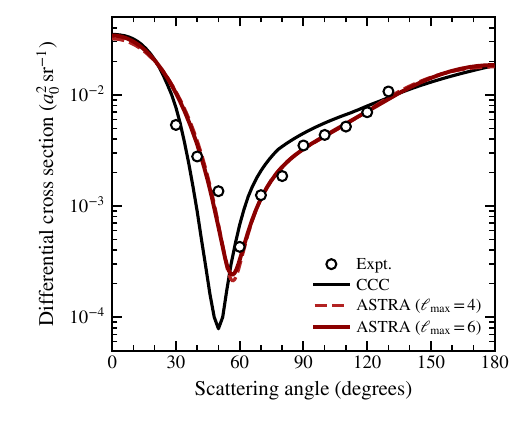}
\end{minipage}

\vspace{0.5em}

\begin{minipage}[t]{0.48\textwidth}
    \centering
    \panelgraphic[0.26,0.94]{\linewidth}{(c)}
    {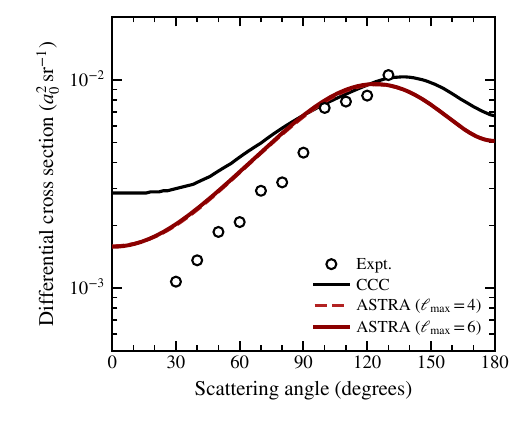}
\end{minipage}
\hfill
\begin{minipage}[t]{0.48\textwidth}
    \centering
    \panelgraphic[0.42,0.94]{\linewidth}{(d)}
    {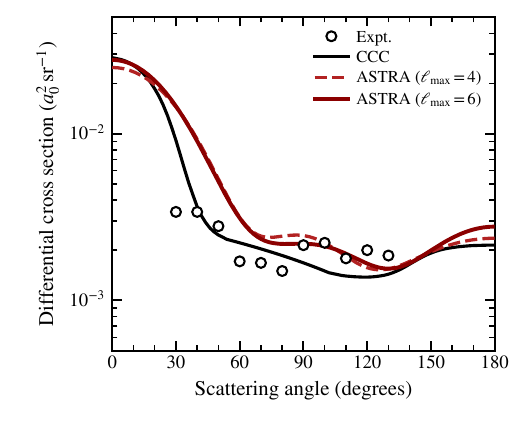}
\end{minipage}

\caption{\label{fig:he-hoshino-dcs-24}
Differential cross sections for electron-impact excitation of helium
at an incident energy of $24.0$~eV:
(a) $1s^2\,{}^1S^e\rightarrow1s2s\,{}^3S^e$,
(b) $1s^2\,{}^1S^e\rightarrow1s2s\,{}^1S^e$,
(c) $1s^2\,{}^1S^e\rightarrow1s2p\,{}^3P^o$,
and (d) $1s^2\,{}^1S^e\rightarrow1s2p\,{}^1P^o$.
The \ASTRA{} results obtained with projectile partial-wave cutoffs
$\ell_{\max}=4$ and $6$ correspond to an incident energy of
$24.0$~eV and are compared with the experimental measurements
and CCC calculations reported by Hoshino \textit{et al.}%
~\cite{Hoshino-JPB-09}. The differential cross sections are shown
on a logarithmic scale.}
\end{figure*}

Hoshino \textit{et al.}~\cite{Hoshino-JPB-09} reported absolute differential cross sections for excitation of the four $n=2$ states of helium over the incident-energy range $23.5$-$35~\mathrm{eV}$ and scattering angles up to approximately $130^\circ$, with estimated uncertainties of $15\%$-$20\%$, together with a new set of CCC calculations converged to within approximately $5\%$ at most scattering angles. Figures~\ref{fig:he-hoshino-dcs-23p5} and \ref{fig:he-hoshino-dcs-24} compare these data with the \ASTRA{} calculations near $23.5$ and $24.0~\mathrm{eV}$, on a logarithmic scale.

Near $23.5~\mathrm{eV}$, Fig.~\ref{fig:he-hoshino-dcs-23p5}(a), \ASTRA{} reproduces the broad maximum near $85^\circ$-$90^\circ$, its magnitude, and the decrease at larger angles in the measured angular distribution of the $1s2s\,{}^3S^e$ transition, whereas CCC gives a broader signal peaking at an angle smaller by about 10$^\circ$ compared to both \ASTRA{} and the experiment. The angular distribution of the $1s2s\,{}^1S^e$ state, Fig.~\ref{fig:he-hoshino-dcs-23p5}(b), is dominated by a deep minimum near $50^\circ$-$60^\circ$. CCC places it at a smaller angle and makes it considerably deeper, whereas the \ASTRA{} minimum, at approximately $55^\circ$, is closer to the experimental one in both position and depth. On the rising side of the minimum, \ASTRA{} initially lies below the measurements and approaches them with increasing angle. 

The measured cross section for the $1s2p\,{}^3P^o$ channel, Fig.~\ref{fig:he-hoshino-dcs-23p5}(c), increases from approximately $50^\circ$ to a broad maximum around $120^\circ$. CCC exhibits a similar increase but in an angular range that is shifted to smaller angles by approximately 30$^\circ$ compared to the experiment. Here \ASTRA{}'s results show a significant improvement over CCC, even if it overestimates somewhat the measured magnitude.  
For the dipole-allowed $1s2p\,{}^1P^o$ transition, Fig.~\ref{fig:he-hoshino-dcs-23p5}(d), \ASTRA{} follows the experiment up to $\sim 90^\circ$ and deviates from it between $100^\circ$ and $130^\circ$, where CCC agrees with the experiment.

At $24.0~\mathrm{eV}$, Fig.~\ref{fig:he-hoshino-dcs-24}, both \ASTRA{} and CCC reproduce the broad maximum of the $1s2s\,{}^3S^e$ distribution, Fig.~\ref{fig:he-hoshino-dcs-24}(a), centered near $85^\circ$-$95^\circ$, and its magnitude. \ASTRA{} lies below CCC at forward angles and closer to the measurements on the rising side of the distribution. Beyond the maximum both calculations decrease more rapidly than the experimental points, most visibly at the largest measured angle. Outside the experimental range, the two calculations predict a minimum near $140^\circ$, much deeper in CCC, followed by a rise toward backward scattering. The distribution of the $1s2s\,{}^1S^e$ state, Fig.~\ref{fig:he-hoshino-dcs-24}(b), is dominated by a sharp interference minimum near $50^\circ$-$60^\circ$. CCC places it at approximately $50^\circ$ with a cross section considerably smaller than observed, whereas the \ASTRA{} minimum occurs at a larger angle and has a depth much closer to the experimental one. On the high-angle side of the minimum both calculations follow the measured increase, \ASTRA{} more closely at intermediate angles, and by approximately $120^\circ$-$130^\circ$ the calculated and measured cross sections are again comparable. %

For the $1s2p\,{}^3P^o$ channel, Fig.~\ref{fig:he-hoshino-dcs-24}(c), the cross section increases steadily over the measured angular range. Both calculations overestimate the measurements at small and intermediate angles, \ASTRA{} by less than CCC. The discrepancy decreases with increasing angle, and near $120^\circ$-$130^\circ$ the three results are comparable. Beyond the measured range, \ASTRA{} reaches a broad maximum at approximately $130^\circ$ and then decreases more rapidly than CCC. For the dipole-allowed $1s2p\,{}^1P^o$ transition, Fig.~\ref{fig:he-hoshino-dcs-24}(d), both \ASTRA{} and CCC overestimate the measured cross section substantially at $30^\circ$-$40^\circ$. The agreement improves rapidly with increasing angle. From approximately $60^\circ$ to $120^\circ$ the measurements lie close to, or between, the two calculations, and \ASTRA{} reproduces the weak structure around $80^\circ$-$100^\circ$. Toward $120^\circ$-$130^\circ$ both calculations have a shallow minimum. The modulation of the experimental signal in this range is comparable with the experimental uncertainty of $15\%$-$20\%$.

The absolute differential cross sections at 23.5 and 24.0~eV thus confirm the comparison of the ratios. \ASTRA{} reproduces the angular structures and, for most channels, the absolute magnitude of the measured cross sections, most closely for the $1s2s\,{}^3S^e$ and $1s2s\,{}^1S^e$ states, including the positions of the maxima and of the interference minima. At $24.0$~eV it improves on CCC in the position and depth of the $1s2s\,{}^1S^e$ minimum and in the magnitude of the $1s2p\,{}^3P^o$ distribution at small and intermediate angles. The two $1s2p$ channels deviate more, especially at forward and at large angles. The largest remaining discrepancy with the experiment, in the ratio for the $1s2p\,^3P^o$ state at angles larger than 90$^\circ$ [Fig.~\ref{fig:he-dcs-ratios-23p48}(c)], is not due to the truncation of the angular expansion, since the $\ell_{\max}=4$ and $\ell_{\max}=6$ results virtually coincide. It may be due to a mismatch with the experimental energy or to the truncation of the close-coupling expansion.

\begin{figure}[!t]
\centering
\includegraphics[width=\columnwidth]
{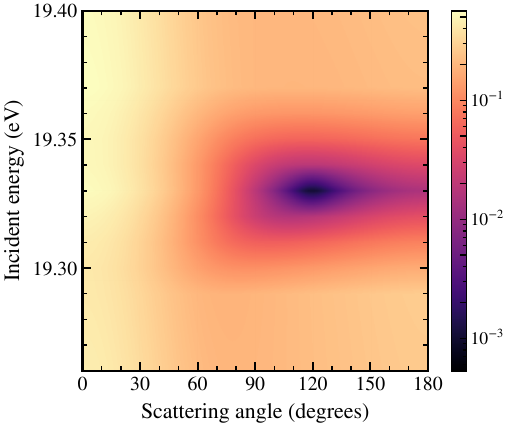}
\caption{\label{fig:he-feshbach-dcs}
Elastic differential cross section for electron scattering from $\mathrm{He}(1s^2\,{}^1S^e)$ as a function of incident energy and scattering angle in the vicinity of the lowest $\mathrm{He}^{-}(1s2s^2\,{}^2S^e)$ Feshbach resonance. The cross section is given in units of $10^{-16}\,\mathrm{cm}^{2}\,\mathrm{sr}^{-1}$. The narrow band of suppressed cross section near $19.33~\mathrm{eV}$ results from interference between the resonant and nonresonant elastic-scattering amplitudes. The recommended resonance energy is $E_r=(19.366\pm0.005)~\mathrm{eV}$~\cite{Buckman-RMP-94}.}
\end{figure}

\subsection{Lowest $\mathrm{He}^{-}$ Feshbach resonance}
\label{subsec:he-feshbach}

The lowest Feshbach resonance of the helium negative ion, $\mathrm{He}^{-}(1s2s^2\,{}^2S^e)$, which can autodetach only into the elastic channel, provides a test of the close-coupling implementation complementary to the differential cross sections. Buckman and Clark~\cite{Buckman-RMP-94} recommend a resonance energy of $(19.366\pm0.005)~\mathrm{eV}$ and a natural width of $(11.0\pm0.5)~\mathrm{meV}$.
To determine the parameters of this resonance, we performed an exterior-complex-scaling (ECS)~\cite{Simon-PLA-79,McCurdy-JPB-04} calculation within \ASTRA{}, with the same close-coupling expansion as in the scattering calculation. The radial coordinate was complex scaled beyond $R_{\mathrm{ECS}}=100\,a_0$ with a scaling angle of $15^\circ$. The resonance is identified from the corresponding complex eigenvalue,
\begin{equation}
E_{\mathrm{res}}
=
E_r-\frac{i}{2}\Gamma ,
\end{equation}
yielding
\begin{equation}
E_r = 19.327~\mathrm{eV},
\qquad
\Gamma = 11~\mathrm{meV}.
\end{equation}
The calculated width agrees with the recommended value, whereas the resonance position lies $39$\,meV below the recommended energy. This offset is close to the difference of approximately $30$\,meV between our excitation energy of the first triplet state and the NIST value in Table~\ref{tab:he-target-states}. In other words, most of the discrepancy can be attributed to the correlation energy missing in the $1s^2$ ground state, whereas the binding energy of the resonance relative to the $1s2s\,{}^3S$ state is reproduced to within about $10$\,meV. 

Figure~\ref{fig:he-feshbach-dcs} shows the elastic differential cross section as a function of energy and angle in the vicinity of the resonance. Away from it, the cross section varies smoothly with both variables. Near $19.33~\mathrm{eV}$ a window resonance with a narrow minimum develops at intermediate and backward angles. The suppression is visible beyond approximately $60^\circ$ and is strongest over the $90^\circ$-$180^\circ$ angular region, whereas the cross section at forward angles is hardly affected. In general, the energy of the minimum need not coincide with the position of the resonance pole, since the profile is characterized by an angle-dependent Fano asymmetry parameter~\cite{Fano-PR-61}. In this case, however, $q\approx0$, and the minimum at $19.33~\mathrm{eV}$ matches the pole position of $19.327~\mathrm{eV}$ obtained with ECS. The close-coupling representation in \ASTRA{} thus captures the coupling between the continuum and the correlated configurations responsible for the formation of transient $\mathrm{He}^{-}$ states.

\section{Molecular hydrogen benchmarks}
\label{sec:results_h2}

\subsection{Target-state description}
\label{subsec:h2-target}

Fixed-nuclei benchmark calculations for electron scattering off molecular hydrogen are available from both the \UKRmolplus{} and the molecular convergent close-coupling (MCCC) methods~\cite{Meltzer-JPB-20}, so that the target description and the scattering observables can be tested independently. For a direct comparison with the benchmark calculations of Meltzer \textit{et al.}~\cite{Meltzer-JPB-20}, all the calculations of this section are performed at the mean vibrational bond length, $R_0=1.448\,a_0$, which was adopted in that study, in place of the equilibrium bond length $R_e=1.40\,a_0$, as the most representative fixed-nuclei geometry for the comparison with measurements. 
\begin{table*}[!t]
\caption{\label{tab:h2-target-states}
Absolute and vertical excitation energies of the $\mathrm{H}_2$ target states at the mean vibrational bond length $R_0=1.448\,a_0$. The \UKRmolplus and MCCC values are taken from Ref.~\cite{Meltzer-JPB-20}. The excitation energies are measured relative to the ground-state energy obtained with the corresponding target model. The \ASTRA{} excitation energies were evaluated from the unrounded target-state energies, whereas the absolute energies displayed in the table have been rounded to three decimal places to match the precision of the published \UKRmolplus{} values. }
\begin{ruledtabular}
\begin{tabular}{lcccccc}
&
\multicolumn{3}{c}{Absolute energy (a.u.)}
&
\multicolumn{3}{c}{Vertical excitation energy (eV)}
\\
State
& \ASTRA{}
& \UKRmolplus{}
& MCCC
& \ASTRA{}
& \UKRmolplus{}
& MCCC
\\
\colrule
$X\,{}^1\Sigma_g^{+}$
& $-1.172$ & $-1.172$ & $-1.161$
& --- & --- & ---
\\
$b\,{}^3\Sigma_u^{+}$
& $-0.796$ & $-0.796$ & $-0.782$
& 10.23 & 10.23 & 10.31
\\
$a\,{}^3\Sigma_g^{+}$
& $-0.718$ & $-0.718$ & $-0.715$
& 12.35 & 12.35 & 12.14
\\
$B\,{}^1\Sigma_u^{+}$
& $-0.712$ & $-0.712$ & $-0.704$
& 12.52 & 12.52 & 12.44
\\
$c\,{}^3\Pi_u$
& $-0.712$ & $-0.712$ & $-0.707$
& 12.52 & 12.52 & 12.35
\\
$EF\,{}^1\Sigma_g^{+}$
& $-0.697$ & $-0.697$ & $-0.693$
& 12.93 & 12.93 & 12.73
\\
$C\,{}^1\Pi_u$
& $-0.694$ & $-0.694$ & $-0.693$
& 13.01 & 13.02 & 12.73
\\
$e\,{}^3\Sigma_u^{+}$
& $-0.650$ & $-0.650$ & $-0.647$
& 14.20 & 14.21 & 13.99
\\
$h\,{}^3\Sigma_g^{+}$
& $-0.636$ & $-0.636$ & $-0.634$
& 14.58 & 14.60 & 14.34
\\
$B'\,{}^1\Sigma_u^{+}$
& $-0.635$ & $-0.635$ & $-0.631$
& 14.62 & 14.63 & 14.42
\\
\end{tabular}
\end{ruledtabular}
\end{table*}

The one-electron orbitals of the \ASTRA{} target calculation are generated at the Hartree--Fock level with \textsc{dalton}~\cite{DALTON2014}, in the triply augmented correlation-consistent polarized valence triple-zeta (t-aug-cc-pVTZ) basis. Meltzer \textit{et al.}~\cite{Meltzer-JPB-20} identified this basis as an effective choice to represent both the low-lying valence states and the diffuse Rydberg states of $\mathrm{H}_2$. In their study, increasing the diffuse augmentation improved the target structure more than increasing the cardinal number of the correlation-consistent basis. As in that work, the \ASTRA{} target calculation is a full CI for the two-electron target in this basis, with 82 spatial orbitals distributed as $(19,9,9,4,19,9,9,4)$ in the $D_{2h}$ irreducible representations $(a_g,b_{3u},b_{2u},b_{1g},b_{1u},b_{2g},b_{3g},a_u)$.
The close-coupling expansion is the one of the \UKRmolplus{} benchmark~\cite{Meltzer-JPB-20}. It comprises the 98 target states with vertical excitation energy below 30~eV, computed with \textsc{lucia} in the active orbital space specified above, including the separate $D_{2h}$ components of degenerate molecular terms. Of these, 49 are singlets and 49 are triplets, distributed among the $D_{2h}$ irreducible representations $(A_g,B_{3u},B_{2u},B_{1g},B_{1u},B_{2g},B_{3g},A_u)$ as $(13,5,5,2,12,5,5,2)$ and $(11,5,5,2,12,6,6,2)$, respectively. Each target is coupled to both internal and external projectile orbitals. Target polarization is represented through coupling to excited target states within this finite-basis expansion. Energetically closed channels contribute through virtual excitation, while open excited-target channels also describe inelastic scattering. The calculations are performed for $\ell_{\max}=4$ and 6.

The energy range considered extends above the vertical ionization threshold of H$_2$. Above-threshold roots of a finite target CI calculation provide a discretized representation of continuum configurations. The present benchmarks concern elastic scattering and selected bound-state electronic excitations. At the geometry used in the scattering calculations ($R_0=1.448\,a_0$), the vertical ionization threshold of the present target model is $I_{\mathrm v}=16.23$~eV. This value is obtained from $I_{\mathrm v}=E_{\mathrm{H}_2^+}(R_0)-E_{\mathrm{H}_2}(R_0)$, using the neutral CAS-CI ground-state energy and the lowest one-electron eigenvalue for $\mathrm{H}_2^+$ in the same t-aug-cc-pVTZ orbital space.

Table~\ref{tab:h2-target-states} lists absolute and vertical excitation energies for ten representative low-lying target states. The common finite-basis full-CI target description used by \ASTRA{} and \UKRmolplus{} makes this a useful consistency check on the target construction and state assignments. At the displayed precision, their absolute energies coincide, and their vertical excitation energies differ by no more than approximately $0.02~\mathrm{eV}$. The MCCC absolute energies are systematically higher, while the vertical excitation energies agree with the Gaussian-basis results to within $0.3~\mathrm{eV}$.

\begin{figure}[!t]
\centering
\includegraphics[width=\columnwidth]
{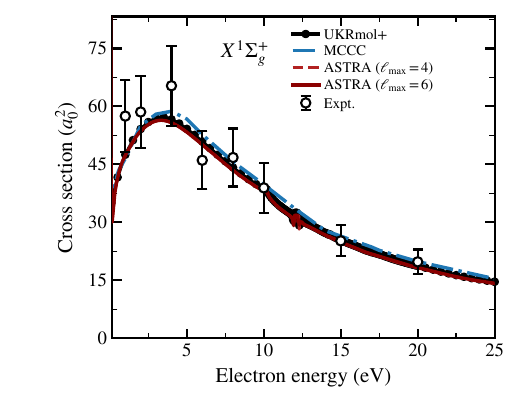}
\caption{\label{fig:h2-elastic-ics}
Elastic integral cross sections for electron scattering from
$\mathrm{H}_2(X\,{}^1\Sigma_g^+)$ at the fixed internuclear separation
$R_0=1.448\,a_0$. The \ASTRA{} results are shown for $\ell_{\max}=4$ and 6. The $\ell_{\max}=4$ calculation uses
$R_{\max}=200\,a_0$, whereas the $\ell_{\max}=6$
calculation uses $R_{\max}=150\,a_0$. See text for details. Fixed-nuclei \UKRmolplus{} results
are from Meltzer \textit{et al.}~\cite{Meltzer-JPB-20}.
The MCCC results of Zammit \textit{et al.}~\cite{Zammit-PRL-16}
were obtained from the CCC database on LXCat~\cite{Scarlett-ADNDT-21,Pitchford-PPP-17}. Measurements
are from Muse \textit{et al.}~\cite{Muse-JPB-08} and error bars represent the experimental uncertainties, approximately 16~\%.}
\end{figure}

\subsection{Elastic integral cross section}
\label{subsec:h2-elastic-ics}
Figure~\ref{fig:h2-elastic-ics} compares the \ASTRA{} elastic integral cross sections with the fixed-nuclei \UKRmolplus{} and MCCC benchmarks and with the measurements of Muse \textit{et al.}~\cite{Muse-JPB-08}. The \ASTRA{} cross section has a broad maximum of approximately $56.4\,a_0^2$ near 3~eV -- 4~eV and is in close agreement with \UKRmolplus{} at all energies. MCCC follows the same trend, with a slightly higher maximum of approximately $59\,a_0^2$.

The \ASTRA{} results reproduce the magnitude and the energy dependence of the measured cross section, within the plotted uncertainties at all the measured energies. The experimental uncertainties, approximately 16\%, do not discriminate between the calculations over much of the measured range. The two \ASTRA{} results nearly coincide. At 25~eV they give $14.13$ and $14.23\,a_0^2$ for $\ell_{\max}=4$ and 6, respectively, compared with $14.61\,a_0^2$ from \UKRmolplus{} and $15.39\,a_0^2$ from MCCC.

\subsection{Inelastic integral cross sections}
\label{subsec:h2-inelastic-ics}

\begin{figure*}[!t]
\centering

\begin{minipage}[t]{0.48\textwidth}
\centering
\panelgraphic[0.25,0.94]{\linewidth}{(a)}
{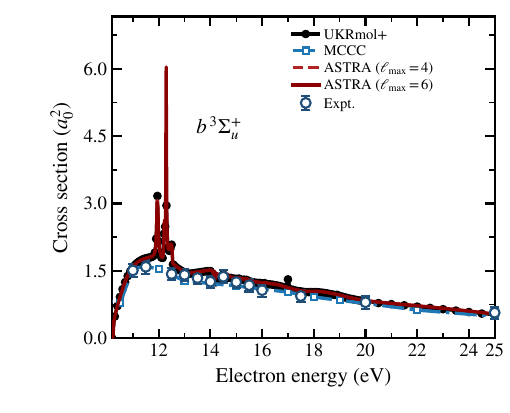}
\end{minipage}
\hfill
\begin{minipage}[t]{0.48\textwidth}
\centering
\panelgraphic[0.47,0.94]{\linewidth}{(b)}
{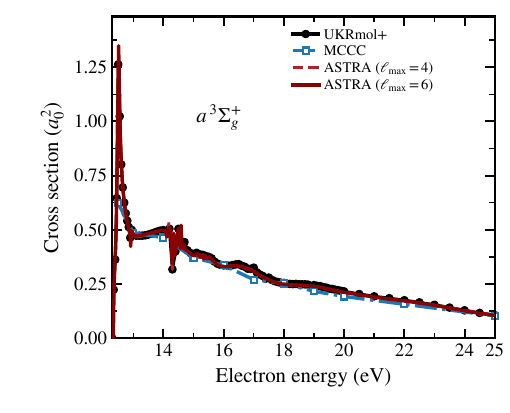}
\end{minipage}

\par\vspace{1ex}

\begin{minipage}[t]{0.48\textwidth}
\centering
\panelgraphic[0.25,0.94]{\linewidth}{(c)}
{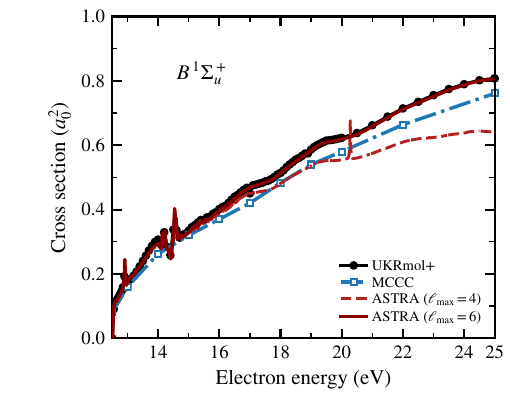}
\end{minipage}
\hfill
\begin{minipage}[t]{0.48\textwidth}
\centering
\panelgraphic[0.25,0.94]{\linewidth}{(d)}
{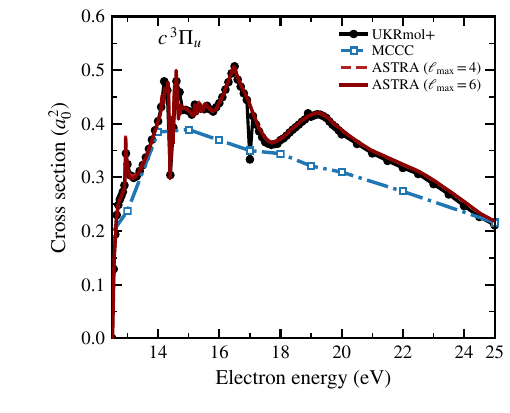}
\end{minipage}

\caption{
Fixed-nuclei integral cross sections for excitation from
$\mathrm{H}_2(X\,{}^1\Sigma_g^+)$ to
(a) $b\,{}^3\Sigma_u^+$,
(b) $a\,{}^3\Sigma_g^+$,
(c) $B\,{}^1\Sigma_u^+$, and
(d) $c\,{}^3\Pi_u$, at $R_0=1.448\,a_0$.
The \ASTRA{} results are shown for
$\ell_{\max}=4$ and 6.
Numerical details are described in the text.
Fixed-nuclei \UKRmolplus{} benchmarks are from
Meltzer \textit{et al.}~\cite{Meltzer-JPB-20}.
The MCCC results of Zammit \textit{et al.}~\cite{Zammit-PRA-17}
were obtained from the CCC database on LXCat~\cite{Scarlett-ADNDT-21,Pitchford-PPP-17}.
The experimental results in panel~(a) are from
Zawadzki \textit{et al.}~\cite{Zawadzki-PRA-18} and error bars in panel (a) represent the combined experimental uncertainties, approximately 16~\%.
}
\label{fig:h2-inelastic-ics}
\end{figure*}

Figure~\ref{fig:h2-inelastic-ics} compares four low-lying excitation channels with the fixed-nuclei \UKRmolplus{} and MCCC benchmarks~\cite{Meltzer-JPB-20,Zammit-PRA-17}. Excitation of the triplet states $b\,{}^3\Sigma_u^+$, $a\,{}^3\Sigma_g^+$, and $c\,{}^3\Pi_u$ from the singlet ground state takes place by exchange, whereas the $B\,{}^1\Sigma_u^+$ transition is dipole allowed, so that these channels test both the exchange and the direct amplitudes. The \ASTRA{} calculations are shown for $\ell_{\max}=4$ and 6.
For the $b\,{}^3\Sigma_u^+$ state, Fig.~\ref{fig:h2-inelastic-ics}(a), the \ASTRA{} calculations reproduce the rapid rise at threshold, the broad maximum below 12~eV, and the subsequent decrease, with narrow resonance structures near 12~eV--13~eV. The $\ell_{\max}=4$ and 6 calculations both give approximately $0.539\,a_0^2$ at 25~eV, about 4\% above the \UKRmolplus{} value of $0.518\,a_0^2$. MCCC has the same rise and fall but lies below the \ASTRA{} and \UKRmolplus{} results over much of the intermediate-energy range, and gives $0.4897\,a_0^2$ at 25~eV.
For the $a\,{}^3\Sigma_g^+$ state, Fig.~\ref{fig:h2-inelastic-ics}(b), the two \ASTRA{} calculations follow \UKRmolplus{}, with a sharp resonance near threshold and a decreasing background. At 25~eV, the $\ell_{\max}=4$ and 6 results are $0.1074$ and $0.1054\,a_0^2$, respectively, and the \UKRmolplus{} and MCCC values are $0.1068$ and $0.1037\,a_0^2$. MCCC follows the decreasing background, with smaller values over parts of the intermediate-energy range.
For the dipole-allowed $B\,{}^1\Sigma_u^+$ excitation, Fig.~\ref{fig:h2-inelastic-ics}(c), the two calculations differ. The $\ell_{\max}=4$ result lies below \UKRmolplus{} at higher energies and gives $0.640\,a_0^2$ at 25~eV, approximately 21\% below the benchmark value of $0.808\,a_0^2$, whereas the $\ell_{\max}=6$ result nearly overlaps \UKRmolplus{} over most of the energy range and gives $0.8058\,a_0^2$ at 25~eV, within about 0.3\% of the benchmark. The stronger model dependence of this dipole-allowed transition is consistent with sensitivity to higher projectile partial waves. While the two calculations also use different outer radii, the dependence on the radius is minimal so the comparison is indicative of the angular-cutoff effect. MCCC follows the same increasing trend but lies below \UKRmolplus{} and the $\ell_{\max}=6$ result at higher energies, reaching $0.7610\,a_0^2$ at 25~eV, about 6\% below \UKRmolplus{}. Finally, in this case the calculations with $\ell_{max}=4$ and 6 do differ, and both \ASTRA{} and \UKRmolplus{} are capped at $\ell_{max}=6$, so we cannot exclude that both calculations, while coinciding, are still unconverged in $\ell$.
For the $c\,{}^3\Pi_u$ state, Fig.~\ref{fig:h2-inelastic-ics}(d), the two \ASTRA{} results nearly coincide and follow \UKRmolplus{}, giving $0.2181$ and $0.2184\,a_0^2$ at 25~eV for $\ell_{\max}=4$ and 6, respectively. MCCC lies below the \ASTRA{} and \UKRmolplus{} results over much of the 14~eV-23~eV range and has no maxima near 16.5 and 19~eV. Around the latter, the difference concerns the magnitude of the cross section as well as the presence of a narrow structure. The results approach one another toward 25~eV, where \UKRmolplus{} and MCCC give $0.211$ and $0.2157\,a_0^2$, respectively.

\begin{figure*}[!t]
\centering

\begin{minipage}[t]{0.48\textwidth}
\centering
\panelgraphic[0.46,0.94]{\linewidth}{(a)}
{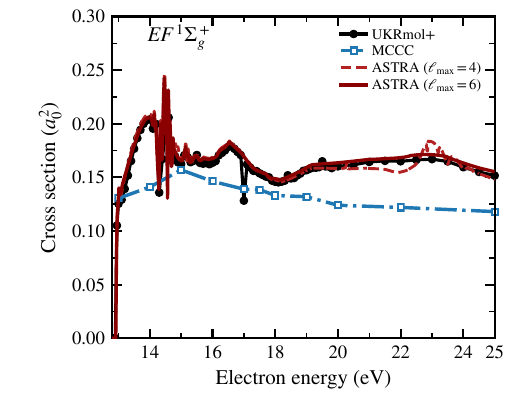}
\end{minipage}
\hfill
\begin{minipage}[t]{0.48\textwidth}
\centering
\panelgraphic[0.68,0.94]{\linewidth}{(b)}
{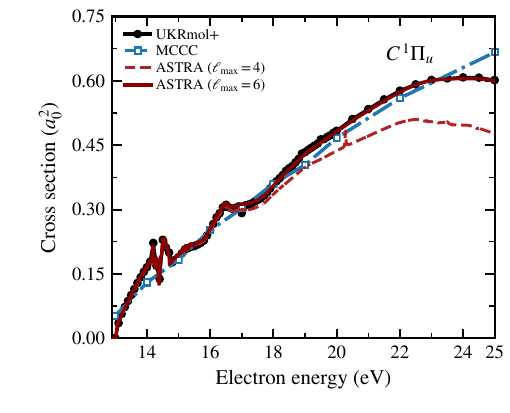}
\end{minipage}

\par\vspace{1ex}

\begin{minipage}[t]{0.48\textwidth}
\centering
\panelgraphic[0.46,0.94]{\linewidth}{(c)}
{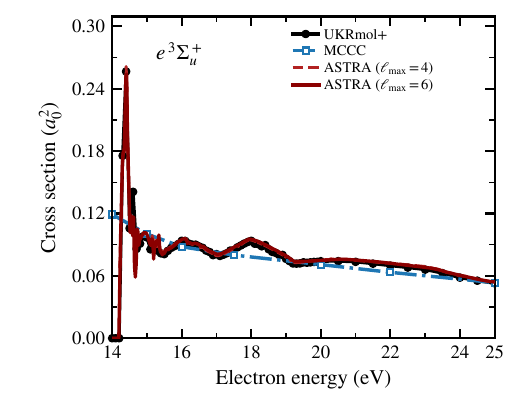}
\end{minipage}
\hfill
\begin{minipage}[t]{0.48\textwidth}
\centering
\panelgraphic[0.46,0.94]{\linewidth}{(d)}
{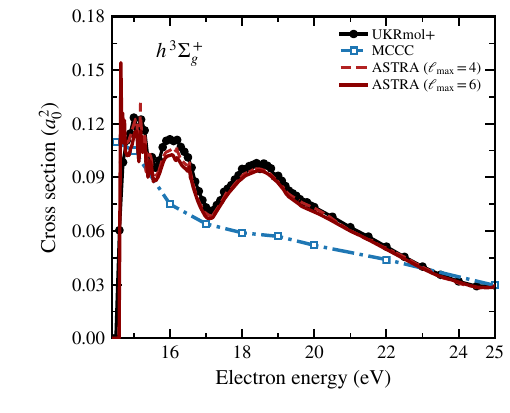}
\end{minipage}

\caption{
Fixed-nuclei integral cross sections for excitation from
$\mathrm{H}_2(X\,{}^1\Sigma_g^+)$ to
(a) $EF\,{}^1\Sigma_g^+$,
(b) $C\,{}^1\Pi_u$,
(c) $e\,{}^3\Sigma_u^+$, and
(d) $h\,{}^3\Sigma_g^+$, at $R_0=1.448\,a_0$.
The \ASTRA{} results are shown for
$\ell_{\max}=4$ and 6.
The \UKRmolplus{} benchmarks are from
Meltzer \textit{et al.}~\cite{Meltzer-JPB-20}.
The fixed-nuclei MCCC results of
Zammit \textit{et al.}~\cite{Zammit-PRL-16,Zammit-PRA-17}
were obtained from the CCC database on LXCat~\cite{Scarlett-ADNDT-21,Pitchford-PPP-17}.
}
\label{fig:h2-higher-inelastic-ics}
\end{figure*}

Figure~\ref{fig:h2-higher-inelastic-ics} extends the comparison to the $EF\,{}^1\Sigma_g^+$, $C\,{}^1\Pi_u$, $e\,{}^3\Sigma_u^+$, and $h\,{}^3\Sigma_g^+$ states, shown in panels~(a)--(d), respectively. For the $EF\,{}^1\Sigma_g^+$ state, both \ASTRA{} calculations follow \UKRmolplus{} and reproduce the main variations of the cross section, with residual differences in localized resonance structures. MCCC is lower over most of the energy range. At 25~eV, the $\ell_{\max}=6$ result is approximately 2.4\% above \UKRmolplus{} and 31\% above MCCC.
The dipole-allowed $C\,{}^1\Pi_u$ transition, Fig.~\ref{fig:h2-higher-inelastic-ics}(b), behaves like the $B\,{}^1\Sigma_u^+$ one in Fig.~\ref{fig:h2-inelastic-ics}(c). The $\ell_{\max}=4$ calculation lies below \UKRmolplus{}, and the $\ell_{\max}=6$ result nearly overlaps it, differing by only about 0.4\% at 25~eV. MCCC has a similar rising background but continues to increase toward the upper end of the range, where the $\ell_{\max}=6$ result is approximately 10\% lower.
For the $e\,{}^3\Sigma_u^+$ state, Fig.~\ref{fig:h2-higher-inelastic-ics}(c), the \ASTRA{} results nearly coincide and follow \UKRmolplus{}, including the sharp resonance near threshold and the variations of the decreasing background. MCCC follows the decrease at higher energies and differs from the $\ell_{\max}=6$ result by about 1\% at 25~eV. Its sparse energy grid does not allow a comparison of the narrow structures near threshold.
For the $h\,{}^3\Sigma_g^+$ state, Fig.~\ref{fig:h2-higher-inelastic-ics}(d), the two \ASTRA{} calculations give similar cross sections and reproduce the main features of \UKRmolplus{}. MCCC lies below these results between approximately 15~eV and 22~eV, and approaches them at higher energies. At 25~eV, the $\ell_{\max}=6$ result is approximately 1.2\% below \UKRmolplus{} and 1.1\% below MCCC.

The agreement with \UKRmolplus{} extends to all eight excitation channels examined. MCCC, which uses a single-centre Laguerre pseudostate expansion rather than the Gaussian-basis target description employed by \ASTRA{} and \UKRmolplus{}~\cite{Masin-CPC-20,Meltzer-JPB-20}, provides an independent comparison. Its expansion also includes the ionization channels, represented by positive-energy pseudostates, to which Meltzer \textit{et al.}~\cite{Meltzer-JPB-20} attribute the lower MCCC cross sections for weak transitions above the ionization threshold. Since \ASTRA{} and \UKRmolplus{} use the same target expansion, the agreement between the two sets of observables shows that our molecular implementation is correct. It is worth pointing out that this agreement is not sufficient to conclude that either of the two calculations is converged. Indeed, the results at high energy may require a still larger orbital angular momentum.

\begin{figure*}[hbtp!]
\centering
\includegraphics[width=\textwidth]
{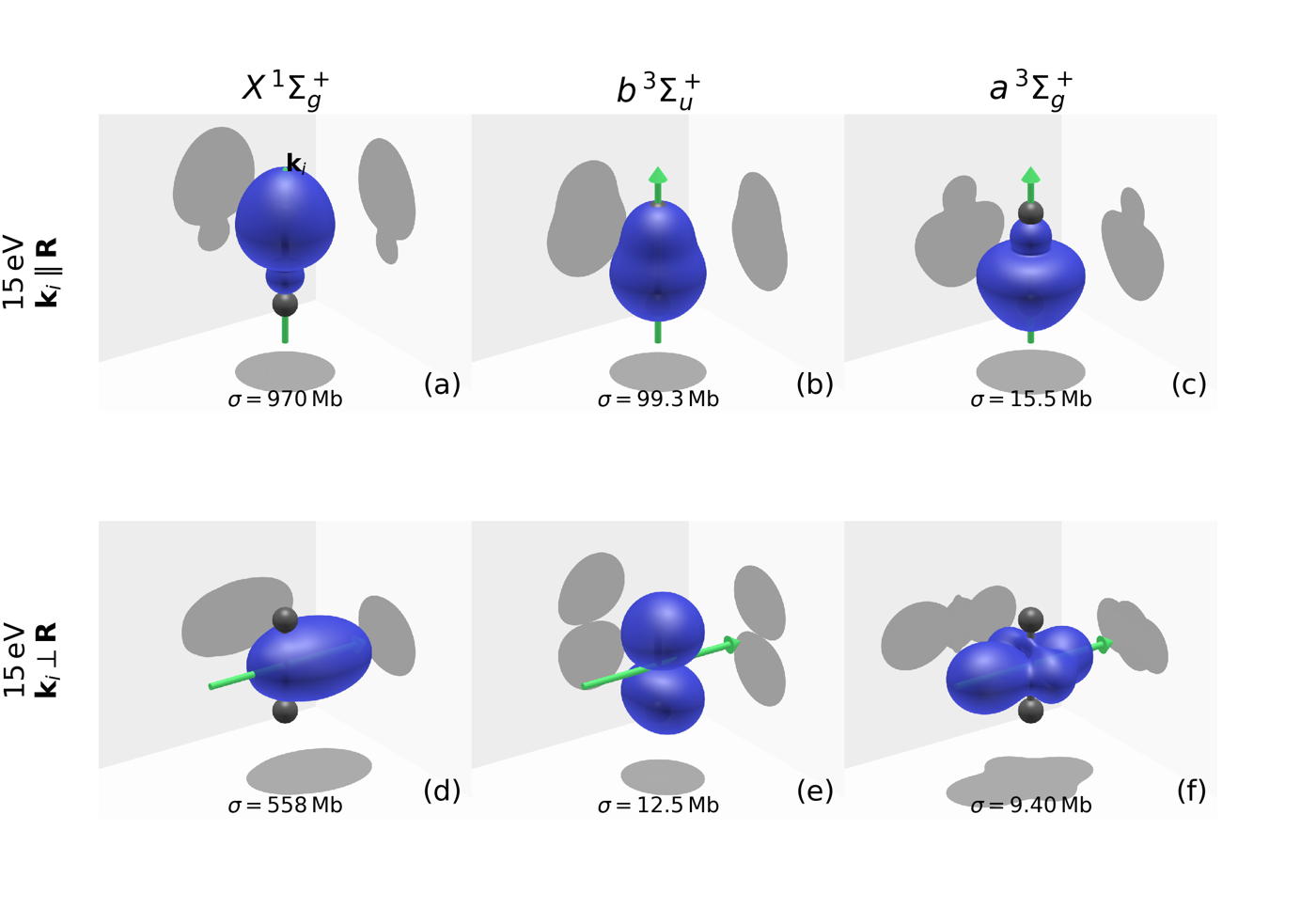}
\caption{\label{fig:h2-molecular-frame-dcs-15ev}
Molecular-frame scattering angular distribution (MFSAD) for electron scattering from $\mathrm{H}_2$ at an incident energy of 15~eV. The columns correspond to elastic scattering from the $X\,{}^1\Sigma_g^+$ ground state and excitation of the $b\,{}^3\Sigma_u^+$ and $a\,{}^3\Sigma_g^+$ states. The upper row shows the parallel geometry, $\mathbf{k}_i\parallel\mathbf{R}$, while the lower row shows the perpendicular geometry, $\mathbf{k}_i\perp\mathbf{R}$. The green arrow indicates the incident momentum $\mathbf{k}_i$, and the molecular axis $\mathbf{R}$ is aligned with the $z$ axis. The blue surfaces represent the molecular-frame differential cross section $d\sigma^{\mathrm{MF}}/d\Omega_f$ for the indicated fixed incident orientation. The value of $\sigma$ shown in each panel is obtained by integrating the corresponding differential cross section over the outgoing direction $\Omega_f$, while keeping the incident direction fixed. Cross sections reported in the panels are given in Mb ($1~\mathrm{Mb}=10^{-18}\,\mathrm{cm}^2$). The gray silhouettes show projections of the three-dimensional angular distributions onto the Cartesian coordinate planes. The distributions are computed with the 98-target close-coupling expansion and $\ell_{\max}=6$.}
\end{figure*}

\subsection{Orientation-resolved molecular-frame scattering angular distributions}
\label{subsec:h2-molecular-frame-dcs}

To illustrate the dependence of molecular-frame electron scattering on the orientation of the molecule, we consider two geometries, $\mathbf{k}_i\parallel\mathbf{R}$ and $\mathbf{k}_i\perp\mathbf{R}$, at an incident energy of 15~eV. At this energy the ten target states of Table~\ref{tab:h2-target-states} are open. We show elastic scattering and the two lowest excitations, to the $b\,{}^3\Sigma_u^+$ and $a\,{}^3\Sigma_g^+$ states. 
Figure~\ref{fig:h2-molecular-frame-dcs-15ev} shows that both the shape and the magnitude of the angular distributions depend strongly on the orientation of the incident momentum relative to the molecular axis. For parallel incidence, the collision is cylindrically symmetric about the common direction of $\mathbf{k}_i$ and $\mathbf{R}$. For perpendicular incidence this symmetry is broken, and the differential cross section depends on both the polar and the azimuthal scattering angles.

The cross section for a fixed incident direction is the integral of the molecular-frame differential cross section over the direction of the scattered electron,
\begin{equation}
\sigma_{f\leftarrow i}^{\mathrm{MF}}
(E;\hat{\mathbf{k}}_i)
=
\int d\Omega_f\,
\frac{d\sigma_{f\leftarrow i}^{\mathrm{MF}}
(E;\hat{\mathbf{k}}_i,\hat{\mathbf{k}}_f)}
{d\Omega_f}.
\label{eq:h2-fixed-incidence-ics}
\end{equation}
The values of $\sigma$ displayed in Fig.~\ref{fig:h2-molecular-frame-dcs-15ev} are those of Eq.~\eqref{eq:h2-fixed-incidence-ics} for the two geometries.

The elastic distributions are shown in Figs.~\ref{fig:h2-molecular-frame-dcs-15ev}(a) and \ref{fig:h2-molecular-frame-dcs-15ev}(d). For $\mathbf{k}_i\parallel\mathbf{R}$, the distribution has a large forward lobe and a smaller backward one. For $\mathbf{k}_i\perp\mathbf{R}$, the forward lobe remains and the backward one disappears, while the fixed-incidence cross section decreases from $970$~Mb to $558$~Mb.

The dependence on the orientation is stronger for the excitation of the $b\,{}^3\Sigma_u^+$ state, Figs.~\ref{fig:h2-molecular-frame-dcs-15ev}(b) and \ref{fig:h2-molecular-frame-dcs-15ev}(e). For parallel incidence, the distribution is strongly asymmetric along the molecular axis, with the larger lobe directed opposite to $\mathbf{k}_i$. For perpendicular incidence, the distribution has a nodal plane that bisects the internuclear axis and is approximately symmetric about that axis, and the integrated cross section drops by about a factor of eight, from $99.3$~Mb for $\mathbf{k}_i\parallel\mathbf{R}$ to only $12.5$~Mb for $\mathbf{k}_i\perp\mathbf{R}$. At this energy, therefore, the excitation of the $b\,{}^3\Sigma_u^+$ state strongly selects molecules aligned with the incident momentum.

The nodal plane for perpendicular incidence follows from reflection symmetry. The incident wave and the $X\,{}^1\Sigma_g^+$ target are even under reflection through the plane perpendicular to the molecular axis at its midpoint, whereas the $b\,{}^3\Sigma_u^+$ target is odd. The outgoing angular amplitude must therefore be odd under this reflection and vanish when $\mathbf{k}_f$ lies in the plane.
The $a\,{}^3\Sigma_g^+$ channel, Figs.~\ref{fig:h2-molecular-frame-dcs-15ev}(c) and \ref{fig:h2-molecular-frame-dcs-15ev}(f), also changes with the orientation. For parallel incidence, its distribution is almost the mirror image of the elastic one. For perpendicular incidence, it has several lobes, mostly along the direction of the incident beam, with an approximate forward-backward symmetry. This large change in shape is accompanied by a moderate change in the integrated cross section, from $\sigma=15.5$~Mb for parallel incidence to $9.40$~Mb for perpendicular incidence. For this transition, the orientation redistributes the scattered electrons more than it changes their number.

In elastic scattering, the direct interaction favors small momentum transfer, which is reflected in the forward lobe present in both geometries. This is to be contrasted with the inelastic case, where the momentum transfer cannot fall below the difference between the incident and outgoing momenta. Furthermore, excitation of the triplet states requires exchange, a short-range process that does not favor small momentum transfer. The inelastic MFSAD, therefore, is not expected to be biased in the forward direction. The approximate forward-backward symmetry of both the $a\,{}^3\Sigma_g^+$ and the $b\,{}^3\Sigma_u^+$ distributions at perpendicular incidence indicates that the backward preference observed at parallel incidence is not intrinsic to these transitions, but is associated with scattering along the internuclear axis, where two-center effects are most pronounced.

The integral cross sections presented above are the average of Eq.~\eqref{eq:h2-fixed-incidence-ics} over the incident direction,
\begin{equation}
\sigma_{f\leftarrow i}(E)
=
\frac{1}{4\pi}
\int d\Omega_i\,
\sigma_{f\leftarrow i}^{\mathrm{MF}}
(E;\hat{\mathbf{k}}_i).
\label{eq:h2-mf-to-ics}
\end{equation}
We evaluated the angular integrals in Eqs.~\eqref{eq:h2-fixed-incidence-ics} and \eqref{eq:h2-mf-to-ics} numerically for the open channels and verified that they reproduce Eq.~\eqref{eq:ics-total} to numerical precision. The agreement between the angularly integrated molecular-frame cross sections and the independently calculated integral cross sections provides an internal consistency check on the normalization, channel summations, and angular quadrature. The present work focuses on orientation-resolved molecular-frame distributions. Orientationally averaged laboratory-frame differential cross sections and their comparison with benchmark data is beyond the scope of the present work and will be presented in a separate study.

\section{Conclusions and Perspectives}
\label{sec:conclusions}

We have extended \ASTRA{}, a close-coupling code for molecular ionization based on TDMs between correlated target states described at the CAS level, to electron-target scattering. The new implementation provides channel-resolved scattering amplitudes, integral cross sections, and differential cross sections. We have validated it for helium and molecular hydrogen, for which extensive benchmark data are available.

For helium, \ASTRA{} reproduces the measured elastic cross section and the benchmark integral cross sections for excitation of the $n=2$ states. The more stringent test is the angular distribution of the inelastically scattered electrons. Here, in most of the cases examined, the \ASTRA{} results agree with experiment markedly better than previous calculations, notably for the position of the minimum and the height of the maximum in the $1s2s\,{}^3S^e$ channel, for the maximum in the $1s2p\,{}^3P^o$ channel, and for the position and depth of the interference minimum in the $1s2s\,{}^1S^e$ channel. The lowest $\mathrm{He}^{-}(1s2s^2\,{}^2S^e)$ Feshbach resonance, which probes the correlated dynamics of the transient negative ion, is reproduced with the recommended width of 11~meV.

For molecular hydrogen in the fixed-nuclei approximation, the \ASTRA{} cross sections for elastic scattering and for the eight excitation channels examined, computed with the close-coupling expansion of the \UKRmolplus{} benchmark, are virtually identical to the \UKRmolplus{} ones. Since the two results are obtained with independent codes, their agreement is a strong validation of the correctness of the \ASTRA{} implementation for a molecular target. For the dipole-allowed $B\,{}^1\Sigma_u^+$ and $C\,{}^1\Pi_u$ transitions, the agreement requires $\ell_{\max}=6$. With $\ell_{\max}=4$, their cross sections at 25~eV are about 20\% lower. The molecular-frame distributions reveal an orientation dependence that would be hidden by rotational averaging. At 15~eV, excitation of $b\,{}^3\Sigma_u^+$ is about eight times stronger for parallel than perpendicular incidence.

The present calculations were designed to establish the correctness and the numerical reliability of the \ASTRA{} electron-scattering formulation. Future work will address more complex molecules and the laboratory-frame angular distribution of the scattered electron, which also gives the momentum-transfer cross section. Of particular interest is the application of \ASTRA{} to electron scattering from $\mathrm{CF}_4$ and larger fluorinated molecules, to generate internally consistent collision datasets. Such datasets, which comprise the momentum-transfer and the inelastic cross sections, are the input of Boltzmann solvers~\cite{Hagelaar-PSST-05}, which yield the electron-transport coefficients and the rate coefficients used in plasma-chemistry models. The energy- and angle-resolved molecular-frame amplitudes are also the field-free scattering parameters required by quantitative rescattering theory and by laser-induced electron diffraction.

On the methodological side, we plan to adopt generalized-active-space
(GAS) target descriptions~\cite{Olsen1988,Fleig-JCP-01,Ma-JCP-11}, enabling larger orbital spaces with controlled
restrictions on the configuration expansion and greater flexibility
in treating target correlation. We also plan to introduce a complementary
$(N+1)$-electron correlation space ($\mathcal Q$ space) to describe
residual target--projectile correlation beyond the
close-coupling channel basis. The inclusion of electron-impact ionization channels remains an open option for the future. Finally, nuclear motion, including vibrational averaging, will be incorporated where it is required for a quantitative comparison near excitation thresholds. The longer-term objective is to make \ASTRA{} a tool that produces accurate electron--molecule collision datasets for transport and plasma modeling and training data for machine-learning models that cover large chemical spaces.

\section*{Acknowledgments}
This work is supported by CHIPS Project 5.07: Coupling Computation and Machine Learning to Develop Novel Methods to Measure and Replace Environmentally Hazardous PFAS Chemicals. LA and JD acknowledge support from DOE Grant No. DE-SC0026377 (Attosecond Molecular Spectroscopy beyond the Fixed-Nuclei Approximation). SC acknowledges support from OpenAI through the ChatGPT for Academic Researchers program, which provided access to some of the AI tools (GPT-5.6 Sol, GPT-6 Astra) used in this work. CM acknowledges support by NSF MPS-Ascend Postdoctoral Research Fellowship under Grant No. 2402225.

\appendix
\section{Derivation of the Molecular-Frame Cross Sections\label{app:mf-derivation}}
We derive the integral and molecular-frame differential cross sections within the fixed-nuclei approximation. Atomic units ($\hbar=m_e=e=1$) are used throughout. Consider a target with a fixed orientation $\mathcal R$ relative to the laboratory axes, as identified for example by the Euler angles~\cite{Varshalovich1988},
\begin{equation}
\mathrm{M}_{a\Sigma_a}(\mathcal R) + e^{-}_{\vec{k}\sigma} \longrightarrow \mathrm{M}_{b\Sigma_b}(\mathcal R) + e^{-}_{\vec{k}'\sigma'},
\end{equation}
where $a$ and $\Sigma_a$ are the initial molecular electronic state and spin projection, $\vec{k}$ and $\sigma$ are the initial electron momentum and spin projection, and $b$, $\Sigma_b$, $\vec{k}'$, $\sigma'$ are the corresponding quantum numbers in the final state. 
The incoming asymptotic state corresponding to this process is the antisymmetrized product of the initial molecular state $\Phi_{a\Sigma_a}(\bar{\mathbf{x}};\mathcal R)$ and a plane wave
\begin{equation}\label{eq:InStatePlaneWave}
\begin{split}
    \psi_{a\Sigma_a\vec{k}\sigma}^{in}(\mathbf{x};\mathcal R) ={}& \frac{1-\sum_{i=1}^{N}\hat{\mathcal{P}}_{i,N+1}}{\sqrt{(N+1)k}}\, \Phi_{a\Sigma_a}(\bar{\mathbf{x}};\mathcal R)\times\\
    &\times{^2\chi_\sigma(\zeta)}\,e^{i\vec{k}\cdot\vec{r}},
\end{split}
\end{equation}
where $N$ is the number of electrons of the target, $\bar{\mathbf{x}}=(x_1,\,x_2,\,\ldots,\,x_{N})$ collects their spatial and spin coordinates, $x_i=(\vec{r}_i,\zeta_i)$, and $\mathbf{x}=(\bar{\mathbf{x}},x_{N+1})$ those of all the electrons. The coordinates of the last electron, the projectile, are written without index, $x_{N+1}=(\vec{r},\zeta)$. Furthermore, $^2\chi_\sigma(\zeta)=\delta_{\sigma\zeta}$ is the doublet spin function, with $\sigma,\,\zeta=\pm\frac{1}{2}$, and $\hat{\mathcal{P}}_{ij}$ swaps the spatial and spin coordinates of electrons $i$ and $j$.
These states are normalized to have unit incident flux density, as can be verified by direct computation using the formula for the electron current, in which $\vec{\nabla}$ acts on $\vec{r}$,
\begin{equation}\begin{split}\label{eq:CurrentFormula}
\vec{j}=\frac{N+1}{2i}\sum_{\zeta}\int dx_1\cdots dx_{N}\big[\psi^*\vec{\nabla}\psi-\psi\vec{\nabla}\psi^*\big].
\end{split}\end{equation}
With this normalization the differential cross section, defined as the ratio of the outgoing electron flux per unit solid angle to the incoming current density, coincides with the former. 
The incoming states can be used as a basis for expanding the system's wave packet $\Psi(\mathbf{x},t;\mathcal R)$ before the collision happens, as $t\to-\infty$, 
\begin{equation}\label{eq:IncomingWavePacket}
\begin{split}
    \Psi^{in}(\mathbf{x},t;\mathcal R) = \int d^3 k\,& \psi^{in}_{a\Sigma_a\vec{k}\sigma} (\mathbf{x};\mathcal R)\,c(\vec{k})\times\\
    &\times e^{-i(E_a+k^2/2)t},
\end{split}
\end{equation}
where $c(\vec{k})$ is a square-integrable amplitude narrowly peaked around $\vec{k}_i$, the nominal initial electron momentum, and $E_a$ is the energy of the molecular electronic state $a$. At sufficiently negative times the projectile electron does not interact with the target molecule, so that the dynamics is governed by the sum of the Hamiltonians of the molecule and of the free projectile, which is the origin of the phase factor $e^{-i(E_a+k^2/2)t}$ in Eq.~\eqref{eq:IncomingWavePacket}.
When the initial state of the system is prepared in this way, the state of the system at any time $t$ can be expressed as~\cite{Newton2002},
\begin{equation}\label{eq:wave-packet-evolution}
    \Psi^+(\mathbf{x},t;\mathcal R) = \lim_{t'\to-\infty}e^{-i\hat{H}(t-t')}\Psi^{in}(\mathbf{x},t';\mathcal R),
\end{equation}
where $\hat{H}$ is the full Hamiltonian of the system. The wave packet $\Psi^+(\mathbf{x},t;\mathcal R)$ satisfies the time-dependent Schrödinger equation, $i\partial_t \Psi^+(\mathbf{x},t;\mathcal R) = \hat{H} \Psi^+(\mathbf{x},t;\mathcal R)$, and converges to $\Psi^{in}(\mathbf{x},t;\mathcal R)$ in norm, as $t\to-\infty$ by construction
\begin{equation}
    \lim_{t\to-\infty} \|\Psi^+(\mathbf{x},t;\mathcal R)-\Psi^{in}(\mathbf{x},t;\mathcal R) \|=0.
\end{equation}
By defining the stationary scattering states $\psi^+$ as
\begin{equation}\label{eq:stationary-scattering-state}
\begin{split}
    \psi^+_{a\Sigma_a\vec{k}\sigma}(\mathbf{x};\mathcal R) = \lim_{t'\to-\infty}&e^{-i(E_a+k^2/2-\hat{H})t'}\times\\
    &\times\psi^{in}_{a\Sigma_a\vec{k}\sigma}(\mathbf{x};\mathcal R),
\end{split}
\end{equation}
we can express the wave packet at any time before, during, or after the collision, by means of the exact relation
\begin{equation}\label{eq:Psi+WavePacket}
\begin{split}
    \Psi^+(\mathbf{x},t;\mathcal R) = \int d^3 k\,& \psi^{+}_{a\Sigma_a\vec{k}\sigma} (\mathbf{x};\mathcal R)\,c(\vec{k})\times\\
    &\times e^{-i(E_a+k^2/2)t}.
\end{split}
\end{equation}
This expression is useful to visualize the collision of a specific wave packet with the molecule, as a function of time.
A $\psi^+$ state is a generalized eigenstate of the full Hamiltonian, $\hat{H}\psi^{+}_{a\Sigma_a\vec{k}\sigma} (\mathbf{x};\mathcal R)=(E_a+k^2/2)\psi^{+}_{a\Sigma_a\vec{k}\sigma} (\mathbf{x};\mathcal R)$, whose asymptotic incoming component coincides with that of the corresponding $in$ state,
$[\psi^{+}_{a\Sigma_a\vec{k}\sigma} (\mathbf{x};\mathcal R)]_{in}=[\psi^{in}_{a\Sigma_a\vec{k}\sigma} (\mathbf{x};\mathcal R)]_{in}$.

At large times, $t\to\infty$, the wave packet takes contributions only from the outgoing part of the scattering states $\psi^+_{a\Sigma_a\vec{k}\sigma} (\mathbf{x};\mathcal R)$, which accounts for both scattered and unscattered electrons. In a scattering experiment, the impinging electrons are collimated in a narrow beam, and the detector is positioned in such a way that unscattered electrons never reach it. These conditions can be theoretically reproduced with Eq.~\eqref{eq:Psi+WavePacket}, by choosing a suitable transverse distribution for the momenta of the incoming electron. When the electron dynamics is not resolved in time and only the asymptotic flux of electrons in a stationary regime is of interest, it is convenient to abandon the wave-packet formulation and adopt the limit of an incoming plane wave, $c(\vec{k})\to \delta^{(3)}(\vec{k}-\vec{k}_i)$. In this limit the incoming wave extends over all space, and the unscattered electron contributes to the outgoing flux wherever the detector is located. This contribution is eliminated by removing from the scattering state the amplitude of the unscattered beam, which leaves the scattering component $\psi^{scat}_{a\Sigma_a\vec{k}_i\sigma}(\mathbf{x};\mathcal R)$ of the $\psi^+$ state, 
\begin{equation}
\psi^{scat}_{a\Sigma_a\vec{k}_i\sigma}(\mathbf{x};\mathcal R)=\psi^{+}_{a\Sigma_a\vec{k}_i\sigma}(\mathbf{x};\mathcal R)-\psi^{in}_{a\Sigma_a\vec{k}_i\sigma}(\mathbf{x};\mathcal R).
\end{equation}
To determine the differential inelastic scattering cross section $d\sigma_{b\Sigma_b\sigma'\hat{\Omega}\gets a\Sigma_a\vec{k}_i\sigma}/d\Omega$, it is sufficient to evaluate, from $\psi^{scat}$, the asymptotic outgoing electron flux per unit solid angle associated with $b$. The two quantities coincide, since the $in$ state is normalized to unit incoming current density. 

\ASTRA{} computes the scattering states of the Hamiltonian with well-defined spin $S$, spin projection $\Sigma$, and spatial symmetry $\Gamma$, within the close-coupling \emph{ansatz}~\cite{Marante2026}. Within our conventions for the asymptotic phase and angular distributions of the continuum channels, the scattering state $\psi^{S\Sigma\Gamma(+)}_{\alpha E}(\mathbf{x};\mathcal R)$ has the following expression, as $r\to\infty$, 
\begin{equation}\begin{split}
    &\psi^{S\Sigma\Gamma(+)}_{\alpha E}(\mathbf{x};\mathcal R) = \Xi_{\alpha}^{S\Sigma\Gamma}(\bar{\mathbf{x}},\hat{r},\zeta;\mathcal R)\times\\
    &\qquad\times\frac{e^{-i(k_i r-\ell_\alpha \pi/2)}}{\sqrt{2\pi k_i (N+1)}\, r}-\\
    &\quad-\sum_\beta \Xi_{\beta}^{S\Sigma\Gamma}(\bar{\mathbf{x}},\hat{r},\zeta;\mathcal R)\times\\
    &\qquad\times\frac{e^{i(k_\beta r-\ell_\beta \pi/2)}}{\sqrt{2\pi k_\beta (N+1)}\, r}\,S^{S\Gamma}_{\beta\alpha}(E;\mathcal R),
\end{split}\end{equation}
where $E = E_a + k_i^2/2 = E_b + k_\beta^2/2$, $\alpha=(a\ell_\alpha m_\alpha)$ [or $\beta = (b\ell_\beta m_\beta)$, etc.] is a set of quantum numbers that uniquely identifies, within the space with spin $S$ and spatial symmetry $\Gamma$, an asymptotic channel in the electronic continuum, and $\mathbf{S}^{S\Gamma}(E)$ is the scattering matrix. The channel functions $\Xi_{\alpha}^{S\Sigma\Gamma}(\bar{\mathbf{x}},\hat{r},\zeta;\mathcal R)$ are defined as
\begin{equation}
\begin{split}
    &\Xi_{\alpha}^{S\Sigma\Gamma}(\bar{\mathbf{x}},\hat{r},\zeta;\mathcal R) = \sum_{\Sigma_a \sigma} C_{S_a\Sigma_a,\frac{1}{2}\sigma}^{S\Sigma}\times\\
    &\qquad\times\Phi_{a\Sigma_a}(\bar{\mathbf{x}};\mathcal R)\,{^2\chi_\sigma(\zeta)}\,X_{\ell_\alpha m_\alpha}(\hat{r}).
\end{split}
\end{equation}
The angular functions $X_{\ell m}(\hat{r})$ are real spherical harmonics, which in \ASTRA{} are expressed in terms of standard spherical harmonics $Y_{\ell m}(\hat{r})$ as follows, where $m>0$:
\begin{equation}\begin{split}\label{eq:real-harmonics-definition}
X_{\ell m} &= (-1)^m\sqrt{2}\Re[Y_{\ell m}],\\
X_{\ell 0} &= Y_{\ell 0},\\
X_{\ell -m} &= (-1)^m\sqrt{2}\Im[Y_{\ell m}],
\end{split}\end{equation}
or, equivalently,
\begin{equation}\begin{split}\label{eq:real-to-complex-harmonics}
X_{\ell 0} &= Y_{\ell 0},\\
X_{\ell m} &= \frac{1}{\sqrt{2}}\left[ (-1)^m Y_{\ell m} + Y_{\ell -m}\right],\\
X_{\ell -m} &= \frac{1}{i\sqrt{2}}\left[(-1)^m Y_{\ell m} - Y_{\ell -m}\right].
\end{split}\end{equation}
Conversely, again for $m>0$,
\begin{equation}\begin{split}\label{eq:YtoX}
Y_{\ell m} &= \frac{(-1)^m}{\sqrt{2}}(X_{\ell m} + i X_{\ell -m}),\\
Y_{\ell 0} &= X_{\ell 0},\\
Y_{\ell -m} &=\frac{1}{\sqrt{2}}(X_{\ell m} - i X_{\ell -m}).
\end{split}
\end{equation}

These spin- and symmetry-adapted scattering states are normalized as
\begin{equation}
\begin{split}
    \langle \psi^{S\Sigma\Gamma(+)}_{\alpha E} |\psi^{S'\Sigma'\Gamma'(+)}_{\beta E'}\rangle ={}& \delta_{SS'}\delta_{\Sigma\Sigma'}\delta_{\Gamma\Gamma'}\times\\
    &\times\delta_{\alpha\beta}\,\delta(E-E').
\end{split}
\end{equation}
For the differential cross sections it is convenient to switch the angular dependence of these states from real to complex spherical harmonics, using Eq.~\eqref{eq:real-to-complex-harmonics}, which defines a new set of scattering states
\begin{equation}\begin{split}\label{eq:Psi+Tilde}
    &\tilde{\psi}^{S\Sigma(+)}_{\alpha E}(\mathbf{x};\mathcal R) = \tilde{\Xi}_{\alpha}^{S\Sigma}(\bar{\mathbf{x}},\hat{r},\zeta;\mathcal R)\times\\
    &\qquad\times\frac{e^{-i(k_i r-\ell_\alpha \pi/2)}}{\sqrt{2\pi k_i (N+1)}\, r}-\\
    &\quad-\sum_\beta \tilde{\Xi}_{\beta}^{S\Sigma}(\bar{\mathbf{x}},\hat{r},\zeta;\mathcal R)\times\\
    &\qquad\times\frac{e^{i(k_\beta r-\ell_\beta \pi/2)}}{\sqrt{2\pi k_\beta (N+1)}\, r}\,\tilde{S}^S_{\beta\alpha}(E;\mathcal R),
\end{split}\end{equation}
where 
\begin{equation}\label{eq:ChannelFunctions}
\begin{split}
    &\tilde{\Xi}_{\alpha}^{S\Sigma}(\bar{\mathbf{x}},\hat{r},\zeta;\mathcal R) = \sum_{\Sigma_a \sigma} C_{S_a\Sigma_a,\frac{1}{2}\sigma}^{S\Sigma}\times\\
    &\qquad\times\Phi_{a\Sigma_a}(\bar{\mathbf{x}};\mathcal R)\,{^2\chi_\sigma(\zeta)}\,Y_{\ell_\alpha m_\alpha}(\hat{r}).
\end{split}
\end{equation}
In this formulation the spatial symmetry $\Gamma$ no longer appears, and the angular quantum numbers of the channel $\alpha=(a,\ell,m)$ are unrestricted.

To determine the asymptotic expression of a $\psi^{+}_{a\Sigma_a\vec{k}\sigma}(\mathbf{x};\mathcal R)$ state in terms of the scattering matrices computed in \ASTRA{}, it is sufficient to find the linear combination of $\psi^{S\Sigma\Gamma(+)}_{\alpha E}(\mathbf{x};\mathcal R)$ states that fulfills the same incoming boundary conditions as $\psi^{in}_{a\Sigma_a\vec{k}\sigma}(\mathbf{x};\mathcal R)$. To find out how the latter state relates to the former, we start from the expansion of a plane wave in spherical waves,
\begin{equation}\begin{split}
    e^{i\vec{k}\cdot\vec{r}} &= 4\pi \sum_{\ell m}i^\ell j_\ell(kr) Y_{\ell m}(\hat{r}) Y_{\ell m}^*(\hat{k}),
\end{split}\end{equation}
where $j_\ell(z)$ are spherical Bessel functions.
Asymptotically, $j_\ell(x) = \sin(x-\ell\pi/2)/x$, and hence, as $r\to\infty$,
\begin{equation}\begin{split}\nonumber
    &\psi_{a\Sigma_a\vec{k}_i\sigma}^{in}(\mathbf{x};\mathcal R) = \frac{2i\pi}{\sqrt{N+1}\,k_i^{3/2}r} \sum_{\ell m} i^{\ell}\, Y_{\ell m}^*(\hat{k}_i)\times\\
&\quad\times\Phi_{a\Sigma_a}(\bar{\mathbf{x}};\mathcal R)\,{^2\chi_\sigma(\zeta)}\, Y_{\ell m}(\hat{r})\times\\
&\quad\times\Big[e^{-i(k_ir-\ell\pi/2)}-e^{i(k_ir-\ell\pi/2)}\Big].
\end{split}\end{equation}
To recover the expression in terms of spin-adapted channel functions, we recouple the spin of the molecular target and of the electron projectile as
\begin{equation}\begin{split}\nonumber
    &\Phi_{a\Sigma_a}(\bar{\mathbf{x}};\mathcal R)\,{^2\chi_\sigma(\zeta)}\, Y_{\ell m}(\hat{r})=\\
    &\quad=\sum_{S\Sigma}C_{S_a\Sigma_a,\frac{1}{2}\sigma}^{S\Sigma}\, \tilde{\Xi}^{S\Sigma}_{\alpha}(\bar{\mathbf{x}},\hat{r},\zeta;\mathcal R),
\end{split}\end{equation}
with which we find
\begin{equation}\begin{split}\nonumber
    \psi_{a\Sigma_a\vec{k}_i\sigma}^{in} ={}& \frac{2i\pi}{\sqrt{N+1}\,k_i^{3/2}r} \sum_{S\Sigma}\sum_{\ell m}C_{S_a\Sigma_a,\frac{1}{2}\sigma}^{S\Sigma}\times\\
    &\times i^{\ell}\, Y_{\ell m}^*(\hat{k}_i)\,\tilde{\Xi}^{S\Sigma}_{a\ell m}(\bar{\mathbf{x}},\hat{r},\zeta;\mathcal R)\times\\
    &\times\Big[e^{-i(k_ir-\ell\pi/2)}-e^{i(k_ir-\ell\pi/2)}\Big].
\end{split}\end{equation}
Comparing this expression with Eq.~\eqref{eq:Psi+Tilde} shows that the scattering state $\psi^{+}_{a\Sigma_a\vec{k}_i\sigma}(\mathbf{x};\mathcal R)$ that matches the same incoming boundary conditions as $\psi_{a\Sigma_a\vec{k}_i\sigma}^{in}$ is
\begin{equation}\begin{split}\nonumber
    \psi^{+}_{a\Sigma_a\vec{k}_i\sigma}(\mathbf{x};\mathcal R) ={}& \frac{i(2\pi)^{3/2}}{k_i} \sum_{S\Sigma\ell m}C_{S_a\Sigma_a,\frac{1}{2}\sigma}^{S\Sigma}\, i^{\ell}\times\\
    &\times Y_{\ell m}^*(\hat{k}_i)\,\tilde{\psi}^{S\Sigma(+)}_{\alpha E}(\mathbf{x};\mathcal R).
\end{split}\end{equation}
Subtracting the incoming state, as prescribed earlier, we find
\begin{equation}\begin{split}\nonumber
    \psi^{scat}_{a\Sigma_a\vec{k}_i\sigma}&(\mathbf{x};\mathcal R)= \psi^{+}_{a\Sigma_a\vec{k}_i\sigma}(\mathbf{x};\mathcal R)-\psi^{in}_{a\Sigma_a\vec{k}_i\sigma}(\mathbf{x};\mathcal R)=\\
    &\frac{2\pi}{ik_i} \sum_{S\Sigma\ell m}C_{S_a\Sigma_a,\frac{1}{2}\sigma}^{S\Sigma} i^{\ell} Y_{\ell m}^*(\hat{k}_i)\times\\
    &\times\sum_\beta \tilde{\Xi}_{\beta}^{S\Sigma}(\bar{\mathbf{x}},\hat{r},\zeta;\mathcal R)\frac{e^{i(k_\beta r-\ell_\beta \pi/2)}}{\sqrt{k_\beta (N+1)}\, r}\times\\
    &\times[\tilde{S}^{S}_{\beta\alpha}(E;\mathcal R)-\delta_{\alpha\beta}].
\end{split}\end{equation}
To identify the channels in which the molecule is in state $b$, with spin projection $\Sigma_b$, and the electron has spin $\sigma'$, 
\begin{equation}\begin{split}
    \psi^{scat}_{a\Sigma_a\vec{k}_i\sigma}&(\mathbf{x};\mathcal R)=  \sum_{b \Sigma_b \sigma'}\psi^{scat}_{b\Sigma_b\sigma'\gets a\Sigma_a\vec{k}_i\sigma}(\mathbf{x};\mathcal R),
\end{split}\end{equation}
we use Eq.~\eqref{eq:ChannelFunctions} to expand the channel functions into their target and projectile components,
\begin{equation}\begin{split}
    &\psi^{scat}_{b\Sigma_b\sigma'\gets a\Sigma_a\vec{k}_i\sigma}(\mathbf{x};\mathcal R)= \frac{2\pi}{ik_i} \Phi_{b\Sigma_b}(\bar{\mathbf{x}};\mathcal R)\, {^2\chi_{\sigma'}(\zeta)}\times\\
    &\quad\times\sum_{S\Sigma} C_{S_a\Sigma_a,\frac{1}{2}\sigma}^{S\Sigma}C_{S_b\Sigma_b,\frac{1}{2}\sigma'}^{S\Sigma}\times\\
    &\quad\times \sum_{\ell m\ell'm'}
     i^{\ell} Y_{\ell m}^*(\hat{k}_i)Y_{\ell' m'}(\hat{r})\times\\
    &\quad\times   \frac{e^{i(k_b r-\ell' \pi/2)}}{\sqrt{k_b (N+1)}\, r}[\tilde{S}^{S}_{b\ell' m',a\ell m}(E)-\delta_{b\ell' m',a\ell m}],
\end{split}\end{equation}
where $k_b = \sqrt{2(E-E_b)}$ is the asymptotic radial momentum common to all channels $\beta$ associated with the same electronic state $b$.
Using Eq.~\eqref{eq:CurrentFormula}, one can now evaluate the partial differential cross section in channel $b\Sigma_b\sigma'$ as the radial electron flux per unit of solid angle
\begin{equation}
\frac{d\sigma^{\{\sigma\}}_{b\gets a}}{d\Omega} = \lim_{r\to\infty} r^2\, \vec{j}(\psi^{scat}_{b\Sigma_b\sigma'\gets a\Sigma_a\vec{k}_i\sigma})\cdot\hat{k}_f,
\end{equation}
where the suffix $\{\sigma\}$ indicates a quantity resolved in the spin orientation of the particles involved and $\vec{k}_f = k_b\hat{k}_f$ is the momentum of the scattered electron. The result is
\begin{equation}\begin{split}\label{eq:DiffXS}
    &\frac{d\sigma^{\{\sigma\}}_{b\gets a}}{d\Omega} = \frac{4\pi^2}{k_i^2}\Bigg|
    \sum_{S\Sigma} C_{S_a\Sigma_a,\frac{1}{2}\sigma}^{S\Sigma}C_{S_b\Sigma_b,\frac{1}{2}\sigma'}^{S\Sigma}\times\\
    &\quad\times\sum_{\ell m\ell'm'}
     i^{\ell-\ell'}\,Y_{\ell m}^*(\hat{k}_i)\,Y_{\ell' m'}(\hat{k}_f)\times\\
    &\quad\times[\tilde{S}^{S}_{b\ell' m',a\ell m}(E;\mathcal R)-\delta_{b\ell' m',a\ell m}]
    \Bigg|^2.
\end{split}\end{equation}
Normally, electrons and the target molecules are not prepared in a spin-polarized state, nor is the spin of the emerging particles measured. Furthermore, in the present work we assume that the Hamiltonian of the system is spin independent. As a result, spin polarizations carry only trivial geometrical information about the collision and hence are of no particular interest. To obtain the differential cross section for spin-insensitive measurements, we need to average Eq.~\eqref{eq:DiffXS} over the initial and sum over the final molecule and electron spin polarizations
\begin{equation}\begin{split}
    &\frac{d\sigma_{b\gets a}}{d\Omega}=\frac{1}{2(2S_a+1)}\sum_{\Sigma_a\sigma}\sum_{\Sigma_b\sigma'}\frac{d\sigma^{\{\sigma\}}_{b\gets a}}{d\Omega} =\\
    &\quad=\frac{4\pi^2}{k_i^2}\frac{1}{2(2S_a+1)}\sum_S(2S+1)\times\\
    &\quad\times\Bigg|
    \sum_{\ell m\ell'm'}
     i^{\ell-\ell'}\,Y_{\ell m}^*(\hat{k}_i)\,Y_{\ell' m'}(\hat{k}_f)\times\\
    &\quad\times
    [\tilde{S}^{S}_{b\ell' m',a\ell m}(E;\mathcal R)-\delta_{b\ell' m',a\ell m}]
    \Bigg|^2.
\end{split}\end{equation}
This expression gives the differential cross section for an electron incident along $\hat{k}_i$ and scattered along $\hat{k}_f$ for a fixed target orientation. To connect it with the real-harmonic expression in Eq.~\eqref{eq:mf-dcs-main}, define the unitary angular-basis transformation by
\begin{equation}
|\alpha\rangle_X=\sum_\mu|\mu\rangle_Y\,U_{\mu\alpha},
\qquad U^\dagger U=I,
\end{equation}
where the coefficients follow from Eq.~\eqref{eq:real-to-complex-harmonics}. The transformation mixes only angular components with the same target state and $\ell$. Hence the channel momenta and radial phase factors are unchanged, and
\begin{equation}
S_X^S=U^\dagger\tilde S^S U
=\bigoplus_\Gamma S^{S\Gamma}.
\label{eq:angular-basis-s-transform}
\end{equation}
Transforming the incident and outgoing harmonics together with the matrix leaves the scattering amplitude invariant. Since the $X_{\ell m}$ are real, the incident angular factor no longer needs complex conjugation. The spatial-symmetry blocks are summed coherently inside each fixed-spin amplitude, giving Eq.~\eqref{eq:mf-dcs-main}. Spin averaging remains outside the squared amplitude as above.

The integral cross section for a randomly oriented molecule is readily obtained by integrating over $\hat{k}_f$ and averaging over $\hat{k}_i$, irrespective of the orientation $\mathcal R$,
\begin{equation}\begin{split}
    &\sigma_{b\gets a}=
    \frac{\pi}{4\varepsilon_i}\sum_S\frac{2S+1}{2S_a+1}\times\\
    &\quad\times\sum_{\ell m\ell'm'}\big|
\tilde{S}^{S}_{b\ell' m',a\ell m}(E)-\delta_{b\ell' m',a\ell m}
    \big|^2,
\end{split}\end{equation}
where $\tilde{\mathbf{S}}(E)$ now refers to a calculation where the molecule has a predefined reference orientation.
In the case of the integral cross section, the specific angular basis used for the channels does not matter. For this reason, we can also use the scattering matrix relative to the $X_{\ell m}$ functions, which allows us to preserve the spatial symmetry of the molecule, and hence to separately evaluate the different symmetry components of $\sigma_{b\gets a}$,
\begin{equation}
\sigma_{b\gets a}=\sum_{S\Gamma}\sigma_{b\gets a}^{S\Gamma},
\end{equation}
\begin{equation}
\begin{split}
\sigma_{b\gets a}^{S\Gamma}
={}&\frac{\pi}{4\varepsilon_i}\frac{2S+1}{2S_a+1}\\
&\times
\sum_{\substack{\alpha\in\mathcal O_a^{S\Gamma}\\
                \beta\in\mathcal O_b^{S\Gamma}}}
\left|S_{\beta\alpha}^{S\Gamma}(E)-\delta_{\beta\alpha}\right|^2,
\end{split}
\end{equation}
where $\mathcal O_a^{S\Gamma}$ and $\mathcal O_b^{S\Gamma}$ contain only open channels of the specified initial and final target states in the $S,\Gamma$ block, and $\varepsilon_i=k_i^2/2$ is the incident electron energy.

The orientational averaging required to obtain laboratory-frame differential cross sections for randomly oriented targets is the subject of a separate work.

\bibliography{ElectronScattering,ElectronScattering_additions}

%apsrev4-2.bst 2019-01-14 (MD) hand-edited version of apsrev4-1.bst
%Control: key (0)
%Control: author (8) initials jnrlst
%Control: editor formatted (1) identically to author
%Control: production of article title (0) allowed
%Control: page (0) single
%Control: year (1) truncated
%Control: production of eprint (0) enabled
\begin{thebibliography}{102}%
\makeatletter
\providecommand \@ifxundefined [1]{%
 \@ifx{#1\undefined}
}%
\providecommand \@ifnum [1]{%
 \ifnum #1\expandafter \@firstoftwo
 \else \expandafter \@secondoftwo
 \fi
}%
\providecommand \@ifx [1]{%
 \ifx #1\expandafter \@firstoftwo
 \else \expandafter \@secondoftwo
 \fi
}%
\providecommand \natexlab [1]{#1}%
\providecommand \enquote  [1]{``#1''}%
\providecommand \bibnamefont  [1]{#1}%
\providecommand \bibfnamefont [1]{#1}%
\providecommand \citenamefont [1]{#1}%
\providecommand \href@noop [0]{\@secondoftwo}%
\providecommand \href [0]{\begingroup \@sanitize@url \@href}%
\providecommand \@href[1]{\@@startlink{#1}\@@href}%
\providecommand \@@href[1]{\endgroup#1\@@endlink}%
\providecommand \@sanitize@url [0]{\catcode `\\12\catcode `\$12\catcode
  `\&12\catcode `\#12\catcode `\^12\catcode `\_12\catcode `\%12\relax}%
\providecommand \@@startlink[1]{}%
\providecommand \@@endlink[0]{}%
\providecommand \url  [0]{\begingroup\@sanitize@url \@url }%
\providecommand \@url [1]{\endgroup\@href {#1}{\urlprefix }}%
\providecommand \urlprefix  [0]{URL }%
\providecommand \Eprint [0]{\href }%
\providecommand \doibase [0]{https://doi.org/}%
\providecommand \selectlanguage [0]{\@gobble}%
\providecommand \bibinfo  [0]{\@secondoftwo}%
\providecommand \bibfield  [0]{\@secondoftwo}%
\providecommand \translation [1]{[#1]}%
\providecommand \BibitemOpen [0]{}%
\providecommand \bibitemStop [0]{}%
\providecommand \bibitemNoStop [0]{.\EOS\space}%
\providecommand \EOS [0]{\spacefactor3000\relax}%
\providecommand \BibitemShut  [1]{\csname bibitem#1\endcsname}%
\let\auto@bib@innerbib\@empty
%</preamble>
\bibitem [{\citenamefont {Christophorou}\ and\ \citenamefont
  {Olthoff}(2012)}]{Christophorou2012}%
  \BibitemOpen
  \bibfield  {author} {\bibinfo {author} {\bibfnamefont {L.}~\bibnamefont
  {Christophorou}}\ and\ \bibinfo {author} {\bibfnamefont {J.}~\bibnamefont
  {Olthoff}},\ }\href {https://books.google.com/books?id=BObpBwAAQBAJ} {\emph
  {\bibinfo {title} {Fundamental Electron Interactions with Plasma Processing
  Gases}}},\ Physics of Atoms and Molecules\ (\bibinfo  {publisher} {Springer
  US},\ \bibinfo {year} {2012})\BibitemShut {NoStop}%
\bibitem [{\citenamefont {Bartschat}\ and\ \citenamefont
  {Kushner}(2016)}]{Bartschat-PNAS-16}%
  \BibitemOpen
  \bibfield  {author} {\bibinfo {author} {\bibfnamefont {K.}~\bibnamefont
  {Bartschat}}\ and\ \bibinfo {author} {\bibfnamefont {M.~J.}\ \bibnamefont
  {Kushner}},\ }\bibfield  {title} {\bibinfo {title} {Electron collisions with
  atoms, ions, molecules, and surfaces: Fundamental science empowering advances
  in technology},\ }\href {https://doi.org/10.1073/pnas.1606132113} {\bibfield
  {journal} {\bibinfo  {journal} {Proc. Natl. Acad. Sci. USA}\ }\textbf
  {\bibinfo {volume} {113}},\ \bibinfo {pages} {7026} (\bibinfo {year}
  {2016})}\BibitemShut {NoStop}%
\bibitem [{\citenamefont {Alizadeh}\ \emph {et~al.}(2015)\citenamefont
  {Alizadeh}, \citenamefont {Orlando},\ and\ \citenamefont
  {Sanche}}]{Alizadeh-ARPC-15}%
  \BibitemOpen
  \bibfield  {author} {\bibinfo {author} {\bibfnamefont {E.}~\bibnamefont
  {Alizadeh}}, \bibinfo {author} {\bibfnamefont {T.~M.}\ \bibnamefont
  {Orlando}},\ and\ \bibinfo {author} {\bibfnamefont {L.}~\bibnamefont
  {Sanche}},\ }\bibfield  {title} {\bibinfo {title} {Biomolecular damage
  induced by ionizing radiation: The direct and indirect effects of low-energy
  electrons on {DNA}},\ }\href
  {https://doi.org/10.1146/annurev-physchem-040513-103605} {\bibfield
  {journal} {\bibinfo  {journal} {Annu. Rev. Phys. Chem.}\ }\textbf {\bibinfo
  {volume} {66}},\ \bibinfo {pages} {379} (\bibinfo {year} {2015})}\BibitemShut
  {NoStop}%
\bibitem [{\citenamefont {Gaches}\ \emph {et~al.}(2024)\citenamefont {Gaches},
  \citenamefont {Grassi}, \citenamefont {Vogt-Geisse}, \citenamefont
  {Bovolenta}, \citenamefont {Vallance}, \citenamefont {Heathcote},
  \citenamefont {Padovani}, \citenamefont {Bovino},\ and\ \citenamefont
  {Gorai}}]{Gaches-AA-24}%
  \BibitemOpen
  \bibfield  {author} {\bibinfo {author} {\bibfnamefont {B.~A.~L.}\
  \bibnamefont {Gaches}}, \bibinfo {author} {\bibfnamefont {T.}~\bibnamefont
  {Grassi}}, \bibinfo {author} {\bibfnamefont {S.}~\bibnamefont {Vogt-Geisse}},
  \bibinfo {author} {\bibfnamefont {G.~M.}\ \bibnamefont {Bovolenta}}, \bibinfo
  {author} {\bibfnamefont {C.}~\bibnamefont {Vallance}}, \bibinfo {author}
  {\bibfnamefont {D.}~\bibnamefont {Heathcote}}, \bibinfo {author}
  {\bibfnamefont {M.}~\bibnamefont {Padovani}}, \bibinfo {author}
  {\bibfnamefont {S.}~\bibnamefont {Bovino}},\ and\ \bibinfo {author}
  {\bibfnamefont {P.}~\bibnamefont {Gorai}},\ }\bibfield  {title} {\bibinfo
  {title} {The astrochemistry low-energy electron cross-section ({ALeCS})
  database. {I}. {Semi-empirical electron-impact ionization cross-section
  calculations and ionization rates}},\ }\href
  {https://doi.org/10.1051/0004-6361/202348293} {\bibfield  {journal} {\bibinfo
   {journal} {Astron. Astrophys.}\ }\textbf {\bibinfo {volume} {684}},\
  \bibinfo {pages} {A41} (\bibinfo {year} {2024})}\BibitemShut {NoStop}%
\bibitem [{\citenamefont {McConkey}\ \emph {et~al.}(2008)\citenamefont
  {McConkey}, \citenamefont {Malone}, \citenamefont {Johnson}, \citenamefont
  {Winstead}, \citenamefont {McKoy},\ and\ \citenamefont
  {Kanik}}]{McConkey-PhysRep-08}%
  \BibitemOpen
  \bibfield  {author} {\bibinfo {author} {\bibfnamefont {J.~W.}\ \bibnamefont
  {McConkey}}, \bibinfo {author} {\bibfnamefont {C.~P.}\ \bibnamefont
  {Malone}}, \bibinfo {author} {\bibfnamefont {P.~V.}\ \bibnamefont {Johnson}},
  \bibinfo {author} {\bibfnamefont {C.}~\bibnamefont {Winstead}}, \bibinfo
  {author} {\bibfnamefont {V.}~\bibnamefont {McKoy}},\ and\ \bibinfo {author}
  {\bibfnamefont {I.}~\bibnamefont {Kanik}},\ }\bibfield  {title} {\bibinfo
  {title} {Electron impact dissociation of oxygen-containing molecules---{A}
  critical review},\ }\href {https://doi.org/10.1016/j.physrep.2008.05.001}
  {\bibfield  {journal} {\bibinfo  {journal} {Phys. Rep.}\ }\textbf {\bibinfo
  {volume} {466}},\ \bibinfo {pages} {1} (\bibinfo {year} {2008})}\BibitemShut
  {NoStop}%
\bibitem [{\citenamefont {Campbell}\ and\ \citenamefont
  {Brunger}(2016)}]{Campbell2016}%
  \BibitemOpen
  \bibfield  {author} {\bibinfo {author} {\bibfnamefont {L.}~\bibnamefont
  {Campbell}}\ and\ \bibinfo {author} {\bibfnamefont {M.}~\bibnamefont
  {Brunger}},\ }\bibfield  {title} {\bibinfo {title} {Electron collisions in
  atmospheres},\ }\href {https://doi.org/10.1080/0144235X.2016.1179002}
  {\bibfield  {journal} {\bibinfo  {journal} {Int. Rev. Phys. Chem.}\ }\textbf
  {\bibinfo {volume} {35}},\ \bibinfo {pages} {297} (\bibinfo {year}
  {2016})}\BibitemShut {NoStop}%
\bibitem [{\citenamefont {Vialetto}\ \emph {et~al.}(2024)\citenamefont
  {Vialetto}, \citenamefont {Sugawara},\ and\ \citenamefont
  {Longo}}]{Vialetto-Plasma-24}%
  \BibitemOpen
  \bibfield  {author} {\bibinfo {author} {\bibfnamefont {L.}~\bibnamefont
  {Vialetto}}, \bibinfo {author} {\bibfnamefont {H.}~\bibnamefont {Sugawara}},\
  and\ \bibinfo {author} {\bibfnamefont {S.}~\bibnamefont {Longo}},\ }\bibfield
   {title} {\bibinfo {title} {Particle propagation and electron transport in
  gases},\ }\href {https://doi.org/10.3390/plasma7010009} {\bibfield  {journal}
  {\bibinfo  {journal} {Plasma}\ }\textbf {\bibinfo {volume} {7}},\ \bibinfo
  {pages} {121} (\bibinfo {year} {2024})}\BibitemShut {NoStop}%
\bibitem [{\citenamefont {Boyle}\ \emph {et~al.}(2025)\citenamefont {Boyle},
  \citenamefont {Garland}, \citenamefont {Muccignat}, \citenamefont
  {Simonovi{\'c}}, \citenamefont {Bo{\v{s}}njakovi{\'c}}, \citenamefont
  {Dujko},\ and\ \citenamefont {White}}]{Boyle-FDST-25}%
  \BibitemOpen
  \bibfield  {author} {\bibinfo {author} {\bibfnamefont {G.~J.}\ \bibnamefont
  {Boyle}}, \bibinfo {author} {\bibfnamefont {N.~A.}\ \bibnamefont {Garland}},
  \bibinfo {author} {\bibfnamefont {D.~L.}\ \bibnamefont {Muccignat}}, \bibinfo
  {author} {\bibfnamefont {I.}~\bibnamefont {Simonovi{\'c}}}, \bibinfo {author}
  {\bibfnamefont {D.}~\bibnamefont {Bo{\v{s}}njakovi{\'c}}}, \bibinfo {author}
  {\bibfnamefont {S.}~\bibnamefont {Dujko}},\ and\ \bibinfo {author}
  {\bibfnamefont {R.~D.}\ \bibnamefont {White}},\ }\bibfield  {title} {\bibinfo
  {title} {Review of the experimental and theoretical landscape of electron
  transport in noble liquids},\ }\href
  {https://doi.org/10.3389/fdest.2025.1616204} {\bibfield  {journal} {\bibinfo
  {journal} {Front. Detect. Sci. Technol.}\ }\textbf {\bibinfo {volume} {3}},\
  \bibinfo {pages} {1616204} (\bibinfo {year} {2025})}\BibitemShut {NoStop}%
\bibitem [{\citenamefont {Solov'yov}\ \emph {et~al.}(2024)\citenamefont
  {Solov'yov}, \citenamefont {Verkhovtsev}, \citenamefont {Mason},
  \citenamefont {Amos}, \citenamefont {Bald}, \citenamefont {Baldacchino},
  \citenamefont {Dromey}, \citenamefont {Falk}, \citenamefont {Fedor},
  \citenamefont {Gerhards}, \citenamefont {Hausmann}, \citenamefont
  {Hildenbrand}, \citenamefont {Hrabovsk{\'y}}, \citenamefont {Kadlec},
  \citenamefont {Ko{\v{c}}i{\v{s}}ek}, \citenamefont {L{\'e}pine},
  \citenamefont {Ming}, \citenamefont {Nisbet}, \citenamefont {Ricketts},
  \citenamefont {Sala}, \citenamefont {Schlath{\"o}lter}, \citenamefont
  {Wheatley},\ and\ \citenamefont {Solov'yov}}]{Solovyov-ChemRev-24}%
  \BibitemOpen
  \bibfield  {author} {\bibinfo {author} {\bibfnamefont {A.~V.}\ \bibnamefont
  {Solov'yov}}, \bibinfo {author} {\bibfnamefont {A.~V.}\ \bibnamefont
  {Verkhovtsev}}, \bibinfo {author} {\bibfnamefont {N.~J.}\ \bibnamefont
  {Mason}}, \bibinfo {author} {\bibfnamefont {R.~A.}\ \bibnamefont {Amos}},
  \bibinfo {author} {\bibfnamefont {I.}~\bibnamefont {Bald}}, \bibinfo {author}
  {\bibfnamefont {G.}~\bibnamefont {Baldacchino}}, \bibinfo {author}
  {\bibfnamefont {B.}~\bibnamefont {Dromey}}, \bibinfo {author} {\bibfnamefont
  {M.}~\bibnamefont {Falk}}, \bibinfo {author} {\bibfnamefont {J.}~\bibnamefont
  {Fedor}}, \bibinfo {author} {\bibfnamefont {L.}~\bibnamefont {Gerhards}},
  \bibinfo {author} {\bibfnamefont {M.}~\bibnamefont {Hausmann}}, \bibinfo
  {author} {\bibfnamefont {G.}~\bibnamefont {Hildenbrand}}, \bibinfo {author}
  {\bibfnamefont {M.}~\bibnamefont {Hrabovsk{\'y}}}, \bibinfo {author}
  {\bibfnamefont {S.}~\bibnamefont {Kadlec}}, \bibinfo {author} {\bibfnamefont
  {J.}~\bibnamefont {Ko{\v{c}}i{\v{s}}ek}}, \bibinfo {author} {\bibfnamefont
  {F.}~\bibnamefont {L{\'e}pine}}, \bibinfo {author} {\bibfnamefont
  {S.}~\bibnamefont {Ming}}, \bibinfo {author} {\bibfnamefont {A.}~\bibnamefont
  {Nisbet}}, \bibinfo {author} {\bibfnamefont {K.}~\bibnamefont {Ricketts}},
  \bibinfo {author} {\bibfnamefont {L.}~\bibnamefont {Sala}}, \bibinfo {author}
  {\bibfnamefont {T.}~\bibnamefont {Schlath{\"o}lter}}, \bibinfo {author}
  {\bibfnamefont {A.~E.~H.}\ \bibnamefont {Wheatley}},\ and\ \bibinfo {author}
  {\bibfnamefont {I.~A.}\ \bibnamefont {Solov'yov}},\ }\bibfield  {title}
  {\bibinfo {title} {Condensed matter systems exposed to radiation: Multiscale
  theory, simulations, and experiment},\ }\href
  {https://doi.org/10.1021/acs.chemrev.3c00902} {\bibfield  {journal} {\bibinfo
   {journal} {Chem. Rev.}\ }\textbf {\bibinfo {volume} {124}},\ \bibinfo
  {pages} {8014} (\bibinfo {year} {2024})}\BibitemShut {NoStop}%
\bibitem [{\citenamefont {Taioli}\ and\ \citenamefont
  {Dapor}(2025)}]{Taioli-SSR-25}%
  \BibitemOpen
  \bibfield  {author} {\bibinfo {author} {\bibfnamefont {S.}~\bibnamefont
  {Taioli}}\ and\ \bibinfo {author} {\bibfnamefont {M.}~\bibnamefont {Dapor}},\
  }\bibfield  {title} {\bibinfo {title} {Advancements in secondary and
  backscattered electron energy spectra and yields analysis: From theory to
  applications},\ }\href {https://doi.org/10.1016/j.surfrep.2024.100646}
  {\bibfield  {journal} {\bibinfo  {journal} {Surf. Sci. Rep.}\ }\textbf
  {\bibinfo {volume} {80}},\ \bibinfo {pages} {100646} (\bibinfo {year}
  {2025})}\BibitemShut {NoStop}%
\bibitem [{\citenamefont {Tennyson}\ \emph {et~al.}(2022)\citenamefont
  {Tennyson}, \citenamefont {Mohr}, \citenamefont {Hanicinec}, \citenamefont
  {Dzarasova}, \citenamefont {Smith}, \citenamefont {Waddington}, \citenamefont
  {Liu}, \citenamefont {Alves}, \citenamefont {Bartschat}, \citenamefont
  {Bogaerts}, \citenamefont {Engelmann}, \citenamefont {Gans}, \citenamefont
  {Gibson}, \citenamefont {Hamaguchi}, \citenamefont {Hamilton}, \citenamefont
  {Hill}, \citenamefont {O’Connell}, \citenamefont {Rauf}, \citenamefont
  {van~’t Veer},\ and\ \citenamefont {Zatsarinny}}]{Tennyson-PSST-22}%
  \BibitemOpen
  \bibfield  {author} {\bibinfo {author} {\bibfnamefont {J.}~\bibnamefont
  {Tennyson}}, \bibinfo {author} {\bibfnamefont {S.}~\bibnamefont {Mohr}},
  \bibinfo {author} {\bibfnamefont {M.}~\bibnamefont {Hanicinec}}, \bibinfo
  {author} {\bibfnamefont {A.}~\bibnamefont {Dzarasova}}, \bibinfo {author}
  {\bibfnamefont {C.}~\bibnamefont {Smith}}, \bibinfo {author} {\bibfnamefont
  {S.}~\bibnamefont {Waddington}}, \bibinfo {author} {\bibfnamefont
  {B.}~\bibnamefont {Liu}}, \bibinfo {author} {\bibfnamefont {L.~L.}\
  \bibnamefont {Alves}}, \bibinfo {author} {\bibfnamefont {K.}~\bibnamefont
  {Bartschat}}, \bibinfo {author} {\bibfnamefont {A.}~\bibnamefont {Bogaerts}},
  \bibinfo {author} {\bibfnamefont {S.~U.}\ \bibnamefont {Engelmann}}, \bibinfo
  {author} {\bibfnamefont {T.}~\bibnamefont {Gans}}, \bibinfo {author}
  {\bibfnamefont {A.~R.}\ \bibnamefont {Gibson}}, \bibinfo {author}
  {\bibfnamefont {S.}~\bibnamefont {Hamaguchi}}, \bibinfo {author}
  {\bibfnamefont {K.~R.}\ \bibnamefont {Hamilton}}, \bibinfo {author}
  {\bibfnamefont {C.}~\bibnamefont {Hill}}, \bibinfo {author} {\bibfnamefont
  {D.}~\bibnamefont {O’Connell}}, \bibinfo {author} {\bibfnamefont
  {S.}~\bibnamefont {Rauf}}, \bibinfo {author} {\bibfnamefont {K.}~\bibnamefont
  {van~’t Veer}},\ and\ \bibinfo {author} {\bibfnamefont {O.}~\bibnamefont
  {Zatsarinny}},\ }\bibfield  {title} {\bibinfo {title} {The 2021 release of
  the quantemol database {(QDB)} of plasma chemistries and reactions},\ }\href
  {https://doi.org/10.1088/1361-6595/ac907e} {\bibfield  {journal} {\bibinfo
  {journal} {Plasma Sources Sci. Technol.}\ }\textbf {\bibinfo {volume} {31}},\
  \bibinfo {pages} {095020} (\bibinfo {year} {2022})}\BibitemShut {NoStop}%
\bibitem [{\citenamefont {Alves}\ \emph {et~al.}(2023)\citenamefont {Alves},
  \citenamefont {Becker}, \citenamefont {van Dijk}, \citenamefont {Gans},
  \citenamefont {Go}, \citenamefont {Stapelmann}, \citenamefont {Tennyson},
  \citenamefont {Turner},\ and\ \citenamefont {Kushner}}]{Alves-PSST-23}%
  \BibitemOpen
  \bibfield  {author} {\bibinfo {author} {\bibfnamefont {L.~L.}\ \bibnamefont
  {Alves}}, \bibinfo {author} {\bibfnamefont {M.~M.}\ \bibnamefont {Becker}},
  \bibinfo {author} {\bibfnamefont {J.}~\bibnamefont {van Dijk}}, \bibinfo
  {author} {\bibfnamefont {T.}~\bibnamefont {Gans}}, \bibinfo {author}
  {\bibfnamefont {D.~B.}\ \bibnamefont {Go}}, \bibinfo {author} {\bibfnamefont
  {K.}~\bibnamefont {Stapelmann}}, \bibinfo {author} {\bibfnamefont
  {J.}~\bibnamefont {Tennyson}}, \bibinfo {author} {\bibfnamefont {M.~M.}\
  \bibnamefont {Turner}},\ and\ \bibinfo {author} {\bibfnamefont {M.~J.}\
  \bibnamefont {Kushner}},\ }\bibfield  {title} {\bibinfo {title} {Foundations
  of plasma standards},\ }\href {https://doi.org/10.1088/1361-6595/acb810}
  {\bibfield  {journal} {\bibinfo  {journal} {Plasma Sources Sci. Technol.}\
  }\textbf {\bibinfo {volume} {32}},\ \bibinfo {pages} {023001} (\bibinfo
  {year} {2023})}\BibitemShut {NoStop}%
\bibitem [{\citenamefont {Pancheshnyi}\ \emph {et~al.}(2012)\citenamefont
  {Pancheshnyi}, \citenamefont {Biagi}, \citenamefont {Bordage}, \citenamefont
  {Hagelaar}, \citenamefont {Morgan}, \citenamefont {Phelps},\ and\
  \citenamefont {Pitchford}}]{Pancheshnyi-ChemPhys-12}%
  \BibitemOpen
  \bibfield  {author} {\bibinfo {author} {\bibfnamefont {S.}~\bibnamefont
  {Pancheshnyi}}, \bibinfo {author} {\bibfnamefont {S.}~\bibnamefont {Biagi}},
  \bibinfo {author} {\bibfnamefont {M.~C.}\ \bibnamefont {Bordage}}, \bibinfo
  {author} {\bibfnamefont {G.~J.~M.}\ \bibnamefont {Hagelaar}}, \bibinfo
  {author} {\bibfnamefont {W.~L.}\ \bibnamefont {Morgan}}, \bibinfo {author}
  {\bibfnamefont {A.~V.}\ \bibnamefont {Phelps}},\ and\ \bibinfo {author}
  {\bibfnamefont {L.~C.}\ \bibnamefont {Pitchford}},\ }\bibfield  {title}
  {\bibinfo {title} {The {LXCat} project: Electron scattering cross sections
  and swarm parameters for low temperature plasma modeling},\ }\href
  {https://doi.org/10.1016/j.chemphys.2011.04.020} {\bibfield  {journal}
  {\bibinfo  {journal} {Chem. Phys.}\ }\textbf {\bibinfo {volume} {398}},\
  \bibinfo {pages} {148} (\bibinfo {year} {2012})}\BibitemShut {NoStop}%
\bibitem [{\citenamefont {Pitchford}\ \emph {et~al.}(2017)\citenamefont
  {Pitchford}, \citenamefont {Alves}, \citenamefont {Bartschat} \emph
  {et~al.}}]{Pitchford-PPP-17}%
  \BibitemOpen
  \bibfield  {author} {\bibinfo {author} {\bibfnamefont {L.~C.}\ \bibnamefont
  {Pitchford}}, \bibinfo {author} {\bibfnamefont {L.~L.}\ \bibnamefont
  {Alves}}, \bibinfo {author} {\bibfnamefont {K.}~\bibnamefont {Bartschat}},
  \emph {et~al.},\ }\bibfield  {title} {\bibinfo {title} {{LXCat}: An
  open-access, web-based platform for data needed for modeling low temperature
  plasmas},\ }\href {https://doi.org/10.1002/ppap.201600098} {\bibfield
  {journal} {\bibinfo  {journal} {Plasma Process. Polym.}\ }\textbf {\bibinfo
  {volume} {14}},\ \bibinfo {pages} {1600098} (\bibinfo {year}
  {2017})}\BibitemShut {NoStop}%
\bibitem [{\citenamefont {Carbone}\ \emph {et~al.}(2021)\citenamefont
  {Carbone}, \citenamefont {Graef}, \citenamefont {Hagelaar}, \citenamefont
  {Boer}, \citenamefont {Hopkins}, \citenamefont {Stephens}, \citenamefont
  {Yee}, \citenamefont {Pancheshnyi}, \citenamefont {van Dijk},\ and\
  \citenamefont {Pitchford}}]{Carbone-Atoms-21}%
  \BibitemOpen
  \bibfield  {author} {\bibinfo {author} {\bibfnamefont {E.}~\bibnamefont
  {Carbone}}, \bibinfo {author} {\bibfnamefont {W.}~\bibnamefont {Graef}},
  \bibinfo {author} {\bibfnamefont {G.}~\bibnamefont {Hagelaar}}, \bibinfo
  {author} {\bibfnamefont {D.}~\bibnamefont {Boer}}, \bibinfo {author}
  {\bibfnamefont {M.~M.}\ \bibnamefont {Hopkins}}, \bibinfo {author}
  {\bibfnamefont {J.~C.}\ \bibnamefont {Stephens}}, \bibinfo {author}
  {\bibfnamefont {B.~T.}\ \bibnamefont {Yee}}, \bibinfo {author} {\bibfnamefont
  {S.}~\bibnamefont {Pancheshnyi}}, \bibinfo {author} {\bibfnamefont
  {J.}~\bibnamefont {van Dijk}},\ and\ \bibinfo {author} {\bibfnamefont
  {L.}~\bibnamefont {Pitchford}},\ }\bibfield  {title} {\bibinfo {title} {Data
  needs for modeling low-temperature non-equilibrium plasmas: The {LXCat}
  project, history, perspectives and a tutorial},\ }\href
  {https://doi.org/10.3390/atoms9010016} {\bibfield  {journal} {\bibinfo
  {journal} {Atoms}\ }\textbf {\bibinfo {volume} {9}},\ \bibinfo {pages} {16}
  (\bibinfo {year} {2021})}\BibitemShut {NoStop}%
\bibitem [{\citenamefont {Bartschat}\ \emph {et~al.}(2017)\citenamefont
  {Bartschat}, \citenamefont {Tennyson},\ and\ \citenamefont
  {Zatsarinny}}]{Bartschat-PPP-17}%
  \BibitemOpen
  \bibfield  {author} {\bibinfo {author} {\bibfnamefont {K.}~\bibnamefont
  {Bartschat}}, \bibinfo {author} {\bibfnamefont {J.}~\bibnamefont
  {Tennyson}},\ and\ \bibinfo {author} {\bibfnamefont {O.}~\bibnamefont
  {Zatsarinny}},\ }\bibfield  {title} {\bibinfo {title} {Quantum-mechanical
  calculations of cross sections for electron collisions with atoms and
  molecules},\ }\href {https://doi.org/10.1002/ppap.201600093} {\bibfield
  {journal} {\bibinfo  {journal} {Plasma Process. Polym.}\ }\textbf {\bibinfo
  {volume} {14}},\ \bibinfo {pages} {1600093} (\bibinfo {year}
  {2017})}\BibitemShut {NoStop}%
\bibitem [{\citenamefont {Song}\ \emph {et~al.}(2020)\citenamefont {Song},
  \citenamefont {Yoon}, \citenamefont {Cho}, \citenamefont {Karwasz},
  \citenamefont {Kokoouline}, \citenamefont {Nakamura},\ and\ \citenamefont
  {Tennyson}}]{Song-EPJD-20}%
  \BibitemOpen
  \bibfield  {author} {\bibinfo {author} {\bibfnamefont {M.-Y.}\ \bibnamefont
  {Song}}, \bibinfo {author} {\bibfnamefont {J.-S.}\ \bibnamefont {Yoon}},
  \bibinfo {author} {\bibfnamefont {H.}~\bibnamefont {Cho}}, \bibinfo {author}
  {\bibfnamefont {G.~P.}\ \bibnamefont {Karwasz}}, \bibinfo {author}
  {\bibfnamefont {V.}~\bibnamefont {Kokoouline}}, \bibinfo {author}
  {\bibfnamefont {Y.}~\bibnamefont {Nakamura}},\ and\ \bibinfo {author}
  {\bibfnamefont {J.}~\bibnamefont {Tennyson}},\ }\bibfield  {title} {\bibinfo
  {title} {``recommended'' cross sections for electron collisions with
  molecules},\ }\href {https://doi.org/10.1140/epjd/e2020-100543-6} {\bibfield
  {journal} {\bibinfo  {journal} {Eur. Phys. J. D}\ }\textbf {\bibinfo {volume}
  {74}},\ \bibinfo {pages} {60} (\bibinfo {year} {2020})}\BibitemShut {NoStop}%
\bibitem [{\citenamefont {Corkum}(1993)}]{Corkum-PRL-93}%
  \BibitemOpen
  \bibfield  {author} {\bibinfo {author} {\bibfnamefont {P.~B.}\ \bibnamefont
  {Corkum}},\ }\bibfield  {title} {\bibinfo {title} {Plasma perspective on
  strong field multiphoton ionization},\ }\href
  {https://doi.org/10.1103/PhysRevLett.71.1994} {\bibfield  {journal} {\bibinfo
   {journal} {Phys. Rev. Lett.}\ }\textbf {\bibinfo {volume} {71}},\ \bibinfo
  {pages} {1994} (\bibinfo {year} {1993})}\BibitemShut {NoStop}%
\bibitem [{\citenamefont {Schafer}\ \emph {et~al.}(1993)\citenamefont
  {Schafer}, \citenamefont {Yang}, \citenamefont {DiMauro},\ and\ \citenamefont
  {Kulander}}]{Schafer-PRL-93}%
  \BibitemOpen
  \bibfield  {author} {\bibinfo {author} {\bibfnamefont {K.~J.}\ \bibnamefont
  {Schafer}}, \bibinfo {author} {\bibfnamefont {B.}~\bibnamefont {Yang}},
  \bibinfo {author} {\bibfnamefont {L.~F.}\ \bibnamefont {DiMauro}},\ and\
  \bibinfo {author} {\bibfnamefont {K.~C.}\ \bibnamefont {Kulander}},\
  }\bibfield  {title} {\bibinfo {title} {Above threshold ionization beyond the
  high harmonic cutoff},\ }\href {https://doi.org/10.1103/PhysRevLett.70.1599}
  {\bibfield  {journal} {\bibinfo  {journal} {Phys. Rev. Lett.}\ }\textbf
  {\bibinfo {volume} {70}},\ \bibinfo {pages} {1599} (\bibinfo {year}
  {1993})}\BibitemShut {NoStop}%
\bibitem [{\citenamefont {Chen}\ \emph {et~al.}(2009)\citenamefont {Chen},
  \citenamefont {Le}, \citenamefont {Morishita},\ and\ \citenamefont
  {Lin}}]{Chen-PRA-09}%
  \BibitemOpen
  \bibfield  {author} {\bibinfo {author} {\bibfnamefont {Z.}~\bibnamefont
  {Chen}}, \bibinfo {author} {\bibfnamefont {A.-T.}\ \bibnamefont {Le}},
  \bibinfo {author} {\bibfnamefont {T.}~\bibnamefont {Morishita}},\ and\
  \bibinfo {author} {\bibfnamefont {C.~D.}\ \bibnamefont {Lin}},\ }\bibfield
  {title} {\bibinfo {title} {Quantitative rescattering theory for laser-induced
  high-energy plateau photoelectron spectra},\ }\href
  {https://doi.org/10.1103/PhysRevA.79.033409} {\bibfield  {journal} {\bibinfo
  {journal} {Phys. Rev. A}\ }\textbf {\bibinfo {volume} {79}},\ \bibinfo
  {pages} {033409} (\bibinfo {year} {2009})}\BibitemShut {NoStop}%
\bibitem [{\citenamefont {Lin}\ \emph {et~al.}(2010)\citenamefont {Lin},
  \citenamefont {Le}, \citenamefont {Chen}, \citenamefont {Morishita},\ and\
  \citenamefont {Lucchese}}]{Lin-JPB-10}%
  \BibitemOpen
  \bibfield  {author} {\bibinfo {author} {\bibfnamefont {C.~D.}\ \bibnamefont
  {Lin}}, \bibinfo {author} {\bibfnamefont {A.-T.}\ \bibnamefont {Le}},
  \bibinfo {author} {\bibfnamefont {Z.}~\bibnamefont {Chen}}, \bibinfo {author}
  {\bibfnamefont {T.}~\bibnamefont {Morishita}},\ and\ \bibinfo {author}
  {\bibfnamefont {R.~R.}\ \bibnamefont {Lucchese}},\ }\bibfield  {title}
  {\bibinfo {title} {Strong-field rescattering physics---self-imaging of a
  molecule by its own electrons},\ }\href
  {https://doi.org/10.1088/0953-4075/43/12/122001} {\bibfield  {journal}
  {\bibinfo  {journal} {J. Phys. B: At. Mol. Opt. Phys.}\ }\textbf {\bibinfo
  {volume} {43}},\ \bibinfo {pages} {122001} (\bibinfo {year}
  {2010})}\BibitemShut {NoStop}%
\bibitem [{\citenamefont {Meckel}\ \emph {et~al.}(2008)\citenamefont {Meckel},
  \citenamefont {Comtois}, \citenamefont {Zeidler}, \citenamefont {Staudte},
  \citenamefont {Pavi{\v{c}}i{\'c}}, \citenamefont {Bandulet}, \citenamefont
  {P{\'e}pin}, \citenamefont {Kieffer}, \citenamefont {D{\"o}rner},
  \citenamefont {Villeneuve},\ and\ \citenamefont
  {Corkum}}]{Meckel-Science-08}%
  \BibitemOpen
  \bibfield  {author} {\bibinfo {author} {\bibfnamefont {M.}~\bibnamefont
  {Meckel}}, \bibinfo {author} {\bibfnamefont {D.}~\bibnamefont {Comtois}},
  \bibinfo {author} {\bibfnamefont {D.}~\bibnamefont {Zeidler}}, \bibinfo
  {author} {\bibfnamefont {A.}~\bibnamefont {Staudte}}, \bibinfo {author}
  {\bibfnamefont {D.}~\bibnamefont {Pavi{\v{c}}i{\'c}}}, \bibinfo {author}
  {\bibfnamefont {H.~C.}\ \bibnamefont {Bandulet}}, \bibinfo {author}
  {\bibfnamefont {H.}~\bibnamefont {P{\'e}pin}}, \bibinfo {author}
  {\bibfnamefont {J.~C.}\ \bibnamefont {Kieffer}}, \bibinfo {author}
  {\bibfnamefont {R.}~\bibnamefont {D{\"o}rner}}, \bibinfo {author}
  {\bibfnamefont {D.~M.}\ \bibnamefont {Villeneuve}},\ and\ \bibinfo {author}
  {\bibfnamefont {P.~B.}\ \bibnamefont {Corkum}},\ }\bibfield  {title}
  {\bibinfo {title} {Laser-induced electron tunneling and diffraction},\ }\href
  {https://doi.org/10.1126/science.1157980} {\bibfield  {journal} {\bibinfo
  {journal} {Science}\ }\textbf {\bibinfo {volume} {320}},\ \bibinfo {pages}
  {1478} (\bibinfo {year} {2008})}\BibitemShut {NoStop}%
\bibitem [{\citenamefont {Blaga}\ \emph {et~al.}(2012)\citenamefont {Blaga},
  \citenamefont {Xu}, \citenamefont {DiChiara}, \citenamefont {Sistrunk},
  \citenamefont {Zhang}, \citenamefont {Agostini}, \citenamefont {Miller},
  \citenamefont {DiMauro},\ and\ \citenamefont {Lin}}]{Blaga-Nature-12}%
  \BibitemOpen
  \bibfield  {author} {\bibinfo {author} {\bibfnamefont {C.~I.}\ \bibnamefont
  {Blaga}}, \bibinfo {author} {\bibfnamefont {J.}~\bibnamefont {Xu}}, \bibinfo
  {author} {\bibfnamefont {A.~D.}\ \bibnamefont {DiChiara}}, \bibinfo {author}
  {\bibfnamefont {E.}~\bibnamefont {Sistrunk}}, \bibinfo {author}
  {\bibfnamefont {K.}~\bibnamefont {Zhang}}, \bibinfo {author} {\bibfnamefont
  {P.}~\bibnamefont {Agostini}}, \bibinfo {author} {\bibfnamefont {T.~A.}\
  \bibnamefont {Miller}}, \bibinfo {author} {\bibfnamefont {L.~F.}\
  \bibnamefont {DiMauro}},\ and\ \bibinfo {author} {\bibfnamefont {C.~D.}\
  \bibnamefont {Lin}},\ }\bibfield  {title} {\bibinfo {title} {Imaging
  ultrafast molecular dynamics with laser-induced electron diffraction},\
  }\href {https://doi.org/10.1038/nature10820} {\bibfield  {journal} {\bibinfo
  {journal} {Nature}\ }\textbf {\bibinfo {volume} {483}},\ \bibinfo {pages}
  {194} (\bibinfo {year} {2012})}\BibitemShut {NoStop}%
\bibitem [{\citenamefont {Wolter}\ \emph {et~al.}(2016)\citenamefont {Wolter},
  \citenamefont {Pullen}, \citenamefont {Le}, \citenamefont {Baudisch},
  \citenamefont {Doblhoff-Dier}, \citenamefont {Senftleben}, \citenamefont
  {Hemmer}, \citenamefont {Schr{\"o}ter}, \citenamefont {Ullrich},
  \citenamefont {Pfeifer}, \citenamefont {Moshammer}, \citenamefont
  {Gr{\"a}fe}, \citenamefont {Vendrell}, \citenamefont {Lin},\ and\
  \citenamefont {Biegert}}]{Wolter-Science-16}%
  \BibitemOpen
  \bibfield  {author} {\bibinfo {author} {\bibfnamefont {B.}~\bibnamefont
  {Wolter}}, \bibinfo {author} {\bibfnamefont {M.~G.}\ \bibnamefont {Pullen}},
  \bibinfo {author} {\bibfnamefont {A.-T.}\ \bibnamefont {Le}}, \bibinfo
  {author} {\bibfnamefont {M.}~\bibnamefont {Baudisch}}, \bibinfo {author}
  {\bibfnamefont {K.}~\bibnamefont {Doblhoff-Dier}}, \bibinfo {author}
  {\bibfnamefont {A.}~\bibnamefont {Senftleben}}, \bibinfo {author}
  {\bibfnamefont {M.}~\bibnamefont {Hemmer}}, \bibinfo {author} {\bibfnamefont
  {C.~D.}\ \bibnamefont {Schr{\"o}ter}}, \bibinfo {author} {\bibfnamefont
  {J.}~\bibnamefont {Ullrich}}, \bibinfo {author} {\bibfnamefont
  {T.}~\bibnamefont {Pfeifer}}, \bibinfo {author} {\bibfnamefont
  {R.}~\bibnamefont {Moshammer}}, \bibinfo {author} {\bibfnamefont
  {S.}~\bibnamefont {Gr{\"a}fe}}, \bibinfo {author} {\bibfnamefont
  {O.}~\bibnamefont {Vendrell}}, \bibinfo {author} {\bibfnamefont {C.~D.}\
  \bibnamefont {Lin}},\ and\ \bibinfo {author} {\bibfnamefont {J.}~\bibnamefont
  {Biegert}},\ }\bibfield  {title} {\bibinfo {title} {Ultrafast electron
  diffraction imaging of bond breaking in di-ionized acetylene},\ }\href
  {https://doi.org/10.1126/science.aah3429} {\bibfield  {journal} {\bibinfo
  {journal} {Science}\ }\textbf {\bibinfo {volume} {354}},\ \bibinfo {pages}
  {308} (\bibinfo {year} {2016})}\BibitemShut {NoStop}%
\bibitem [{\citenamefont {Lindroth}\ and\ \citenamefont
  {Dahlstr{\"o}m}(2017)}]{Lindroth-PRA-17}%
  \BibitemOpen
  \bibfield  {author} {\bibinfo {author} {\bibfnamefont {E.}~\bibnamefont
  {Lindroth}}\ and\ \bibinfo {author} {\bibfnamefont {J.~M.}\ \bibnamefont
  {Dahlstr{\"o}m}},\ }\bibfield  {title} {\bibinfo {title} {Attosecond delays
  in laser-assisted photodetachment from closed-shell negative ions},\ }\href
  {https://doi.org/10.1103/PhysRevA.96.013420} {\bibfield  {journal} {\bibinfo
  {journal} {Phys. Rev. A}\ }\textbf {\bibinfo {volume} {96}},\ \bibinfo
  {pages} {013420} (\bibinfo {year} {2017})}\BibitemShut {NoStop}%
\bibitem [{\citenamefont {Rescigno}\ \emph {et~al.}(2024)\citenamefont
  {Rescigno}, \citenamefont {Trevisan}, \citenamefont {Lucchese},\ and\
  \citenamefont {McCurdy}}]{Rescigno-PRA-24}%
  \BibitemOpen
  \bibfield  {author} {\bibinfo {author} {\bibfnamefont {T.~N.}\ \bibnamefont
  {Rescigno}}, \bibinfo {author} {\bibfnamefont {C.~S.}\ \bibnamefont
  {Trevisan}}, \bibinfo {author} {\bibfnamefont {R.~R.}\ \bibnamefont
  {Lucchese}},\ and\ \bibinfo {author} {\bibfnamefont {C.~W.}\ \bibnamefont
  {McCurdy}},\ }\bibfield  {title} {\bibinfo {title} {Attosecond time delays in
  the body frame for photodetachment from molecular anions},\ }\href
  {https://doi.org/10.1103/PhysRevA.110.013122} {\bibfield  {journal} {\bibinfo
   {journal} {Phys. Rev. A}\ }\textbf {\bibinfo {volume} {110}},\ \bibinfo
  {pages} {013122} (\bibinfo {year} {2024})}\BibitemShut {NoStop}%
\bibitem [{\citenamefont {Tennyson}(2024)}]{Tennyson-JPB-24}%
  \BibitemOpen
  \bibfield  {author} {\bibinfo {author} {\bibfnamefont {J.}~\bibnamefont
  {Tennyson}},\ }\bibfield  {title} {\bibinfo {title} {Electron–molecule
  collision calculations: a primer},\ }\href
  {https://doi.org/10.1088/1361-6455/ad4243} {\bibfield  {journal} {\bibinfo
  {journal} {J. Phys. B: At. Mol. Opt. Phys.}\ }\textbf {\bibinfo {volume}
  {57}},\ \bibinfo {pages} {233001} (\bibinfo {year} {2024})}\BibitemShut
  {NoStop}%
\bibitem [{\citenamefont {Tennyson}(2010)}]{Tennyson-PhysRep-10}%
  \BibitemOpen
  \bibfield  {author} {\bibinfo {author} {\bibfnamefont {J.}~\bibnamefont
  {Tennyson}},\ }\bibfield  {title} {\bibinfo {title} {Electron--molecule
  collision calculations using the {$R$}-matrix method},\ }\href
  {https://doi.org/10.1016/j.physrep.2010.02.001} {\bibfield  {journal}
  {\bibinfo  {journal} {Phys. Rep.}\ }\textbf {\bibinfo {volume} {491}},\
  \bibinfo {pages} {29} (\bibinfo {year} {2010})}\BibitemShut {NoStop}%
\bibitem [{\citenamefont {da~Costa}\ \emph {et~al.}(2015)\citenamefont
  {da~Costa}, \citenamefont {Varella}, \citenamefont {Bettega},\ and\
  \citenamefont {Lima}}]{daCosta-EPJD-15}%
  \BibitemOpen
  \bibfield  {author} {\bibinfo {author} {\bibfnamefont {R.~F.}\ \bibnamefont
  {da~Costa}}, \bibinfo {author} {\bibfnamefont {M.~T. d.~N.}\ \bibnamefont
  {Varella}}, \bibinfo {author} {\bibfnamefont {M.~H.~F.}\ \bibnamefont
  {Bettega}},\ and\ \bibinfo {author} {\bibfnamefont {M.~A.~P.}\ \bibnamefont
  {Lima}},\ }\bibfield  {title} {\bibinfo {title} {Recent advances in the
  application of the {Schwinger} multichannel method with pseudopotentials to
  electron-molecule collisions},\ }\href
  {https://doi.org/10.1140/epjd/e2015-60192-6} {\bibfield  {journal} {\bibinfo
  {journal} {The Eur. Phys. J. D}\ }\textbf {\bibinfo {volume} {69}},\ \bibinfo
  {pages} {159} (\bibinfo {year} {2015})}\BibitemShut {NoStop}%
\bibitem [{\citenamefont {Mašín}\ \emph {et~al.}(2020)\citenamefont
  {Mašín}, \citenamefont {Benda}, \citenamefont {Gorfinkiel}, \citenamefont
  {Harvey},\ and\ \citenamefont {Tennyson}}]{Masin-CPC-20}%
  \BibitemOpen
  \bibfield  {author} {\bibinfo {author} {\bibfnamefont {Z.}~\bibnamefont
  {Mašín}}, \bibinfo {author} {\bibfnamefont {J.}~\bibnamefont {Benda}},
  \bibinfo {author} {\bibfnamefont {J.~D.}\ \bibnamefont {Gorfinkiel}},
  \bibinfo {author} {\bibfnamefont {A.~G.}\ \bibnamefont {Harvey}},\ and\
  \bibinfo {author} {\bibfnamefont {J.}~\bibnamefont {Tennyson}},\ }\bibfield
  {title} {\bibinfo {title} {{UKRmol+}: A suite for modelling electronic
  processes in molecules interacting with electrons, positrons and photons
  using the {R-matrix} method},\ }\href
  {https://doi.org/https://doi.org/10.1016/j.cpc.2019.107092} {\bibfield
  {journal} {\bibinfo  {journal} {Comput. Phys. Commun.}\ }\textbf {\bibinfo
  {volume} {249}},\ \bibinfo {pages} {107092} (\bibinfo {year}
  {2020})}\BibitemShut {NoStop}%
\bibitem [{\citenamefont {Domcke}(1991)}]{Domcke-PhysRep-91}%
  \BibitemOpen
  \bibfield  {author} {\bibinfo {author} {\bibfnamefont {W.}~\bibnamefont
  {Domcke}},\ }\bibfield  {title} {\bibinfo {title} {Theory of resonance and
  threshold effects in electron--molecule collisions: The projection-operator
  approach},\ }\href {https://doi.org/10.1016/0370-1573(91)90125-6} {\bibfield
  {journal} {\bibinfo  {journal} {Phys. Rep.}\ }\textbf {\bibinfo {volume}
  {208}},\ \bibinfo {pages} {97} (\bibinfo {year} {1991})}\BibitemShut
  {NoStop}%
\bibitem [{\citenamefont {Zammit}\ \emph
  {et~al.}(2017{\natexlab{a}})\citenamefont {Zammit}, \citenamefont {Fursa},
  \citenamefont {Savage},\ and\ \citenamefont {Bray}}]{Zammit-JPB-17}%
  \BibitemOpen
  \bibfield  {author} {\bibinfo {author} {\bibfnamefont {M.~C.}\ \bibnamefont
  {Zammit}}, \bibinfo {author} {\bibfnamefont {D.~V.}\ \bibnamefont {Fursa}},
  \bibinfo {author} {\bibfnamefont {J.~S.}\ \bibnamefont {Savage}},\ and\
  \bibinfo {author} {\bibfnamefont {I.}~\bibnamefont {Bray}},\ }\bibfield
  {title} {\bibinfo {title} {Electron– and positron–molecule scattering:
  development of the molecular convergent close-coupling method},\ }\href
  {https://doi.org/10.1088/1361-6455/aa6e74} {\bibfield  {journal} {\bibinfo
  {journal} {J. Phys. B: At. Mol. Opt. Phys.}\ }\textbf {\bibinfo {volume}
  {50}},\ \bibinfo {pages} {123001} (\bibinfo {year}
  {2017}{\natexlab{a}})}\BibitemShut {NoStop}%
\bibitem [{\citenamefont {Burke}(2011)}]{Burke2011}%
  \BibitemOpen
  \bibfield  {author} {\bibinfo {author} {\bibfnamefont {P.~G.}\ \bibnamefont
  {Burke}},\ }\href@noop {} {\emph {\bibinfo {title} {R-matrix theory of atomic
  collisions: Application to atomic, molecular and optical processes}}}\
  (\bibinfo  {publisher} {Springer Science \& Business Media},\ \bibinfo {year}
  {2011})\BibitemShut {NoStop}%
\bibitem [{\citenamefont {Zatsarinny}\ and\ \citenamefont
  {Bartschat}(2013)}]{Zatsarinny-JPB-13}%
  \BibitemOpen
  \bibfield  {author} {\bibinfo {author} {\bibfnamefont {O.}~\bibnamefont
  {Zatsarinny}}\ and\ \bibinfo {author} {\bibfnamefont {K.}~\bibnamefont
  {Bartschat}},\ }\bibfield  {title} {\bibinfo {title} {The {B-spline}
  {R-matrix} method for atomic processes: application to atomic structure,
  electron collisions and photoionization},\ }\href
  {https://doi.org/10.1088/0953-4075/46/11/112001} {\bibfield  {journal}
  {\bibinfo  {journal} {J. Phys. B: At. Mol. Opt. Phys.}\ }\textbf {\bibinfo
  {volume} {46}},\ \bibinfo {pages} {112001} (\bibinfo {year}
  {2013})}\BibitemShut {NoStop}%
\bibitem [{\citenamefont {Mašín}\ \emph {et~al.}(2026)\citenamefont
  {Mašín}, \citenamefont {Benda}, \citenamefont {Crhán}, \citenamefont
  {Armstrong}, \citenamefont {Nelson}, \citenamefont {Mohr},\ and\
  \citenamefont {Tennyson}}]{Masin-JCTC-26}%
  \BibitemOpen
  \bibfield  {author} {\bibinfo {author} {\bibfnamefont {Z.}~\bibnamefont
  {Mašín}}, \bibinfo {author} {\bibfnamefont {J.}~\bibnamefont {Benda}},
  \bibinfo {author} {\bibfnamefont {M.}~\bibnamefont {Crhán}}, \bibinfo
  {author} {\bibfnamefont {G.~S.~J.}\ \bibnamefont {Armstrong}}, \bibinfo
  {author} {\bibfnamefont {A.}~\bibnamefont {Nelson}}, \bibinfo {author}
  {\bibfnamefont {S.}~\bibnamefont {Mohr}},\ and\ \bibinfo {author}
  {\bibfnamefont {J.}~\bibnamefont {Tennyson}},\ }\bibfield  {title} {\bibinfo
  {title} {Effective core potentials for calculations of continuum spectra of
  molecules using the molecular {R‑Matrix} method},\ }\href
  {https://doi.org/10.1021/acs.jctc.5c01729} {\bibfield  {journal} {\bibinfo
  {journal} {J. Chem. Theory Comput.}\ }\textbf {\bibinfo {volume} {22}},\
  \bibinfo {pages} {1563} (\bibinfo {year} {2026})}\BibitemShut {NoStop}%
\bibitem [{\citenamefont {Gorfinkiel}\ and\ \citenamefont
  {Tennyson}(2005)}]{Gorfinkiel-JPB-05}%
  \BibitemOpen
  \bibfield  {author} {\bibinfo {author} {\bibfnamefont {J.~D.}\ \bibnamefont
  {Gorfinkiel}}\ and\ \bibinfo {author} {\bibfnamefont {J.}~\bibnamefont
  {Tennyson}},\ }\bibfield  {title} {\bibinfo {title} {Electron impact
  ionization of small molecules at intermediate energies: The molecular
  {$R$}-matrix with pseudostates method},\ }\href
  {https://doi.org/10.1088/0953-4075/38/11/003} {\bibfield  {journal} {\bibinfo
   {journal} {J. Phys. B: At. Mol. Opt. Phys.}\ }\textbf {\bibinfo {volume}
  {38}},\ \bibinfo {pages} {1607} (\bibinfo {year} {2005})}\BibitemShut
  {NoStop}%
\bibitem [{\citenamefont {Bray}(1994)}]{Bray-PRA-94}%
  \BibitemOpen
  \bibfield  {author} {\bibinfo {author} {\bibfnamefont {I.}~\bibnamefont
  {Bray}},\ }\bibfield  {title} {\bibinfo {title} {Convergent close-coupling
  method for the calculation of electron scattering on hydrogenlike targets},\
  }\href {https://doi.org/10.1103/PhysRevA.49.1066} {\bibfield  {journal}
  {\bibinfo  {journal} {Phys. Rev. A}\ }\textbf {\bibinfo {volume} {49}},\
  \bibinfo {pages} {1066} (\bibinfo {year} {1994})}\BibitemShut {NoStop}%
\bibitem [{\citenamefont {Fursa}\ and\ \citenamefont
  {Bray}(1997)}]{Fursa-JPB-97}%
  \BibitemOpen
  \bibfield  {author} {\bibinfo {author} {\bibfnamefont {D.~V.}\ \bibnamefont
  {Fursa}}\ and\ \bibinfo {author} {\bibfnamefont {I.}~\bibnamefont {Bray}},\
  }\bibfield  {title} {\bibinfo {title} {Convergent close-coupling calculations
  of electron - helium scattering},\ }\href
  {https://doi.org/10.1088/0953-4075/30/4/003} {\bibfield  {journal} {\bibinfo
  {journal} {J. Phys. B: At. Mol. Opt. Phys.}\ }\textbf {\bibinfo {volume}
  {30}},\ \bibinfo {pages} {757} (\bibinfo {year} {1997})}\BibitemShut
  {NoStop}%
\bibitem [{\citenamefont {Bray}\ \emph {et~al.}(2002)\citenamefont {Bray},
  \citenamefont {Fursa}, \citenamefont {Kheifets},\ and\ \citenamefont
  {Stelbovics}}]{Bray-JPB-02}%
  \BibitemOpen
  \bibfield  {author} {\bibinfo {author} {\bibfnamefont {I.}~\bibnamefont
  {Bray}}, \bibinfo {author} {\bibfnamefont {D.~V.}\ \bibnamefont {Fursa}},
  \bibinfo {author} {\bibfnamefont {A.~S.}\ \bibnamefont {Kheifets}},\ and\
  \bibinfo {author} {\bibfnamefont {A.~T.}\ \bibnamefont {Stelbovics}},\
  }\bibfield  {title} {\bibinfo {title} {Electrons and photons colliding with
  atoms: development and application of the convergent close-coupling method},\
  }\href {https://doi.org/10.1088/0953-4075/35/15/201} {\bibfield  {journal}
  {\bibinfo  {journal} {J. Phys. B: At. Mol. Opt. Phys.}\ }\textbf {\bibinfo
  {volume} {35}},\ \bibinfo {pages} {R117} (\bibinfo {year}
  {2002})}\BibitemShut {NoStop}%
\bibitem [{\citenamefont {Bray}\ and\ \citenamefont
  {Fursa}(1996)}]{Bray-PRA-96}%
  \BibitemOpen
  \bibfield  {author} {\bibinfo {author} {\bibfnamefont {I.}~\bibnamefont
  {Bray}}\ and\ \bibinfo {author} {\bibfnamefont {D.~V.}\ \bibnamefont
  {Fursa}},\ }\bibfield  {title} {\bibinfo {title} {Calculation of ionization
  within the close-coupling formalism},\ }\href
  {https://doi.org/10.1103/PhysRevA.54.2991} {\bibfield  {journal} {\bibinfo
  {journal} {Phys. Rev. A}\ }\textbf {\bibinfo {volume} {54}},\ \bibinfo
  {pages} {2991} (\bibinfo {year} {1996})}\BibitemShut {NoStop}%
\bibitem [{\citenamefont {Bray}(1997)}]{Bray-PRL-97}%
  \BibitemOpen
  \bibfield  {author} {\bibinfo {author} {\bibfnamefont {I.}~\bibnamefont
  {Bray}},\ }\bibfield  {title} {\bibinfo {title} {Close-coupling theory of
  ionization: Successes and failures},\ }\href
  {https://doi.org/10.1103/PhysRevLett.78.4721} {\bibfield  {journal} {\bibinfo
   {journal} {Phys. Rev. Lett.}\ }\textbf {\bibinfo {volume} {78}},\ \bibinfo
  {pages} {4721} (\bibinfo {year} {1997})}\BibitemShut {NoStop}%
\bibitem [{\citenamefont {Bray}\ \emph {et~al.}(2001)\citenamefont {Bray},
  \citenamefont {Fursa},\ and\ \citenamefont {Stelbovics}}]{Bray-PRA-01}%
  \BibitemOpen
  \bibfield  {author} {\bibinfo {author} {\bibfnamefont {I.}~\bibnamefont
  {Bray}}, \bibinfo {author} {\bibfnamefont {D.~V.}\ \bibnamefont {Fursa}},\
  and\ \bibinfo {author} {\bibfnamefont {A.~T.}\ \bibnamefont {Stelbovics}},\
  }\bibfield  {title} {\bibinfo {title} {Close-coupling approach to
  electron-impact ionization of helium},\ }\href
  {https://doi.org/10.1103/PhysRevA.63.040702} {\bibfield  {journal} {\bibinfo
  {journal} {Phys. Rev. A}\ }\textbf {\bibinfo {volume} {63}},\ \bibinfo
  {pages} {040702} (\bibinfo {year} {2001})}\BibitemShut {NoStop}%
\bibitem [{\citenamefont {Zammit}\ \emph {et~al.}(2013)\citenamefont {Zammit},
  \citenamefont {Fursa},\ and\ \citenamefont {Bray}}]{Zammit-PRA-13}%
  \BibitemOpen
  \bibfield  {author} {\bibinfo {author} {\bibfnamefont {M.~C.}\ \bibnamefont
  {Zammit}}, \bibinfo {author} {\bibfnamefont {D.~V.}\ \bibnamefont {Fursa}},\
  and\ \bibinfo {author} {\bibfnamefont {I.}~\bibnamefont {Bray}},\ }\bibfield
  {title} {\bibinfo {title} {Calculations of electron scattering from
  {H${{}_{2}}^{+}$}},\ }\href {https://doi.org/10.1103/PhysRevA.88.062709}
  {\bibfield  {journal} {\bibinfo  {journal} {Phys. Rev. A}\ }\textbf {\bibinfo
  {volume} {88}},\ \bibinfo {pages} {062709} (\bibinfo {year}
  {2013})}\BibitemShut {NoStop}%
\bibitem [{\citenamefont {Zammit}\ \emph {et~al.}(2014)\citenamefont {Zammit},
  \citenamefont {Fursa},\ and\ \citenamefont {Bray}}]{Zammit-PRA-14}%
  \BibitemOpen
  \bibfield  {author} {\bibinfo {author} {\bibfnamefont {M.~C.}\ \bibnamefont
  {Zammit}}, \bibinfo {author} {\bibfnamefont {D.~V.}\ \bibnamefont {Fursa}},\
  and\ \bibinfo {author} {\bibfnamefont {I.}~\bibnamefont {Bray}},\ }\bibfield
  {title} {\bibinfo {title} {Electron scattering from the molecular hydrogen
  ion and its isotopologues},\ }\href
  {https://doi.org/10.1103/PhysRevA.90.022711} {\bibfield  {journal} {\bibinfo
  {journal} {Phys. Rev. A}\ }\textbf {\bibinfo {volume} {90}},\ \bibinfo
  {pages} {022711} (\bibinfo {year} {2014})}\BibitemShut {NoStop}%
\bibitem [{\citenamefont {Zammit}\ \emph {et~al.}(2016)\citenamefont {Zammit},
  \citenamefont {Savage}, \citenamefont {Fursa},\ and\ \citenamefont
  {Bray}}]{Zammit-PRL-16}%
  \BibitemOpen
  \bibfield  {author} {\bibinfo {author} {\bibfnamefont {M.~C.}\ \bibnamefont
  {Zammit}}, \bibinfo {author} {\bibfnamefont {J.~S.}\ \bibnamefont {Savage}},
  \bibinfo {author} {\bibfnamefont {D.~V.}\ \bibnamefont {Fursa}},\ and\
  \bibinfo {author} {\bibfnamefont {I.}~\bibnamefont {Bray}},\ }\bibfield
  {title} {\bibinfo {title} {Complete solution of electronic excitation and
  ionization in electron-hydrogen molecule scattering},\ }\href
  {https://doi.org/10.1103/PhysRevLett.116.233201} {\bibfield  {journal}
  {\bibinfo  {journal} {Phys. Rev. Lett.}\ }\textbf {\bibinfo {volume} {116}},\
  \bibinfo {pages} {233201} (\bibinfo {year} {2016})}\BibitemShut {NoStop}%
\bibitem [{\citenamefont {Zammit}\ \emph
  {et~al.}(2017{\natexlab{b}})\citenamefont {Zammit}, \citenamefont {Savage},
  \citenamefont {Fursa},\ and\ \citenamefont {Bray}}]{Zammit-PRA-17}%
  \BibitemOpen
  \bibfield  {author} {\bibinfo {author} {\bibfnamefont {M.~C.}\ \bibnamefont
  {Zammit}}, \bibinfo {author} {\bibfnamefont {J.~S.}\ \bibnamefont {Savage}},
  \bibinfo {author} {\bibfnamefont {D.~V.}\ \bibnamefont {Fursa}},\ and\
  \bibinfo {author} {\bibfnamefont {I.}~\bibnamefont {Bray}},\ }\bibfield
  {title} {\bibinfo {title} {Electron-impact excitation of molecular
  hydrogen},\ }\href {https://doi.org/10.1103/PhysRevA.95.022708} {\bibfield
  {journal} {\bibinfo  {journal} {Phys. Rev. A}\ }\textbf {\bibinfo {volume}
  {95}},\ \bibinfo {pages} {022708} (\bibinfo {year}
  {2017}{\natexlab{b}})}\BibitemShut {NoStop}%
\bibitem [{\citenamefont {Takatsuka}\ and\ \citenamefont
  {McKoy}(1981)}]{Takatsuka-PRA-81}%
  \BibitemOpen
  \bibfield  {author} {\bibinfo {author} {\bibfnamefont {K.}~\bibnamefont
  {Takatsuka}}\ and\ \bibinfo {author} {\bibfnamefont {V.}~\bibnamefont
  {McKoy}},\ }\bibfield  {title} {\bibinfo {title} {Extension of the
  {Schwinger} variational principle beyond the static-exchange approximation},\
  }\href {https://doi.org/10.1103/PhysRevA.24.2473} {\bibfield  {journal}
  {\bibinfo  {journal} {Phys. Rev. A}\ }\textbf {\bibinfo {volume} {24}},\
  \bibinfo {pages} {2473} (\bibinfo {year} {1981})}\BibitemShut {NoStop}%
\bibitem [{\citenamefont {Takatsuka}\ and\ \citenamefont
  {McKoy}(1984)}]{Takatsuka-PRA-84}%
  \BibitemOpen
  \bibfield  {author} {\bibinfo {author} {\bibfnamefont {K.}~\bibnamefont
  {Takatsuka}}\ and\ \bibinfo {author} {\bibfnamefont {V.}~\bibnamefont
  {McKoy}},\ }\bibfield  {title} {\bibinfo {title} {Theory of electronically
  inelastic scattering of electrons by molecules},\ }\href
  {https://doi.org/10.1103/PhysRevA.30.1734} {\bibfield  {journal} {\bibinfo
  {journal} {Phys. Rev. A}\ }\textbf {\bibinfo {volume} {30}},\ \bibinfo
  {pages} {1734} (\bibinfo {year} {1984})}\BibitemShut {NoStop}%
\bibitem [{\citenamefont {Lucchese}\ \emph {et~al.}(1986)\citenamefont
  {Lucchese}, \citenamefont {Takatsuka},\ and\ \citenamefont
  {McKoy}}]{Lucchese-PhysRep-86}%
  \BibitemOpen
  \bibfield  {author} {\bibinfo {author} {\bibfnamefont {R.~R.}\ \bibnamefont
  {Lucchese}}, \bibinfo {author} {\bibfnamefont {K.}~\bibnamefont
  {Takatsuka}},\ and\ \bibinfo {author} {\bibfnamefont {V.}~\bibnamefont
  {McKoy}},\ }\bibfield  {title} {\bibinfo {title} {Applications of the
  {Schwinger} variational principle to electron-molecule collisions and
  molecular photoionization},\ }\href
  {https://doi.org/10.1016/0370-1573(86)90147-X} {\bibfield  {journal}
  {\bibinfo  {journal} {Phys. Rep.}\ }\textbf {\bibinfo {volume} {131}},\
  \bibinfo {pages} {147} (\bibinfo {year} {1986})}\BibitemShut {NoStop}%
\bibitem [{\citenamefont {Schneider}\ and\ \citenamefont
  {Rescigno}(1988)}]{Schneider-PRA-88}%
  \BibitemOpen
  \bibfield  {author} {\bibinfo {author} {\bibfnamefont {B.~I.}\ \bibnamefont
  {Schneider}}\ and\ \bibinfo {author} {\bibfnamefont {T.~N.}\ \bibnamefont
  {Rescigno}},\ }\bibfield  {title} {\bibinfo {title} {Complex {Kohn}
  variational method: Application to low-energy electron-molecule collisions},\
  }\href {https://doi.org/10.1103/PhysRevA.37.3749} {\bibfield  {journal}
  {\bibinfo  {journal} {Phys. Rev. A}\ }\textbf {\bibinfo {volume} {37}},\
  \bibinfo {pages} {3749} (\bibinfo {year} {1988})}\BibitemShut {NoStop}%
\bibitem [{\citenamefont {Rescigno}\ and\ \citenamefont
  {Schneider}(1988)}]{Rescigno-JPB-88}%
  \BibitemOpen
  \bibfield  {author} {\bibinfo {author} {\bibfnamefont {T.~N.}\ \bibnamefont
  {Rescigno}}\ and\ \bibinfo {author} {\bibfnamefont {B.~I.}\ \bibnamefont
  {Schneider}},\ }\bibfield  {title} {\bibinfo {title} {Electron-impact
  excitation of the {$b\,{}^3\Sigma_u^+$} state of {$\mathrm{H}_2$} using the
  complex {Kohn} method: {$R$} dependence of the cross section},\ }\href
  {https://doi.org/10.1088/0953-4075/21/22/003} {\bibfield  {journal} {\bibinfo
   {journal} {J. Phys. B: At. Mol. Opt. Phys.}\ }\textbf {\bibinfo {volume}
  {21}},\ \bibinfo {pages} {L691} (\bibinfo {year} {1988})}\BibitemShut
  {NoStop}%
\bibitem [{\citenamefont {McCurdy}\ and\ \citenamefont
  {Rescigno}(1989)}]{McCurdy-PRA-89}%
  \BibitemOpen
  \bibfield  {author} {\bibinfo {author} {\bibfnamefont {C.~W.}\ \bibnamefont
  {McCurdy}}\ and\ \bibinfo {author} {\bibfnamefont {T.~N.}\ \bibnamefont
  {Rescigno}},\ }\bibfield  {title} {\bibinfo {title} {Collisions of electrons
  with polyatomic molecules: Electron-methane scattering by the complex {Kohn}
  variational method},\ }\href {https://doi.org/10.1103/PhysRevA.39.4487}
  {\bibfield  {journal} {\bibinfo  {journal} {Phys. Rev. A}\ }\textbf {\bibinfo
  {volume} {39}},\ \bibinfo {pages} {4487} (\bibinfo {year}
  {1989})}\BibitemShut {NoStop}%
\bibitem [{\citenamefont {Rescigno}\ \emph {et~al.}(1989)\citenamefont
  {Rescigno}, \citenamefont {McCurdy},\ and\ \citenamefont
  {Schneider}}]{Rescigno-PRL-89}%
  \BibitemOpen
  \bibfield  {author} {\bibinfo {author} {\bibfnamefont {T.~N.}\ \bibnamefont
  {Rescigno}}, \bibinfo {author} {\bibfnamefont {C.~W.}\ \bibnamefont
  {McCurdy}},\ and\ \bibinfo {author} {\bibfnamefont {B.~I.}\ \bibnamefont
  {Schneider}},\ }\bibfield  {title} {\bibinfo {title} {Accurate ab initio
  treatment of low-energy electron collisions with polyatomic molecules:
  Resonant electron-formaldehyde scattering},\ }\href
  {https://doi.org/10.1103/PhysRevLett.63.248} {\bibfield  {journal} {\bibinfo
  {journal} {Phys. Rev. Lett.}\ }\textbf {\bibinfo {volume} {63}},\ \bibinfo
  {pages} {248} (\bibinfo {year} {1989})}\BibitemShut {NoStop}%
\bibitem [{\citenamefont {Schneider}\ \emph {et~al.}(1991)\citenamefont
  {Schneider}, \citenamefont {Rescigno}, \citenamefont {Lengsfield},\ and\
  \citenamefont {McCurdy}}]{Schneider-PRL-91}%
  \BibitemOpen
  \bibfield  {author} {\bibinfo {author} {\bibfnamefont {B.~I.}\ \bibnamefont
  {Schneider}}, \bibinfo {author} {\bibfnamefont {T.~N.}\ \bibnamefont
  {Rescigno}}, \bibinfo {author} {\bibfnamefont {B.~H.}\ \bibnamefont
  {Lengsfield}},\ and\ \bibinfo {author} {\bibfnamefont {C.~W.}\ \bibnamefont
  {McCurdy}},\ }\bibfield  {title} {\bibinfo {title} {Accurate ab initio
  treatment of low-energy electron collisions with ethylene},\ }\href
  {https://doi.org/10.1103/PhysRevLett.66.2728} {\bibfield  {journal} {\bibinfo
   {journal} {Phys. Rev. Lett.}\ }\textbf {\bibinfo {volume} {66}},\ \bibinfo
  {pages} {2728} (\bibinfo {year} {1991})}\BibitemShut {NoStop}%
\bibitem [{\citenamefont {Rescigno}\ and\ \citenamefont
  {Schneider}(1992)}]{Rescigno-PRA-92}%
  \BibitemOpen
  \bibfield  {author} {\bibinfo {author} {\bibfnamefont {T.~N.}\ \bibnamefont
  {Rescigno}}\ and\ \bibinfo {author} {\bibfnamefont {B.~I.}\ \bibnamefont
  {Schneider}},\ }\bibfield  {title} {\bibinfo {title} {Electron-impact
  excitation of the {$T$} and {$V$} states of ethylene: An ab initio study},\
  }\href {https://doi.org/10.1103/PhysRevA.45.2894} {\bibfield  {journal}
  {\bibinfo  {journal} {Phys. Rev. A}\ }\textbf {\bibinfo {volume} {45}},\
  \bibinfo {pages} {2894} (\bibinfo {year} {1992})}\BibitemShut {NoStop}%
\bibitem [{\citenamefont {Greenman}\ \emph {et~al.}(2017)\citenamefont
  {Greenman}, \citenamefont {Lucchese},\ and\ \citenamefont
  {McCurdy}}]{Greenman-PRA-17}%
  \BibitemOpen
  \bibfield  {author} {\bibinfo {author} {\bibfnamefont {L.}~\bibnamefont
  {Greenman}}, \bibinfo {author} {\bibfnamefont {R.~R.}\ \bibnamefont
  {Lucchese}},\ and\ \bibinfo {author} {\bibfnamefont {C.~W.}\ \bibnamefont
  {McCurdy}},\ }\bibfield  {title} {\bibinfo {title} {Variational treatment of
  electron--polyatomic-molecule scattering calculations using adaptive overset
  grids},\ }\href {https://doi.org/10.1103/PhysRevA.96.052706} {\bibfield
  {journal} {\bibinfo  {journal} {Phys. Rev. A}\ }\textbf {\bibinfo {volume}
  {96}},\ \bibinfo {pages} {052706} (\bibinfo {year} {2017})}\BibitemShut
  {NoStop}%
\bibitem [{\citenamefont {Gianturco}\ and\ \citenamefont
  {Jain}(1986)}]{Gianturco-PhysRep-86}%
  \BibitemOpen
  \bibfield  {author} {\bibinfo {author} {\bibfnamefont {F.~A.}\ \bibnamefont
  {Gianturco}}\ and\ \bibinfo {author} {\bibfnamefont {A.}~\bibnamefont
  {Jain}},\ }\bibfield  {title} {\bibinfo {title} {The theory of electron
  scattering from polyatomic molecules},\ }\href
  {https://doi.org/10.1016/0370-1573(86)90125-0} {\bibfield  {journal}
  {\bibinfo  {journal} {Phys. Rep.}\ }\textbf {\bibinfo {volume} {143}},\
  \bibinfo {pages} {347} (\bibinfo {year} {1986})}\BibitemShut {NoStop}%
\bibitem [{\citenamefont {Staszewska}\ \emph {et~al.}(1983)\citenamefont
  {Staszewska}, \citenamefont {Schwenke}, \citenamefont {Thirumalai},\ and\
  \citenamefont {Truhlar}}]{Staszewska-PRA-83}%
  \BibitemOpen
  \bibfield  {author} {\bibinfo {author} {\bibfnamefont {G.}~\bibnamefont
  {Staszewska}}, \bibinfo {author} {\bibfnamefont {D.~W.}\ \bibnamefont
  {Schwenke}}, \bibinfo {author} {\bibfnamefont {D.}~\bibnamefont
  {Thirumalai}},\ and\ \bibinfo {author} {\bibfnamefont {D.~G.}\ \bibnamefont
  {Truhlar}},\ }\bibfield  {title} {\bibinfo {title} {Quasifree-scattering
  model for the imaginary part of the optical potential for electron
  scattering},\ }\href {https://doi.org/10.1103/PhysRevA.28.2740} {\bibfield
  {journal} {\bibinfo  {journal} {Phys. Rev. A}\ }\textbf {\bibinfo {volume}
  {28}},\ \bibinfo {pages} {2740} (\bibinfo {year} {1983})}\BibitemShut
  {NoStop}%
\bibitem [{\citenamefont {Fliflet}\ and\ \citenamefont
  {McKoy}(1980)}]{Fliflet-PRA-80}%
  \BibitemOpen
  \bibfield  {author} {\bibinfo {author} {\bibfnamefont {A.~W.}\ \bibnamefont
  {Fliflet}}\ and\ \bibinfo {author} {\bibfnamefont {V.}~\bibnamefont
  {McKoy}},\ }\bibfield  {title} {\bibinfo {title}
  {Distorted-wave-approximation cross sections for excitation of the
  {$b\,{}^3\Sigma_u^+$} and {$B\,{}^1\Sigma_u^+$} states of {H$_2$} by
  low-energy-electron impact},\ }\href
  {https://doi.org/10.1103/PhysRevA.21.1863} {\bibfield  {journal} {\bibinfo
  {journal} {Phys. Rev. A}\ }\textbf {\bibinfo {volume} {21}},\ \bibinfo
  {pages} {1863} (\bibinfo {year} {1980})}\BibitemShut {NoStop}%
\bibitem [{Note1()}]{Note1}%
  \BibitemOpen
  \bibinfo {note} {Certain software is identified in this paper to foster
  understanding. Such identification does not imply recommendation or
  endorsement by the National Institute of Standards and Technology, nor does
  it imply that the materials or equipment identified are necessarily the best
  available for the purpose.}\BibitemShut {Stop}%
\bibitem [{\citenamefont {Randazzo}\ \emph {et~al.}(2023)\citenamefont
  {Randazzo}, \citenamefont {Marante}, \citenamefont {Chattopadhyay},
  \citenamefont {Schneider}, \citenamefont {Olsen},\ and\ \citenamefont
  {Argenti}}]{Randazzo-PRR-23}%
  \BibitemOpen
  \bibfield  {author} {\bibinfo {author} {\bibfnamefont {J.~M.}\ \bibnamefont
  {Randazzo}}, \bibinfo {author} {\bibfnamefont {C.}~\bibnamefont {Marante}},
  \bibinfo {author} {\bibfnamefont {S.}~\bibnamefont {Chattopadhyay}}, \bibinfo
  {author} {\bibfnamefont {B.~I.}\ \bibnamefont {Schneider}}, \bibinfo {author}
  {\bibfnamefont {J.}~\bibnamefont {Olsen}},\ and\ \bibinfo {author}
  {\bibfnamefont {L.}~\bibnamefont {Argenti}},\ }\bibfield  {title} {\bibinfo
  {title} {{ASTRA}: Transition-density-matrix approach to molecular
  ionization},\ }\href {https://doi.org/10.1103/PhysRevResearch.5.043115}
  {\bibfield  {journal} {\bibinfo  {journal} {Phys. Rev. Res.}\ }\textbf
  {\bibinfo {volume} {5}},\ \bibinfo {pages} {043115} (\bibinfo {year}
  {2023})}\BibitemShut {NoStop}%
\bibitem [{\citenamefont {Marante}\ \emph {et~al.}(2026)\citenamefont
  {Marante}, \citenamefont {Borr{\`a}s}, \citenamefont {{Gonz{\'a}lez
  V{\'a}zquez}}, \citenamefont {Mart{\'i}n},\ and\ \citenamefont
  {Argenti}}]{Marante2026}%
  \BibitemOpen
  \bibfield  {author} {\bibinfo {author} {\bibfnamefont {C.}~\bibnamefont
  {Marante}}, \bibinfo {author} {\bibfnamefont {V.~J.}\ \bibnamefont
  {Borr{\`a}s}}, \bibinfo {author} {\bibfnamefont {J.}~\bibnamefont
  {{Gonz{\'a}lez V{\'a}zquez}}}, \bibinfo {author} {\bibfnamefont
  {F.}~\bibnamefont {Mart{\'i}n}},\ and\ \bibinfo {author} {\bibfnamefont
  {L.}~\bibnamefont {Argenti}},\ }\bibfield  {title} {\bibinfo {title}
  {Variational surface matching for multichannel molecular photoionization},\
  }\Eprint {https://arxiv.org/abs/2610.40953} {arXiv:2610.40953
  [physics.atom-ph]}  (\bibinfo {year} {2026}),\ \bibinfo {note}
  {arXiv:2610.40953 [physics.atom-ph]}\BibitemShut {NoStop}%
\bibitem [{\citenamefont {Buckman}\ and\ \citenamefont
  {Lohmann}(1986)}]{Buckman-JPB-86}%
  \BibitemOpen
  \bibfield  {author} {\bibinfo {author} {\bibfnamefont {S.~J.}\ \bibnamefont
  {Buckman}}\ and\ \bibinfo {author} {\bibfnamefont {B.}~\bibnamefont
  {Lohmann}},\ }\bibfield  {title} {\bibinfo {title} {Low-energy total cross
  section measurements for electron scattering from helium and argon},\ }\href
  {https://doi.org/10.1088/0022-3700/19/16/013} {\bibfield  {journal} {\bibinfo
   {journal} {J. Phys. B: At. Mol. Phys.}\ }\textbf {\bibinfo {volume} {19}},\
  \bibinfo {pages} {2547} (\bibinfo {year} {1986})}\BibitemShut {NoStop}%
\bibitem [{\citenamefont {Shigemura}\ \emph {et~al.}(2014)\citenamefont
  {Shigemura}, \citenamefont {Kitajima}, \citenamefont {Kurokawa},
  \citenamefont {Toyoshima}, \citenamefont {Odagiri}, \citenamefont {Suga},
  \citenamefont {Kato}, \citenamefont {Hoshino}, \citenamefont {Tanaka},\ and\
  \citenamefont {Ito}}]{Shigemura-PRA-14}%
  \BibitemOpen
  \bibfield  {author} {\bibinfo {author} {\bibfnamefont {K.}~\bibnamefont
  {Shigemura}}, \bibinfo {author} {\bibfnamefont {M.}~\bibnamefont {Kitajima}},
  \bibinfo {author} {\bibfnamefont {M.}~\bibnamefont {Kurokawa}}, \bibinfo
  {author} {\bibfnamefont {K.}~\bibnamefont {Toyoshima}}, \bibinfo {author}
  {\bibfnamefont {T.}~\bibnamefont {Odagiri}}, \bibinfo {author} {\bibfnamefont
  {A.}~\bibnamefont {Suga}}, \bibinfo {author} {\bibfnamefont {H.}~\bibnamefont
  {Kato}}, \bibinfo {author} {\bibfnamefont {M.}~\bibnamefont {Hoshino}},
  \bibinfo {author} {\bibfnamefont {H.}~\bibnamefont {Tanaka}},\ and\ \bibinfo
  {author} {\bibfnamefont {K.}~\bibnamefont {Ito}},\ }\bibfield  {title}
  {\bibinfo {title} {Total cross sections for electron scattering from {He} and
  {Ne} at very low energies},\ }\href
  {https://doi.org/10.1103/PhysRevA.89.022709} {\bibfield  {journal} {\bibinfo
  {journal} {Phys. Rev. A}\ }\textbf {\bibinfo {volume} {89}},\ \bibinfo
  {pages} {022709} (\bibinfo {year} {2014})}\BibitemShut {NoStop}%
\bibitem [{\citenamefont {Bartschat}(1998)}]{Bartschat-JPB-98}%
  \BibitemOpen
  \bibfield  {author} {\bibinfo {author} {\bibfnamefont {K.}~\bibnamefont
  {Bartschat}},\ }\bibfield  {title} {\bibinfo {title} {Electron-impact
  excitation of helium from the ${1~^1S}$ and ${2~^3S}$ states},\ }\href
  {https://doi.org/10.1088/0953-4075/31/10/005} {\bibfield  {journal} {\bibinfo
   {journal} {J. Phys. B: At. Mol. Opt. Phys.}\ }\textbf {\bibinfo {volume}
  {31}},\ \bibinfo {pages} {L469} (\bibinfo {year} {1998})}\BibitemShut
  {NoStop}%
\bibitem [{\citenamefont {McEachran}\ \emph {et~al.}(2021)\citenamefont
  {McEachran}, \citenamefont {Hamilton},\ and\ \citenamefont
  {Bartschat}}]{McEachran-Atoms-21}%
  \BibitemOpen
  \bibfield  {author} {\bibinfo {author} {\bibfnamefont {R.~P.}\ \bibnamefont
  {McEachran}}, \bibinfo {author} {\bibfnamefont {K.~R.}\ \bibnamefont
  {Hamilton}},\ and\ \bibinfo {author} {\bibfnamefont {K.}~\bibnamefont
  {Bartschat}},\ }\bibfield  {title} {\bibinfo {title} {Low-energy elastic
  electron scattering from helium atoms},\ }\href
  {https://doi.org/10.3390/atoms9040082} {\bibfield  {journal} {\bibinfo
  {journal} {Atoms}\ }\textbf {\bibinfo {volume} {9}},\ \bibinfo {pages} {82}
  (\bibinfo {year} {2021})}\BibitemShut {NoStop}%
\bibitem [{\citenamefont {Bartschat}\ \emph {et~al.}(2025)\citenamefont
  {Bartschat}, \citenamefont {Bray},\ and\ \citenamefont
  {Fursa}}]{Bartschat-Atoms-25}%
  \BibitemOpen
  \bibfield  {author} {\bibinfo {author} {\bibfnamefont {K.}~\bibnamefont
  {Bartschat}}, \bibinfo {author} {\bibfnamefont {I.}~\bibnamefont {Bray}},\
  and\ \bibinfo {author} {\bibfnamefont {D.~V.}\ \bibnamefont {Fursa}},\
  }\bibfield  {title} {\bibinfo {title} {Benchmark calculations for
  near-threshold electron-impact excitation of the {$(1s3s)\,{}^{3,1}S$} states
  of helium},\ }\href {https://doi.org/10.3390/atoms13040027} {\bibfield
  {journal} {\bibinfo  {journal} {Atoms}\ }\textbf {\bibinfo {volume} {13}},\
  \bibinfo {pages} {27} (\bibinfo {year} {2025})}\BibitemShut {NoStop}%
\bibitem [{\citenamefont {Lange}\ \emph {et~al.}(2006)\citenamefont {Lange},
  \citenamefont {Matsumoto}, \citenamefont {Lower}, \citenamefont {Buckman},
  \citenamefont {Zatsarinny}, \citenamefont {Bartschat}, \citenamefont {Bray},\
  and\ \citenamefont {Fursa}}]{Lange-JPB-06}%
  \BibitemOpen
  \bibfield  {author} {\bibinfo {author} {\bibfnamefont {M.}~\bibnamefont
  {Lange}}, \bibinfo {author} {\bibfnamefont {J.}~\bibnamefont {Matsumoto}},
  \bibinfo {author} {\bibfnamefont {J.}~\bibnamefont {Lower}}, \bibinfo
  {author} {\bibfnamefont {S.}~\bibnamefont {Buckman}}, \bibinfo {author}
  {\bibfnamefont {O.}~\bibnamefont {Zatsarinny}}, \bibinfo {author}
  {\bibfnamefont {K.}~\bibnamefont {Bartschat}}, \bibinfo {author}
  {\bibfnamefont {I.}~\bibnamefont {Bray}},\ and\ \bibinfo {author}
  {\bibfnamefont {D.}~\bibnamefont {Fursa}},\ }\bibfield  {title} {\bibinfo
  {title} {Benchmark experiment and theory for near-threshold excitation of
  helium by electron impact},\ }\href
  {https://doi.org/10.1088/0953-4075/39/20/016} {\bibfield  {journal} {\bibinfo
   {journal} {J. Phys. B: At. Mol. Opt. Phys.}\ }\textbf {\bibinfo {volume}
  {39}},\ \bibinfo {pages} {4179} (\bibinfo {year} {2006})}\BibitemShut
  {NoStop}%
\bibitem [{\citenamefont {Hoshino}\ \emph {et~al.}(2009)\citenamefont
  {Hoshino}, \citenamefont {Kato}, \citenamefont {Tanaka}, \citenamefont
  {Bray}, \citenamefont {Fursa}, \citenamefont {Buckman}, \citenamefont
  {Ingólfsson},\ and\ \citenamefont {Brunger}}]{Hoshino-JPB-09}%
  \BibitemOpen
  \bibfield  {author} {\bibinfo {author} {\bibfnamefont {M.}~\bibnamefont
  {Hoshino}}, \bibinfo {author} {\bibfnamefont {H.}~\bibnamefont {Kato}},
  \bibinfo {author} {\bibfnamefont {H.}~\bibnamefont {Tanaka}}, \bibinfo
  {author} {\bibfnamefont {I.}~\bibnamefont {Bray}}, \bibinfo {author}
  {\bibfnamefont {D.~V.}\ \bibnamefont {Fursa}}, \bibinfo {author}
  {\bibfnamefont {S.~J.}\ \bibnamefont {Buckman}}, \bibinfo {author}
  {\bibfnamefont {O.}~\bibnamefont {Ingólfsson}},\ and\ \bibinfo {author}
  {\bibfnamefont {M.~J.}\ \bibnamefont {Brunger}},\ }\bibfield  {title}
  {\bibinfo {title} {Benchmark differential cross sections for electron impact
  excitation of the n = 2 states in helium at near-ionization- threshold
  energies},\ }\href {https://doi.org/10.1088/0953-4075/42/14/145202}
  {\bibfield  {journal} {\bibinfo  {journal} {J. Phys. B: At. Mol. Opt. Phys.}\
  }\textbf {\bibinfo {volume} {42}},\ \bibinfo {pages} {145202} (\bibinfo
  {year} {2009})}\BibitemShut {NoStop}%
\bibitem [{\citenamefont {Zeng}\ \emph {et~al.}(2015)\citenamefont {Zeng},
  \citenamefont {Gao}, \citenamefont {Han},\ and\ \citenamefont
  {Li}}]{Zeng-PRA-15}%
  \BibitemOpen
  \bibfield  {author} {\bibinfo {author} {\bibfnamefont {D.-L.}\ \bibnamefont
  {Zeng}}, \bibinfo {author} {\bibfnamefont {X.}~\bibnamefont {Gao}}, \bibinfo
  {author} {\bibfnamefont {X.-Y.}\ \bibnamefont {Han}},\ and\ \bibinfo {author}
  {\bibfnamefont {J.-M.}\ \bibnamefont {Li}},\ }\bibfield  {title} {\bibinfo
  {title} {Precision calculation of low-energy electron-impact excitation cross
  sections of helium among the ground and excited states},\ }\href
  {https://doi.org/10.1103/PhysRevA.91.022707} {\bibfield  {journal} {\bibinfo
  {journal} {Phys. Rev. A}\ }\textbf {\bibinfo {volume} {91}},\ \bibinfo
  {pages} {022707} (\bibinfo {year} {2015})}\BibitemShut {NoStop}%
\bibitem [{\citenamefont {Branchett}\ \emph {et~al.}(1990)\citenamefont
  {Branchett}, \citenamefont {Tennyson},\ and\ \citenamefont
  {Morgan}}]{Branchett-JPB-90}%
  \BibitemOpen
  \bibfield  {author} {\bibinfo {author} {\bibfnamefont {S.~E.}\ \bibnamefont
  {Branchett}}, \bibinfo {author} {\bibfnamefont {J.}~\bibnamefont
  {Tennyson}},\ and\ \bibinfo {author} {\bibfnamefont {L.~A.}\ \bibnamefont
  {Morgan}},\ }\bibfield  {title} {\bibinfo {title} {Electronic excitation of
  molecular hydrogen using the {$R$}-matrix method},\ }\href
  {https://doi.org/10.1088/0953-4075/23/24/017} {\bibfield  {journal} {\bibinfo
   {journal} {J. Phys. B: At. Mol. Opt. Phys.}\ }\textbf {\bibinfo {volume}
  {23}},\ \bibinfo {pages} {4625} (\bibinfo {year} {1990})}\BibitemShut
  {NoStop}%
\bibitem [{\citenamefont {Wrkich}\ \emph {et~al.}(2002)\citenamefont {Wrkich},
  \citenamefont {Mathews}, \citenamefont {Kanik}, \citenamefont {Trajmar},\
  and\ \citenamefont {Khakoo}}]{Wrkich-JPB-02}%
  \BibitemOpen
  \bibfield  {author} {\bibinfo {author} {\bibfnamefont {J.}~\bibnamefont
  {Wrkich}}, \bibinfo {author} {\bibfnamefont {D.}~\bibnamefont {Mathews}},
  \bibinfo {author} {\bibfnamefont {I.}~\bibnamefont {Kanik}}, \bibinfo
  {author} {\bibfnamefont {S.}~\bibnamefont {Trajmar}},\ and\ \bibinfo {author}
  {\bibfnamefont {M.~A.}\ \bibnamefont {Khakoo}},\ }\bibfield  {title}
  {\bibinfo {title} {Differential cross-sections for the electron impact
  excitation of the {$B,{}^{1}\Sigma_{u}^{+}$}, {$c,{}^{3}\Pi_{u}$},
  {$a,{}^{3}\Sigma_{g}^{+}$}, {$C,{}^{1}\Pi_{u}$},
  {$E,F,{}^{1}\Sigma_{g}^{+}$}, and {$e,{}^{3}\Sigma_{u}^{+}$} states of
  molecular hydrogen},\ }\href {https://doi.org/10.1088/0953-4075/35/22/311}
  {\bibfield  {journal} {\bibinfo  {journal} {J. Phys. B: At. Mol. Opt. Phys.}\
  }\textbf {\bibinfo {volume} {35}},\ \bibinfo {pages} {4695} (\bibinfo {year}
  {2002})}\BibitemShut {NoStop}%
\bibitem [{\citenamefont {Yoon}\ \emph {et~al.}(2008)\citenamefont {Yoon},
  \citenamefont {Song}, \citenamefont {Han}, \citenamefont {Hwang},
  \citenamefont {Chang}, \citenamefont {Lee},\ and\ \citenamefont
  {Itikawa}}]{Yoon-JPCRD-08}%
  \BibitemOpen
  \bibfield  {author} {\bibinfo {author} {\bibfnamefont {J.-S.}\ \bibnamefont
  {Yoon}}, \bibinfo {author} {\bibfnamefont {M.-Y.}\ \bibnamefont {Song}},
  \bibinfo {author} {\bibfnamefont {J.-M.}\ \bibnamefont {Han}}, \bibinfo
  {author} {\bibfnamefont {S.~H.}\ \bibnamefont {Hwang}}, \bibinfo {author}
  {\bibfnamefont {W.-S.}\ \bibnamefont {Chang}}, \bibinfo {author}
  {\bibfnamefont {B.}~\bibnamefont {Lee}},\ and\ \bibinfo {author}
  {\bibfnamefont {Y.}~\bibnamefont {Itikawa}},\ }\bibfield  {title} {\bibinfo
  {title} {Cross sections for electron collisions with hydrogen molecules},\
  }\href {https://doi.org/10.1063/1.2838023} {\bibfield  {journal} {\bibinfo
  {journal} {J. Phys. Chem. Ref. Data}\ }\textbf {\bibinfo {volume} {37}},\
  \bibinfo {pages} {913} (\bibinfo {year} {2008})}\BibitemShut {NoStop}%
\bibitem [{\citenamefont {Zawadzki}\ \emph {et~al.}(2018)\citenamefont
  {Zawadzki}, \citenamefont {Wright}, \citenamefont {Dolmat}, \citenamefont
  {Martin}, \citenamefont {Hargreaves}, \citenamefont {Fursa}, \citenamefont
  {Zammit}, \citenamefont {Scarlett}, \citenamefont {Tapley}, \citenamefont
  {Savage}, \citenamefont {Bray},\ and\ \citenamefont
  {Khakoo}}]{Zawadzki-PRA-18}%
  \BibitemOpen
  \bibfield  {author} {\bibinfo {author} {\bibfnamefont {M.}~\bibnamefont
  {Zawadzki}}, \bibinfo {author} {\bibfnamefont {R.}~\bibnamefont {Wright}},
  \bibinfo {author} {\bibfnamefont {G.}~\bibnamefont {Dolmat}}, \bibinfo
  {author} {\bibfnamefont {M.~F.}\ \bibnamefont {Martin}}, \bibinfo {author}
  {\bibfnamefont {L.}~\bibnamefont {Hargreaves}}, \bibinfo {author}
  {\bibfnamefont {D.~V.}\ \bibnamefont {Fursa}}, \bibinfo {author}
  {\bibfnamefont {M.~C.}\ \bibnamefont {Zammit}}, \bibinfo {author}
  {\bibfnamefont {L.~H.}\ \bibnamefont {Scarlett}}, \bibinfo {author}
  {\bibfnamefont {J.~K.}\ \bibnamefont {Tapley}}, \bibinfo {author}
  {\bibfnamefont {J.~S.}\ \bibnamefont {Savage}}, \bibinfo {author}
  {\bibfnamefont {I.}~\bibnamefont {Bray}},\ and\ \bibinfo {author}
  {\bibfnamefont {M.~A.}\ \bibnamefont {Khakoo}},\ }\bibfield  {title}
  {\bibinfo {title} {Time-of-flight electron scattering from molecular
  hydrogen: Benchmark cross sections for excitation of the
  {$X{\phantom{\rule{0.16em}{0ex}}}^{1}{\mathrm{\ensuremath{\Sigma}}}_{g}^{+}\ensuremath{\rightarrow}b{\phantom{\rule{0.16em}{0ex}}}^{3}{\mathrm{\ensuremath{\Sigma}}}_{u}^{+}$}
  transition},\ }\href {https://doi.org/10.1103/PhysRevA.97.050702} {\bibfield
  {journal} {\bibinfo  {journal} {Phys. Rev. A}\ }\textbf {\bibinfo {volume}
  {97}},\ \bibinfo {pages} {050702(R)} (\bibinfo {year} {2018})}\BibitemShut
  {NoStop}%
\bibitem [{\citenamefont {Meltzer}\ \emph {et~al.}(2020)\citenamefont
  {Meltzer}, \citenamefont {Tennyson}, \citenamefont {Mašín}, \citenamefont
  {Zammit}, \citenamefont {Scarlett}, \citenamefont {Fursa},\ and\
  \citenamefont {Bray}}]{Meltzer-JPB-20}%
  \BibitemOpen
  \bibfield  {author} {\bibinfo {author} {\bibfnamefont {T.}~\bibnamefont
  {Meltzer}}, \bibinfo {author} {\bibfnamefont {J.}~\bibnamefont {Tennyson}},
  \bibinfo {author} {\bibfnamefont {Z.}~\bibnamefont {Mašín}}, \bibinfo
  {author} {\bibfnamefont {M.~C.}\ \bibnamefont {Zammit}}, \bibinfo {author}
  {\bibfnamefont {L.~H.}\ \bibnamefont {Scarlett}}, \bibinfo {author}
  {\bibfnamefont {D.~V.}\ \bibnamefont {Fursa}},\ and\ \bibinfo {author}
  {\bibfnamefont {I.}~\bibnamefont {Bray}},\ }\bibfield  {title} {\bibinfo
  {title} {Benchmark calculations of electron impact electronic excitation of
  the hydrogen molecule},\ }\href {https://doi.org/10.1088/1361-6455/ab8c58}
  {\bibfield  {journal} {\bibinfo  {journal} {J. Phys. B: At. Mol. Opt. Phys.}\
  }\textbf {\bibinfo {volume} {53}},\ \bibinfo {pages} {145204} (\bibinfo
  {year} {2020})}\BibitemShut {NoStop}%
\bibitem [{\citenamefont {Scarlett}\ \emph {et~al.}(2021)\citenamefont
  {Scarlett}, \citenamefont {Fursa}, \citenamefont {Zammit}, \citenamefont
  {Bray}, \citenamefont {Ralchenko},\ and\ \citenamefont
  {Davie}}]{Scarlett-ADNDT-21}%
  \BibitemOpen
  \bibfield  {author} {\bibinfo {author} {\bibfnamefont {L.~H.}\ \bibnamefont
  {Scarlett}}, \bibinfo {author} {\bibfnamefont {D.~V.}\ \bibnamefont {Fursa}},
  \bibinfo {author} {\bibfnamefont {M.~C.}\ \bibnamefont {Zammit}}, \bibinfo
  {author} {\bibfnamefont {I.}~\bibnamefont {Bray}}, \bibinfo {author}
  {\bibfnamefont {Y.}~\bibnamefont {Ralchenko}},\ and\ \bibinfo {author}
  {\bibfnamefont {K.~D.}\ \bibnamefont {Davie}},\ }\bibfield  {title} {\bibinfo
  {title} {Complete collision data set for electrons scattering on molecular
  hydrogen and its isotopologues: {I}. fully vibrationally-resolved electronic
  excitation of {H$_2$}({$X\,{}^1\Sigma_g^+$})},\ }\href
  {https://doi.org/10.1016/j.adt.2020.101361} {\bibfield  {journal} {\bibinfo
  {journal} {At. Data Nucl. Data Tables}\ }\textbf {\bibinfo {volume} {137}},\
  \bibinfo {pages} {101361} (\bibinfo {year} {2021})}\BibitemShut {NoStop}%
\bibitem [{\citenamefont {Scarlett}\ \emph {et~al.}(2023)\citenamefont
  {Scarlett}, \citenamefont {Rehill}, \citenamefont {Zammit}, \citenamefont
  {Mori}, \citenamefont {Bray},\ and\ \citenamefont {Fursa}}]{Scarlett-PRA-23}%
  \BibitemOpen
  \bibfield  {author} {\bibinfo {author} {\bibfnamefont {L.~H.}\ \bibnamefont
  {Scarlett}}, \bibinfo {author} {\bibfnamefont {U.~S.}\ \bibnamefont
  {Rehill}}, \bibinfo {author} {\bibfnamefont {M.~C.}\ \bibnamefont {Zammit}},
  \bibinfo {author} {\bibfnamefont {N.~A.}\ \bibnamefont {Mori}}, \bibinfo
  {author} {\bibfnamefont {I.}~\bibnamefont {Bray}},\ and\ \bibinfo {author}
  {\bibfnamefont {D.~V.}\ \bibnamefont {Fursa}},\ }\bibfield  {title} {\bibinfo
  {title} {Elastic scattering and rotational excitation of {$\mathrm{H}_2$} by
  electron impact: Convergent close-coupling calculations},\ }\href
  {https://doi.org/10.1103/PhysRevA.107.062804} {\bibfield  {journal} {\bibinfo
   {journal} {Phys. Rev. A}\ }\textbf {\bibinfo {volume} {107}},\ \bibinfo
  {pages} {062804} (\bibinfo {year} {2023})}\BibitemShut {NoStop}%
\bibitem [{\citenamefont {Colgan}\ \emph {et~al.}(2008)\citenamefont {Colgan},
  \citenamefont {Pindzola}, \citenamefont {Robicheaux}, \citenamefont {Kaiser},
  \citenamefont {Murray},\ and\ \citenamefont {Madison}}]{Colgan-PRL-08}%
  \BibitemOpen
  \bibfield  {author} {\bibinfo {author} {\bibfnamefont {J.}~\bibnamefont
  {Colgan}}, \bibinfo {author} {\bibfnamefont {M.~S.}\ \bibnamefont
  {Pindzola}}, \bibinfo {author} {\bibfnamefont {F.}~\bibnamefont
  {Robicheaux}}, \bibinfo {author} {\bibfnamefont {C.}~\bibnamefont {Kaiser}},
  \bibinfo {author} {\bibfnamefont {A.~J.}\ \bibnamefont {Murray}},\ and\
  \bibinfo {author} {\bibfnamefont {D.~H.}\ \bibnamefont {Madison}},\
  }\bibfield  {title} {\bibinfo {title} {Differential cross sections for the
  ionization of oriented {$\mathrm{H}_2$} molecules by electron impact},\
  }\href {https://doi.org/10.1103/PhysRevLett.101.233201} {\bibfield  {journal}
  {\bibinfo  {journal} {Phys. Rev. Lett.}\ }\textbf {\bibinfo {volume} {101}},\
  \bibinfo {pages} {233201} (\bibinfo {year} {2008})}\BibitemShut {NoStop}%
\bibitem [{\citenamefont {Bellm}\ \emph {et~al.}(2010)\citenamefont {Bellm},
  \citenamefont {Lower}, \citenamefont {Weigold},\ and\ \citenamefont
  {Mueller}}]{Bellm-PRL-10}%
  \BibitemOpen
  \bibfield  {author} {\bibinfo {author} {\bibfnamefont {S.}~\bibnamefont
  {Bellm}}, \bibinfo {author} {\bibfnamefont {J.}~\bibnamefont {Lower}},
  \bibinfo {author} {\bibfnamefont {E.}~\bibnamefont {Weigold}},\ and\ \bibinfo
  {author} {\bibfnamefont {D.~W.}\ \bibnamefont {Mueller}},\ }\bibfield
  {title} {\bibinfo {title} {Fully differential molecular-frame measurements
  for the electron-impact dissociative ionization of {$\mathrm{H}_2$}},\ }\href
  {https://doi.org/10.1103/PhysRevLett.104.023202} {\bibfield  {journal}
  {\bibinfo  {journal} {Phys. Rev. Lett.}\ }\textbf {\bibinfo {volume} {104}},\
  \bibinfo {pages} {023202} (\bibinfo {year} {2010})}\BibitemShut {NoStop}%
\bibitem [{\citenamefont {Olsen}(2000)}]{Olsen2000a}%
  \BibitemOpen
  \bibfield  {author} {\bibinfo {author} {\bibfnamefont {J.}~\bibnamefont
  {Olsen}},\ }\href@noop {} {\bibinfo {title} {{LUCIA, a general CI code
  written by J. Olsen, University of Aarhus with contributions from H. Larsen
  and M. F{\"{u}}lscher.}}} (\bibinfo {year} {2000})\BibitemShut {NoStop}%
\bibitem [{\citenamefont {Dvo{\v{r}}{\'a}k}\ \emph {et~al.}(2026)\citenamefont
  {Dvo{\v{r}}{\'a}k}, \citenamefont {Earwood}, \citenamefont {Marante},
  \citenamefont {Chattopadhyay}, \citenamefont {Allison},\ and\ \citenamefont
  {Argenti}}]{PRISM-repository}%
  \BibitemOpen
  \bibfield  {author} {\bibinfo {author} {\bibfnamefont {J.}~\bibnamefont
  {Dvo{\v{r}}{\'a}k}}, \bibinfo {author} {\bibfnamefont {W.}~\bibnamefont
  {Earwood}}, \bibinfo {author} {\bibfnamefont {C.}~\bibnamefont {Marante}},
  \bibinfo {author} {\bibfnamefont {S.}~\bibnamefont {Chattopadhyay}}, \bibinfo
  {author} {\bibfnamefont {T.~C.}\ \bibnamefont {Allison}},\ and\ \bibinfo
  {author} {\bibfnamefont {L.}~\bibnamefont {Argenti}},\ }\bibfield  {title}
  {\bibinfo {title} {{PRISM}: {P}olycentric {R}outines for {I}onization and
  {S}cattering in {M}olecules}} (\bibinfo {year} {2026}),\ \bibinfo {note}
  {forthcoming arXiv preprint, to be submitted to Comput. Phys.
  Commun.}\BibitemShut {Stop}%
\bibitem [{\citenamefont {Dunning}(1989)}]{Dunning-JCP-89}%
  \BibitemOpen
  \bibfield  {author} {\bibinfo {author} {\bibfnamefont {T.~H.}\ \bibnamefont
  {Dunning}},\ }\bibfield  {title} {\bibinfo {title} {Gaussian basis sets for
  use in correlated molecular calculations. {I}. {The} atoms boron through neon
  and hydrogen},\ }\href {https://doi.org/10.1063/1.456153} {\bibfield
  {journal} {\bibinfo  {journal} {J. Chem. Phys.}\ }\textbf {\bibinfo {volume}
  {90}},\ \bibinfo {pages} {1007} (\bibinfo {year} {1989})}\BibitemShut
  {NoStop}%
\bibitem [{\citenamefont {Kendall}\ \emph {et~al.}(1992)\citenamefont
  {Kendall}, \citenamefont {Dunning},\ and\ \citenamefont
  {Harrison}}]{Kendall-JCP-92}%
  \BibitemOpen
  \bibfield  {author} {\bibinfo {author} {\bibfnamefont {R.~A.}\ \bibnamefont
  {Kendall}}, \bibinfo {author} {\bibfnamefont {T.~H.}\ \bibnamefont
  {Dunning}},\ and\ \bibinfo {author} {\bibfnamefont {R.~J.}\ \bibnamefont
  {Harrison}},\ }\bibfield  {title} {\bibinfo {title} {Electron affinities of
  the first-row atoms revisited. {Systematic} basis sets and wave functions},\
  }\href {https://doi.org/10.1063/1.462569} {\bibfield  {journal} {\bibinfo
  {journal} {J. Chem. Phys.}\ }\textbf {\bibinfo {volume} {96}},\ \bibinfo
  {pages} {6796} (\bibinfo {year} {1992})}\BibitemShut {NoStop}%
\bibitem [{\citenamefont {Woon}\ and\ \citenamefont
  {Dunning}(1994)}]{Woon-JCP-94}%
  \BibitemOpen
  \bibfield  {author} {\bibinfo {author} {\bibfnamefont {D.~E.}\ \bibnamefont
  {Woon}}\ and\ \bibinfo {author} {\bibfnamefont {T.~H.}\ \bibnamefont
  {Dunning}},\ }\bibfield  {title} {\bibinfo {title} {Gaussian basis sets for
  use in correlated molecular calculations. {IV}. {Calculation} of static
  electrical response properties},\ }\href {https://doi.org/10.1063/1.466439}
  {\bibfield  {journal} {\bibinfo  {journal} {J. Chem. Phys.}\ }\textbf
  {\bibinfo {volume} {100}},\ \bibinfo {pages} {2975} (\bibinfo {year}
  {1994})}\BibitemShut {NoStop}%
\bibitem [{\citenamefont {Bachau}\ \emph {et~al.}(2001)\citenamefont {Bachau},
  \citenamefont {Cormier}, \citenamefont {Decleva}, \citenamefont {Hansen},\
  and\ \citenamefont {Mart{\'i}n}}]{Bachau-RPP-01}%
  \BibitemOpen
  \bibfield  {author} {\bibinfo {author} {\bibfnamefont {H.}~\bibnamefont
  {Bachau}}, \bibinfo {author} {\bibfnamefont {E.}~\bibnamefont {Cormier}},
  \bibinfo {author} {\bibfnamefont {P.}~\bibnamefont {Decleva}}, \bibinfo
  {author} {\bibfnamefont {J.~E.}\ \bibnamefont {Hansen}},\ and\ \bibinfo
  {author} {\bibfnamefont {F.}~\bibnamefont {Mart{\'i}n}},\ }\bibfield  {title}
  {\bibinfo {title} {Applications of {B}-splines in atomic and molecular
  physics},\ }\href {https://doi.org/10.1088/0034-4885/64/12/205} {\bibfield
  {journal} {\bibinfo  {journal} {Rep. Prog. Phys.}\ }\textbf {\bibinfo
  {volume} {64}},\ \bibinfo {pages} {1815} (\bibinfo {year}
  {2001})}\BibitemShut {NoStop}%
\bibitem [{Note2()}]{Note2}%
  \BibitemOpen
  \bibinfo {note} {1 Bohr = 1 $a_0$ = $5.291\protect \,772\protect
  \,105\protect \,44\times 10^{-11}$ m}\BibitemShut {NoStop}%
\bibitem [{\citenamefont {Aidas}\ \emph {et~al.}(2014)\citenamefont {Aidas},
  \citenamefont {Angeli}, \citenamefont {Bak}, \citenamefont {Bakken},
  \citenamefont {Bast}, \citenamefont {Boman}, \citenamefont {Christiansen},
  \citenamefont {Cimiraglia}, \citenamefont {Coriani}, \citenamefont {Dahle},
  \citenamefont {Dalskov}, \citenamefont {Ekstr{\"{o}}m}, \citenamefont
  {Enevoldsen}, \citenamefont {Eriksen}, \citenamefont {Ettenhuber},
  \citenamefont {Fern{\'{a}}ndez}, \citenamefont {Ferrighi}, \citenamefont
  {Fliegl}, \citenamefont {Frediani}, \citenamefont {Hald}, \citenamefont
  {Halkier}, \citenamefont {H{\"{a}}ttig}, \citenamefont {Heiberg},
  \citenamefont {Helgaker}, \citenamefont {Hennum}, \citenamefont {Hettema},
  \citenamefont {Hjertenaes}, \citenamefont {H{\o}st}, \citenamefont
  {H{\o}yvik}, \citenamefont {Iozzi}, \citenamefont {Jans{\'{i}}k},
  \citenamefont {Jensen}, \citenamefont {Jonsson}, \citenamefont
  {J{\o}rgensen}, \citenamefont {Kauczor}, \citenamefont {Kirpekar},
  \citenamefont {Kjaergaard}, \citenamefont {Klopper}, \citenamefont {Knecht},
  \citenamefont {Kobayashi}, \citenamefont {Koch}, \citenamefont {Kongsted},
  \citenamefont {Krapp}, \citenamefont {Kristensen}, \citenamefont {Ligabue},
  \citenamefont {Lutnaes}, \citenamefont {Melo}, \citenamefont {Mikkelsen},
  \citenamefont {Myhre}, \citenamefont {Neiss}, \citenamefont {Nielsen},
  \citenamefont {Norman}, \citenamefont {Olsen}, \citenamefont {Olsen},
  \citenamefont {Osted}, \citenamefont {Packer}, \citenamefont {Pawlowski},
  \citenamefont {Pedersen}, \citenamefont {Provasi}, \citenamefont {Reine},
  \citenamefont {Rinkevicius}, \citenamefont {Ruden}, \citenamefont {Ruud},
  \citenamefont {Rybkin}, \citenamefont {Sa{\l}ek}, \citenamefont {Samson},
  \citenamefont {de~Mer{\'{a}}s}, \citenamefont {Saue}, \citenamefont {Sauer},
  \citenamefont {Schimmelpfennig}, \citenamefont {Sneskov}, \citenamefont
  {Steindal}, \citenamefont {Sylvester-Hvid}, \citenamefont {Taylor},
  \citenamefont {Teale}, \citenamefont {Tellgren}, \citenamefont {Tew},
  \citenamefont {Thorvaldsen}, \citenamefont {Th{\o}gersen}, \citenamefont
  {Vahtras}, \citenamefont {Watson}, \citenamefont {Wilson}, \citenamefont
  {Ziolkowski},\ and\ \citenamefont {{\AA}gren}}]{DALTON2014}%
  \BibitemOpen
  \bibfield  {author} {\bibinfo {author} {\bibfnamefont {K.}~\bibnamefont
  {Aidas}}, \bibinfo {author} {\bibfnamefont {C.}~\bibnamefont {Angeli}},
  \bibinfo {author} {\bibfnamefont {K.~L.}\ \bibnamefont {Bak}}, \bibinfo
  {author} {\bibfnamefont {V.}~\bibnamefont {Bakken}}, \bibinfo {author}
  {\bibfnamefont {R.}~\bibnamefont {Bast}}, \bibinfo {author} {\bibfnamefont
  {L.}~\bibnamefont {Boman}}, \bibinfo {author} {\bibfnamefont
  {O.}~\bibnamefont {Christiansen}}, \bibinfo {author} {\bibfnamefont
  {R.}~\bibnamefont {Cimiraglia}}, \bibinfo {author} {\bibfnamefont
  {S.}~\bibnamefont {Coriani}}, \bibinfo {author} {\bibfnamefont
  {P.}~\bibnamefont {Dahle}}, \bibinfo {author} {\bibfnamefont {E.~K.}\
  \bibnamefont {Dalskov}}, \bibinfo {author} {\bibfnamefont {U.}~\bibnamefont
  {Ekstr{\"{o}}m}}, \bibinfo {author} {\bibfnamefont {T.}~\bibnamefont
  {Enevoldsen}}, \bibinfo {author} {\bibfnamefont {J.~J.}\ \bibnamefont
  {Eriksen}}, \bibinfo {author} {\bibfnamefont {P.}~\bibnamefont {Ettenhuber}},
  \bibinfo {author} {\bibfnamefont {B.}~\bibnamefont {Fern{\'{a}}ndez}},
  \bibinfo {author} {\bibfnamefont {L.}~\bibnamefont {Ferrighi}}, \bibinfo
  {author} {\bibfnamefont {H.}~\bibnamefont {Fliegl}}, \bibinfo {author}
  {\bibfnamefont {L.}~\bibnamefont {Frediani}}, \bibinfo {author}
  {\bibfnamefont {K.}~\bibnamefont {Hald}}, \bibinfo {author} {\bibfnamefont
  {A.}~\bibnamefont {Halkier}}, \bibinfo {author} {\bibfnamefont
  {C.}~\bibnamefont {H{\"{a}}ttig}}, \bibinfo {author} {\bibfnamefont
  {H.}~\bibnamefont {Heiberg}}, \bibinfo {author} {\bibfnamefont
  {T.}~\bibnamefont {Helgaker}}, \bibinfo {author} {\bibfnamefont {A.~C.}\
  \bibnamefont {Hennum}}, \bibinfo {author} {\bibfnamefont {H.}~\bibnamefont
  {Hettema}}, \bibinfo {author} {\bibfnamefont {E.}~\bibnamefont {Hjertenaes}},
  \bibinfo {author} {\bibfnamefont {S.}~\bibnamefont {H{\o}st}}, \bibinfo
  {author} {\bibfnamefont {I.-M.}\ \bibnamefont {H{\o}yvik}}, \bibinfo {author}
  {\bibfnamefont {M.~F.}\ \bibnamefont {Iozzi}}, \bibinfo {author}
  {\bibfnamefont {B.}~\bibnamefont {Jans{\'{i}}k}}, \bibinfo {author}
  {\bibfnamefont {H.~J.~A.}\ \bibnamefont {Jensen}}, \bibinfo {author}
  {\bibfnamefont {D.}~\bibnamefont {Jonsson}}, \bibinfo {author} {\bibfnamefont
  {P.}~\bibnamefont {J{\o}rgensen}}, \bibinfo {author} {\bibfnamefont
  {J.}~\bibnamefont {Kauczor}}, \bibinfo {author} {\bibfnamefont
  {S.}~\bibnamefont {Kirpekar}}, \bibinfo {author} {\bibfnamefont
  {T.}~\bibnamefont {Kjaergaard}}, \bibinfo {author} {\bibfnamefont
  {W.}~\bibnamefont {Klopper}}, \bibinfo {author} {\bibfnamefont
  {S.}~\bibnamefont {Knecht}}, \bibinfo {author} {\bibfnamefont
  {R.}~\bibnamefont {Kobayashi}}, \bibinfo {author} {\bibfnamefont
  {H.}~\bibnamefont {Koch}}, \bibinfo {author} {\bibfnamefont {J.}~\bibnamefont
  {Kongsted}}, \bibinfo {author} {\bibfnamefont {A.}~\bibnamefont {Krapp}},
  \bibinfo {author} {\bibfnamefont {K.}~\bibnamefont {Kristensen}}, \bibinfo
  {author} {\bibfnamefont {A.}~\bibnamefont {Ligabue}}, \bibinfo {author}
  {\bibfnamefont {O.~B.}\ \bibnamefont {Lutnaes}}, \bibinfo {author}
  {\bibfnamefont {J.~I.}\ \bibnamefont {Melo}}, \bibinfo {author}
  {\bibfnamefont {K.~V.}\ \bibnamefont {Mikkelsen}}, \bibinfo {author}
  {\bibfnamefont {R.~H.}\ \bibnamefont {Myhre}}, \bibinfo {author}
  {\bibfnamefont {C.}~\bibnamefont {Neiss}}, \bibinfo {author} {\bibfnamefont
  {C.~B.}\ \bibnamefont {Nielsen}}, \bibinfo {author} {\bibfnamefont
  {P.}~\bibnamefont {Norman}}, \bibinfo {author} {\bibfnamefont
  {J.}~\bibnamefont {Olsen}}, \bibinfo {author} {\bibfnamefont {J.~M.~H.}\
  \bibnamefont {Olsen}}, \bibinfo {author} {\bibfnamefont {A.}~\bibnamefont
  {Osted}}, \bibinfo {author} {\bibfnamefont {M.~J.}\ \bibnamefont {Packer}},
  \bibinfo {author} {\bibfnamefont {F.}~\bibnamefont {Pawlowski}}, \bibinfo
  {author} {\bibfnamefont {T.~B.}\ \bibnamefont {Pedersen}}, \bibinfo {author}
  {\bibfnamefont {P.~F.}\ \bibnamefont {Provasi}}, \bibinfo {author}
  {\bibfnamefont {S.}~\bibnamefont {Reine}}, \bibinfo {author} {\bibfnamefont
  {Z.}~\bibnamefont {Rinkevicius}}, \bibinfo {author} {\bibfnamefont {T.~A.}\
  \bibnamefont {Ruden}}, \bibinfo {author} {\bibfnamefont {K.}~\bibnamefont
  {Ruud}}, \bibinfo {author} {\bibfnamefont {V.~V.}\ \bibnamefont {Rybkin}},
  \bibinfo {author} {\bibfnamefont {P.}~\bibnamefont {Sa{\l}ek}}, \bibinfo
  {author} {\bibfnamefont {C.~C.~M.}\ \bibnamefont {Samson}}, \bibinfo {author}
  {\bibfnamefont {A.~S.}\ \bibnamefont {de~Mer{\'{a}}s}}, \bibinfo {author}
  {\bibfnamefont {T.}~\bibnamefont {Saue}}, \bibinfo {author} {\bibfnamefont
  {S.~P.~A.}\ \bibnamefont {Sauer}}, \bibinfo {author} {\bibfnamefont
  {B.}~\bibnamefont {Schimmelpfennig}}, \bibinfo {author} {\bibfnamefont
  {K.}~\bibnamefont {Sneskov}}, \bibinfo {author} {\bibfnamefont {A.~H.}\
  \bibnamefont {Steindal}}, \bibinfo {author} {\bibfnamefont {K.~O.}\
  \bibnamefont {Sylvester-Hvid}}, \bibinfo {author} {\bibfnamefont {P.~R.}\
  \bibnamefont {Taylor}}, \bibinfo {author} {\bibfnamefont {A.~M.}\
  \bibnamefont {Teale}}, \bibinfo {author} {\bibfnamefont {E.~I.}\ \bibnamefont
  {Tellgren}}, \bibinfo {author} {\bibfnamefont {D.~P.}\ \bibnamefont {Tew}},
  \bibinfo {author} {\bibfnamefont {A.~J.}\ \bibnamefont {Thorvaldsen}},
  \bibinfo {author} {\bibfnamefont {L.}~\bibnamefont {Th{\o}gersen}}, \bibinfo
  {author} {\bibfnamefont {O.}~\bibnamefont {Vahtras}}, \bibinfo {author}
  {\bibfnamefont {M.~A.}\ \bibnamefont {Watson}}, \bibinfo {author}
  {\bibfnamefont {D.~J.~D.}\ \bibnamefont {Wilson}}, \bibinfo {author}
  {\bibfnamefont {M.}~\bibnamefont {Ziolkowski}},\ and\ \bibinfo {author}
  {\bibfnamefont {H.}~\bibnamefont {{\AA}gren}},\ }\bibfield  {title} {\bibinfo
  {title} {{The Dalton quantum chemistry program system}},\ }\href
  {https://doi.org/10.1002/wcms.1172} {\bibfield  {journal} {\bibinfo
  {journal} {Wiley Interdiscip. Rev. Comput. Mol. Sci.}\ }\textbf {\bibinfo
  {volume} {4}},\ \bibinfo {pages} {269} (\bibinfo {year} {2014})}\BibitemShut
  {NoStop}%
\bibitem [{\citenamefont {Olsen}\ \emph {et~al.}(1988)\citenamefont {Olsen},
  \citenamefont {Roos}, \citenamefont {J{\o}rgensen},\ and\ \citenamefont
  {Jensen}}]{Olsen1988}%
  \BibitemOpen
  \bibfield  {author} {\bibinfo {author} {\bibfnamefont {J.}~\bibnamefont
  {Olsen}}, \bibinfo {author} {\bibfnamefont {B.~O.}\ \bibnamefont {Roos}},
  \bibinfo {author} {\bibfnamefont {P.}~\bibnamefont {J{\o}rgensen}},\ and\
  \bibinfo {author} {\bibfnamefont {H.~J.~A.}\ \bibnamefont {Jensen}},\
  }\bibfield  {title} {\bibinfo {title} {{Determinant based configuration
  interaction algorithms for complete and restricted configuration interaction
  spaces}},\ }\href {https://doi.org/10.1063/1.455063} {\bibfield  {journal}
  {\bibinfo  {journal} {The J. Chem. Phys.}\ }\textbf {\bibinfo {volume}
  {89}},\ \bibinfo {pages} {2185} (\bibinfo {year} {1988})}\BibitemShut
  {NoStop}%
\bibitem [{\citenamefont {Stepanović}\ \emph {et~al.}(2006)\citenamefont
  {Stepanović}, \citenamefont {Minić}, \citenamefont {Cvejanović},
  \citenamefont {Jureta}, \citenamefont {Kurepa}, \citenamefont {Cvejanović},
  \citenamefont {Zatsarinny},\ and\ \citenamefont
  {Bartschat}}]{Stepanovic-JPB-06}%
  \BibitemOpen
  \bibfield  {author} {\bibinfo {author} {\bibfnamefont {M.}~\bibnamefont
  {Stepanović}}, \bibinfo {author} {\bibfnamefont {M.}~\bibnamefont {Minić}},
  \bibinfo {author} {\bibfnamefont {D.}~\bibnamefont {Cvejanović}}, \bibinfo
  {author} {\bibfnamefont {J.}~\bibnamefont {Jureta}}, \bibinfo {author}
  {\bibfnamefont {J.}~\bibnamefont {Kurepa}}, \bibinfo {author} {\bibfnamefont
  {S.}~\bibnamefont {Cvejanović}}, \bibinfo {author} {\bibfnamefont
  {O.}~\bibnamefont {Zatsarinny}},\ and\ \bibinfo {author} {\bibfnamefont
  {K.}~\bibnamefont {Bartschat}},\ }\bibfield  {title} {\bibinfo {title}
  {Integral cross sections for electron-impact excitation of the 33s and 31s
  states of he near threshold},\ }\href
  {https://doi.org/10.1088/0953-4075/39/6/021} {\bibfield  {journal} {\bibinfo
  {journal} {J. Phys. B: At. Mol. Opt. Phys.}\ }\textbf {\bibinfo {volume}
  {39}},\ \bibinfo {pages} {1547} (\bibinfo {year} {2006})}\BibitemShut
  {NoStop}%
\bibitem [{\citenamefont {Kramida}\ \emph {et~al.}(2024)\citenamefont
  {Kramida}, \citenamefont {Ralchenko}, \citenamefont {Reader},\ and\
  \citenamefont {{{NIST ASD Team}}}}]{NIST_ASD}%
  \BibitemOpen
  \bibfield  {author} {\bibinfo {author} {\bibfnamefont {A.}~\bibnamefont
  {Kramida}}, \bibinfo {author} {\bibfnamefont {Y.}~\bibnamefont {Ralchenko}},
  \bibinfo {author} {\bibfnamefont {J.}~\bibnamefont {Reader}},\ and\ \bibinfo
  {author} {\bibnamefont {{{NIST ASD Team}}}},\ }\href
  {https://doi.org/10.18434/T4W30F} {\bibinfo {title} {{NIST Atomic Spectra
  Database}}} (\bibinfo {year} {2024}),\ \bibinfo {note} {version 5.12,
  National Institute of Standards and Technology, Gaithersburg, MD, USA;
  accessed 14 August 2026}\BibitemShut {NoStop}%
\bibitem [{\citenamefont {Nesbet}(1979)}]{Nesbet-PRA-79}%
  \BibitemOpen
  \bibfield  {author} {\bibinfo {author} {\bibfnamefont {R.~K.}\ \bibnamefont
  {Nesbet}},\ }\bibfield  {title} {\bibinfo {title} {Variational calculations
  of accurate {${e}^{\ensuremath{-}}$-He} cross sections below 19 {eV}},\
  }\href {https://doi.org/10.1103/PhysRevA.20.58} {\bibfield  {journal}
  {\bibinfo  {journal} {Phys. Rev. A}\ }\textbf {\bibinfo {volume} {20}},\
  \bibinfo {pages} {58} (\bibinfo {year} {1979})}\BibitemShut {NoStop}%
\bibitem [{\citenamefont {Fursa}\ and\ \citenamefont
  {Bray}(1995)}]{Fursa-PRA-95}%
  \BibitemOpen
  \bibfield  {author} {\bibinfo {author} {\bibfnamefont {D.~V.}\ \bibnamefont
  {Fursa}}\ and\ \bibinfo {author} {\bibfnamefont {I.}~\bibnamefont {Bray}},\
  }\bibfield  {title} {\bibinfo {title} {Calculation of electron-helium
  scattering},\ }\href {https://doi.org/10.1103/PhysRevA.52.1279} {\bibfield
  {journal} {\bibinfo  {journal} {Phys. Rev. A}\ }\textbf {\bibinfo {volume}
  {52}},\ \bibinfo {pages} {1279} (\bibinfo {year} {1995})}\BibitemShut
  {NoStop}%
\bibitem [{\citenamefont {Buckman}\ and\ \citenamefont
  {Clark}(1994)}]{Buckman-RMP-94}%
  \BibitemOpen
  \bibfield  {author} {\bibinfo {author} {\bibfnamefont {S.~J.}\ \bibnamefont
  {Buckman}}\ and\ \bibinfo {author} {\bibfnamefont {C.~W.}\ \bibnamefont
  {Clark}},\ }\bibfield  {title} {\bibinfo {title} {Atomic negative-ion
  resonances},\ }\href {https://doi.org/10.1103/RevModPhys.66.539} {\bibfield
  {journal} {\bibinfo  {journal} {Rev. Mod. Phys.}\ }\textbf {\bibinfo {volume}
  {66}},\ \bibinfo {pages} {539} (\bibinfo {year} {1994})}\BibitemShut
  {NoStop}%
\bibitem [{\citenamefont {Simon}(1979)}]{Simon-PLA-79}%
  \BibitemOpen
  \bibfield  {author} {\bibinfo {author} {\bibfnamefont {B.}~\bibnamefont
  {Simon}},\ }\bibfield  {title} {\bibinfo {title} {The definition of molecular
  resonance curves by the method of exterior complex scaling},\ }\href
  {https://doi.org/10.1016/0375-9601(79)90165-8} {\bibfield  {journal}
  {\bibinfo  {journal} {Phys. Lett. A}\ }\textbf {\bibinfo {volume} {71}},\
  \bibinfo {pages} {211} (\bibinfo {year} {1979})}\BibitemShut {NoStop}%
\bibitem [{\citenamefont {McCurdy}\ \emph {et~al.}(2004)\citenamefont
  {McCurdy}, \citenamefont {Baertschy},\ and\ \citenamefont
  {Rescigno}}]{McCurdy-JPB-04}%
  \BibitemOpen
  \bibfield  {author} {\bibinfo {author} {\bibfnamefont {C.~W.}\ \bibnamefont
  {McCurdy}}, \bibinfo {author} {\bibfnamefont {M.}~\bibnamefont {Baertschy}},\
  and\ \bibinfo {author} {\bibfnamefont {T.~N.}\ \bibnamefont {Rescigno}},\
  }\bibfield  {title} {\bibinfo {title} {Solving the three-body {Coulomb}
  breakup problem using exterior complex scaling},\ }\href
  {https://doi.org/10.1088/0953-4075/37/17/R01} {\bibfield  {journal} {\bibinfo
   {journal} {J. Phys. B: At. Mol. Opt. Phys.}\ }\textbf {\bibinfo {volume}
  {37}},\ \bibinfo {pages} {R137} (\bibinfo {year} {2004})}\BibitemShut
  {NoStop}%
\bibitem [{\citenamefont {Fano}(1961)}]{Fano-PR-61}%
  \BibitemOpen
  \bibfield  {author} {\bibinfo {author} {\bibfnamefont {U.}~\bibnamefont
  {Fano}},\ }\bibfield  {title} {\bibinfo {title} {Effects of configuration
  interaction on intensities and phase shifts},\ }\href
  {https://doi.org/10.1103/PhysRev.124.1866} {\bibfield  {journal} {\bibinfo
  {journal} {Phys. Rev.}\ }\textbf {\bibinfo {volume} {124}},\ \bibinfo {pages}
  {1866} (\bibinfo {year} {1961})}\BibitemShut {NoStop}%
\bibitem [{\citenamefont {Muse}\ \emph {et~al.}(2008)\citenamefont {Muse},
  \citenamefont {Silva}, \citenamefont {Lopes},\ and\ \citenamefont
  {Khakoo}}]{Muse-JPB-08}%
  \BibitemOpen
  \bibfield  {author} {\bibinfo {author} {\bibfnamefont {J.}~\bibnamefont
  {Muse}}, \bibinfo {author} {\bibfnamefont {H.}~\bibnamefont {Silva}},
  \bibinfo {author} {\bibfnamefont {M.~C.~A.}\ \bibnamefont {Lopes}},\ and\
  \bibinfo {author} {\bibfnamefont {M.~A.}\ \bibnamefont {Khakoo}},\ }\bibfield
   {title} {\bibinfo {title} {Low energy elastic scattering of electrons from
  {H${{}_{2}}$} and {N${{}_{2}}$}},\ }\href
  {https://doi.org/10.1088/0953-4075/41/9/095203} {\bibfield  {journal}
  {\bibinfo  {journal} {J. Phys. B: At. Mol. Opt. Phys.}\ }\textbf {\bibinfo
  {volume} {41}},\ \bibinfo {pages} {095203} (\bibinfo {year}
  {2008})}\BibitemShut {NoStop}%
\bibitem [{\citenamefont {Hagelaar}\ and\ \citenamefont
  {Pitchford}(2005)}]{Hagelaar-PSST-05}%
  \BibitemOpen
  \bibfield  {author} {\bibinfo {author} {\bibfnamefont {G.~J.~M.}\
  \bibnamefont {Hagelaar}}\ and\ \bibinfo {author} {\bibfnamefont {L.~C.}\
  \bibnamefont {Pitchford}},\ }\bibfield  {title} {\bibinfo {title} {Solving
  the {Boltzmann} equation to obtain electron transport coefficients and rate
  coefficients for fluid models},\ }\href
  {https://doi.org/10.1088/0963-0252/14/4/011} {\bibfield  {journal} {\bibinfo
  {journal} {Plasma Sources Sci. Technol.}\ }\textbf {\bibinfo {volume} {14}},\
  \bibinfo {pages} {722} (\bibinfo {year} {2005})}\BibitemShut {NoStop}%
\bibitem [{\citenamefont {Fleig}\ \emph {et~al.}(2001)\citenamefont {Fleig},
  \citenamefont {Olsen},\ and\ \citenamefont {Marian}}]{Fleig-JCP-01}%
  \BibitemOpen
  \bibfield  {author} {\bibinfo {author} {\bibfnamefont {T.}~\bibnamefont
  {Fleig}}, \bibinfo {author} {\bibfnamefont {J.}~\bibnamefont {Olsen}},\ and\
  \bibinfo {author} {\bibfnamefont {C.~M.}\ \bibnamefont {Marian}},\ }\bibfield
   {title} {\bibinfo {title} {The generalized active space concept for the
  relativistic treatment of electron correlation. {I}. {Kramers}-restricted
  two-component configuration interaction},\ }\href
  {https://doi.org/10.1063/1.1349076} {\bibfield  {journal} {\bibinfo
  {journal} {J. Chem. Phys.}\ }\textbf {\bibinfo {volume} {114}},\ \bibinfo
  {pages} {4775} (\bibinfo {year} {2001})}\BibitemShut {NoStop}%
\bibitem [{\citenamefont {Ma}\ \emph {et~al.}(2011)\citenamefont {Ma},
  \citenamefont {Li~Manni},\ and\ \citenamefont {Gagliardi}}]{Ma-JCP-11}%
  \BibitemOpen
  \bibfield  {author} {\bibinfo {author} {\bibfnamefont {D.}~\bibnamefont
  {Ma}}, \bibinfo {author} {\bibfnamefont {G.}~\bibnamefont {Li~Manni}},\ and\
  \bibinfo {author} {\bibfnamefont {L.}~\bibnamefont {Gagliardi}},\ }\bibfield
  {title} {\bibinfo {title} {The generalized active space concept in
  multiconfigurational self-consistent field methods},\ }\href
  {https://doi.org/10.1063/1.3611401} {\bibfield  {journal} {\bibinfo
  {journal} {J. Chem. Phys.}\ }\textbf {\bibinfo {volume} {135}},\ \bibinfo
  {pages} {044128} (\bibinfo {year} {2011})}\BibitemShut {NoStop}%
\bibitem [{\citenamefont {Varshalovich}\ \emph {et~al.}(1988)\citenamefont
  {Varshalovich}, \citenamefont {Moskalev},\ and\ \citenamefont
  {Khersonskii}}]{Varshalovich1988}%
  \BibitemOpen
  \bibfield  {author} {\bibinfo {author} {\bibfnamefont {D.~A.}\ \bibnamefont
  {Varshalovich}}, \bibinfo {author} {\bibfnamefont {A.~N.}\ \bibnamefont
  {Moskalev}},\ and\ \bibinfo {author} {\bibfnamefont {V.~K.}\ \bibnamefont
  {Khersonskii}},\ }\href {https://doi.org/10.1142/0270} {\emph {\bibinfo
  {title} {Quantum Theory of Angular Momentum}}}\ (\bibinfo  {publisher} {WORLD
  SCIENTIFIC},\ \bibinfo {year} {1988})\BibitemShut {NoStop}%
\bibitem [{\citenamefont {Newton}(2002)}]{Newton2002}%
  \BibitemOpen
  \bibfield  {author} {\bibinfo {author} {\bibfnamefont {R.~G.}\ \bibnamefont
  {Newton}},\ }\href {https://books.google.com/books?id=uYmtAQAAQBAJ} {\emph
  {\bibinfo {title} {Scattering Theory of Waves and Particles}}}\ (\bibinfo
  {publisher} {Dover Publications},\ \bibinfo {year} {2002})\BibitemShut
  {NoStop}%
\end{thebibliography}%
\end{document}